\documentclass[12pt]{revtex4}
\usepackage{epsfig}

\newcommand{\cerr}[3]   {\mbox{${{#1}^{+ #2}_{- #3}}$}}

\begin{document}
%


\title{Searching for New Physics in $b \to s$ Hadronic Penguin Decays}

\author{Luca Silvestrini}
\affiliation{Dip. di Fisica, Univ. di Roma ``La Sapienza'' and INFN,
  Sez. di Roma\\
P.le A. Moro, 2,
I-00185 Rome,
Italy}

\begin{abstract}
  We review the theoretical status of $b \to s$ hadronic penguin
  decays in the Standard Model and beyond. We summarize the main
  theoretical tools to compute Branching Ratios and CP asymmetries for
  $b \to s$ penguin dominated nonleptonic decays, and discuss the
  theoretical uncertainties in the prediction of time-dependent CP
  asymmetries in this processes. We consider general aspects of $b \to
  s$ transitions beyond the Standard Model. Then we present detailed
  predictions in supersymmetric models with new sources of flavor and
  CP violation.
\end{abstract}

\maketitle

\section{INTRODUCTION}

New Physics (NP) can be searched for in two ways: either by raising
the available energy at colliders to produce new particles and reveal
them directly, or by increasing the experimental precision on certain
processes involving Standard Model (SM) particles as external states.
The latter option, indirect search for NP, should be pursued using
processes that are forbidden, very rare or precisely calculable in the
SM. In this respect, Flavor Changing Neutral Current (FCNC) and
CP-violating processes are among the most powerful probes of NP, since
in the SM they cannot arise at the tree-level and even at the loop
level they are strongly suppressed by the GIM mechanism. Furthermore,
in the quark sector they are all calculable in terms of the CKM
matrix, and in particular of the parameters $\bar \rho$ and $\bar
\eta$ in the generalized Wolfenstein
parametrization~\cite{Wolfenstein:1983yz}.
Unfortunately, in many cases a deep understanding of hadronic dynamics
is required in order to be able to extract the relevant short-distance
information from measured processes. Lattice QCD allows us to compute
the necessary hadronic parameters in many processes, for example in
$\Delta F=2$ amplitudes.  Indeed, the Unitarity Triangle Analysis
(UTA) with Lattice QCD input is extremely successful in determining
$\bar \rho$ and $\bar \eta$ and in constraining NP contributions to
$\Delta F=2$
amplitudes~\cite{hep-ph/0501199,hep-ph/0606167,hep-ph/0509219,hep-ph/0605213,hep-ph/0406184}.

Once the CKM matrix is precisely determined by means of the UTA
(either within the SM or allowing for generic NP in $\Delta F=2$
processes), it is possible to search for NP contributions to $\Delta
F=1$ transitions. FCNC and CP-violating hadronic decays are indeed the
most sensitive probes of NP contributions to penguin operators. In
particular, penguin-dominated nonleptonic $B$ decays can reveal the
presence of NP in decay
amplitudes~\cite{hep-ph/9612269,hep-ph/9704274,hep-ph/9704277}. The
dominance of penguin operators is realized in $b \to s q \bar q$
transitions.

Thanks to the efforts of the BaBar and Belle collaborations,
$B$-factories have been able to measure CP violation in several $b \to
s$ penguin-dominated channels with an impressive
accuracy~\cite{hep-ex/0608039,hep-ex/0607112,hep-ex/0609052,hep-ex/0702046,hep-ex/0609006,hep-ex/0607096,hep-ex/0608051,hep-ex/0607101,hep-ex/0408095,hep-ex/0702010,hep-ex/0507016}.
To fully exploit this rich experimental information to test the SM and
look for NP, we need to determine the SM predictions for each channel.
As we shall see in the following, computing the uncertainty in the SM
predictions is an extremely delicate task. Only in very few cases it
is possible to control this uncertainty using only experimental data;
in general, one has to use some dynamical information, either from
flavor symmetries or from factorization.  Computing CP violation in $b
\to s$ penguins beyond the SM is even harder: additional operators
arise, and in many cases the dominant contribution is expected to come
from new operators or from operators that are subdominant in the SM.
In the near future, say before the start of the LHC, we can aim at
establishing possible hints of NP in $b \to s$ penguins.  With the
advent of the LHC, two scenarios are possible. If new particles are
revealed, $b \to s$ penguin decays will help us identify the flavor
structure of the underlying NP model. If no new particles are seen, $b
\to s$ penguins can either indirectly reveal the presence of NP, if
the present hints are confirmed, or allow us to push further the lower
bound on the scale of NP.  In all cases, experimental and theoretical
progress in $b \to s$ hadronic penguins is crucial for our
understanding of flavor physics beyond the SM.

This review is organized as follows. In Sec.~\ref{sec:basic} we
quickly review the basic formalism for $b \to s$ nonleptonic decays,
and the different approaches to the calculation of decay amplitudes
present in the literature. In Sec.~\ref{sec:SM}, we present the
predictions for Branching Ratios (BR's) and CP violation within the SM
following the various approaches, and compare them with the
experimental data. In Sec.~\ref{sec:NP}, we discuss the possible
sources of NP contributions to $b \to s$ penguins and how these NP
contributions are constrained by experimental data on other $b \to s$
transitions. In Sec.~\ref{sec:SUSY}, we concentrate on SUSY
extensions of the SM, discuss the present constraints and present
detailed predictions for CP violation in $b \to s$ penguins. In
Sec.~\ref{sec:non-SUSY} we briefly discuss $b \to s$ penguins in the
context of non-SUSY extensions of the SM. Finally, in
Sec.~\ref{sec:concl} we summarize the present status and discuss
future prospects.

\section{BASIC FORMALISM}
\label{sec:basic}

\subsection{Generalities}
\label{sec:general}

The basic theoretical framework for non-leptonic $B$ decays is based
on the Operator Product Expansion (OPE) and renormalization group
methods which allow to write the amplitude for a decay of a given meson
$B$=$B_d$, $B_s$, $B^+$ into a final state $F$ generally as follows:
\begin{eqnarray}
  \mathcal{A}(B \to F) = \langle F \vert \mathcal{H}_\mathrm{eff} \vert B
  \rangle &=&
  \left(
    \frac {G_F}{\sqrt{2}} \sum_{i=1}^{12} V^\mathrm{CKM}_i C_i(\mu) +
    C^\mathrm{NP}_i(\mu) 
  \right)
  \langle F \vert Q_i(\mu) \vert B \rangle \nonumber \\
  &&+ \sum_{i=1}^{N_\mathrm{NP}} \tilde C^\mathrm{NP}_i(\mu)
  \langle F \vert \tilde Q_i(\mu) \vert B \rangle. 
  \label{eq:ope}
\end{eqnarray}
Here $\mathcal{H}_\mathrm{eff}$ is the effective weak Hamiltonian,
with $Q_i$ denoting the relevant local operators which govern the
decays in question within the SM, and $\tilde Q_i$ denoting the ones
possibly arising beyond the SM. The CKM factors $V^\mathrm{CKM}_i$ and
the Wilson coefficients $C_i(\mu)$ describe the strength with which a
given operator enters the Hamiltonian; for NP contributions, we denote
with $C^\mathrm{NP}_i(\mu)$ and $\tilde C^\mathrm{NP}_i(\mu)$ the
Wilson coefficients arising within a given NP model, which can in
general be complex. In a more intuitive language, the operators
$Q_i(\mu)$ can be regarded as effective vertices and the coefficients
$C_i(\mu)$ as the corresponding effective couplings. The latter can be
calculated in renormalization-group improved perturbation theory and
are known including Next-to-Leading order (NLO) QCD corrections within
the SM and in a few SUSY models
\cite{hep-ph/9806308,hep-ph/9904413,hep-ph/0009337}. The scale $\mu$
separates the contributions to ${\cal A}(B\to F)$ into short-distance
contributions with energy scales higher than $\mu$ contained in
$C_i(\mu)$ and long-distance contributions with energy scales lower
than $\mu$ contained in the hadronic matrix elements $\langle Q_i(\mu)
\rangle$.  The scale $\mu$ is usually chosen to be $O(m_b)$ but is
otherwise arbitrary.

The effective weak Hamiltonian for non-leptonic $b \to s$ decays
within the SM is given by:
\begin{eqnarray} 
{\cal H}_{\rm eff} &=& \frac {4 G_F} {\sqrt{2}} 
\Biggl\{V_{ub} V^*_{us} \biggl[C_1(\mu)\Bigl( Q^{u}_1(\mu) 
  - Q_1^{c}(\mu) \Bigr) + C_2(\mu)\Bigl( Q^{u}_2(\mu) 
  - Q_2^{c}(\mu) \Bigr)  \biggr] \nonumber  \\
&& -V_{tb} V^*_{ts} \, \biggl[C_1(\mu) Q_1^{c}(\mu) + 
C_2(\mu) Q_2^{c}(\mu) + \sum_{i=3,12} C_i(\mu) Q_i(\mu)
\biggr] \Biggr\}\,,
\protect\label{eq:eh} 
\end{eqnarray}
with
\begin{equation}
  \begin{array}{ll}
    Q^{u^i}_{ 1}=({\bar b_L}\gamma^\mu u^i_L)
    ({\bar u^i_L }\gamma_\mu s_L)\,,
    &
    Q^{u^i}_{ 2}=({\bar b_L}\gamma^\mu s_L)
    ({\bar u^i_L }\gamma_\mu u^i_L)\,,
    \\
    Q_{ 3,5} = \sum_q ({\bar b_{L}} \gamma^\mu s_L)
    ({\bar q_{L,R}}\gamma_\mu q_{L,R})\,,
    &
    Q_{ 4} = \sum_q ({\bar b_L}\gamma^\mu q_L)
    ({\bar q_L}\gamma_\mu s_L)\,,
    \\
    Q_{6} = -2 \sum_q ({\bar b_L}q_R)
    ({\bar q_R}s_L)\,,
    &
    Q_{ 7,9} = \frac{3}{2}\sum_q ({\bar b_L}\gamma^\mu s_L) e_{ q}({\bar
      q_L}\gamma_\mu q_L)\,,  
    \\
    Q_{8} = - 3 \sum_q e_q({\bar b_L}q_R) ({\bar q_R}s_L)\,, 
    &
    Q_{ 10} = \frac{3}{2} \sum_q e_{ q}({\bar b_L}\gamma^\mu q_L)
    ({\bar q_L}\gamma_\mu s_L)\,,\\
    Q_{11} =  \frac{e}{16 \pi^2}m_b (\bar b_R \sigma^{\mu\nu} s_L)
    F_{\mu\nu} \,, & 
    Q_{12} =  \frac{g}{16 \pi^2}m_b (\bar b_R \sigma^{\mu\nu} T^a s_L)
    G_{\mu\nu}^a \,, \\
  \end{array}  
  \label{eq:basis}
\end{equation}
where $q_{L,R}\equiv (1\mp\gamma_5)/2 q$, $u^i=\{u,c\}$ and $e_q$ denotes
the quark electric charge ($e_u=2/3$, $e_d=-1/3$, etc.).  The sum over
the quarks $q$ runs over the active flavors at the scale $\mu$.

$Q_1$ and $Q_2$ are the so-called current-current operators, $Q_{3-6}$
the QCD-penguin operators, $Q_{7-10}$ the electroweak penguin
operators and $Q_{11,12}$ the (chromo)-magnetic penguin operators.
$C_i(\mu)$ are the Wilson coefficients evaluated at $\mu = O(m_b)$.
They depend generally on the renormalization scheme for the operators.
The scale and scheme dependence of the coefficients is canceled by the
analogous dependence in the matrix elements. It is therefore
convenient to identify the basic renormalization group invariant
parameters (RGI's) and to express the decay amplitudes in terms of
RGI's. This exercise was performed in ref.~\cite{hep-ph/9812392}, where the
RGI's were identified and the decay amplitudes for several two-body
nonleptonic $B$ decays were written down. For our purpose, we just
need to recall a few basic facts about the classification of RGI's.
First of all, we have six non-penguin parameters, containing only
non-penguin contractions of the current-current operators $Q_{1,2}$:
emission parameters $E_{1,2}$, annihilation parameters $A_{1,2}$ and
Zweig-suppressed emission-annihilation parameters
$\mathit{EA}_{1,2}$. Then, we have four parameters containing only
penguin contractions of the current-current operators $Q_{1,2}$ in the
GIM-suppressed combination $Q_{1,2}^{c}-Q_{1,2}^{u}$: $P_1^\mathrm{GIM}$ and
Zweig suppressed $P_{2-4}^\mathrm{GIM}$. Finally, we have four parameters
containing penguin contractions of current-current operators
$Q_{1,2}^{c}$ (the so-called charming penguins \cite{hep-ph/9703353})
and all possible contractions of penguin operators $Q_{3-12}$:
$P_{1,2}$ and the Zweig-suppressed $P_{3,4}$.

Let us now discuss some important aspects of $b \to s$ penguin
nonleptonic decays. First of all, we define as pure penguin channels
the ones that are generated only by $P_i$ and $P_i^\mathrm{GIM}$
parameters. Pure penguin $b \to s$ decays can be written schematically
as:
\begin{equation}
  \label{eq:purepenguins}
  \mathcal{A}(B \to F) = - V^*_{ub} V_{us} \sum P_i^\mathrm{GIM} -
  V^*_{tb} V_{ts} \sum P_i \,. 
\end{equation}
Neglecting doubly Cabibbo suppressed terms, the decay amplitude has
vanishing weak phase. Therefore, there is no direct CP violation and
the coefficient $S_F$ of the $\sin \Delta m t$ term in the
time-dependent CP asymmetry (for $F$ a CP eigenstate with eigenvalue
$\eta_F$) measures the phase of the mixing amplitude: $S_F = \eta_F
\mathrm{Im}\,\lambda_F = -\eta_F \sin 2 \phi_M$, where
$\lambda_F\equiv \frac{q}{p} \frac{\bar A}{A}=e^{-2 i \phi_M}$, $A =
\mathcal{A}(B \to F)$, $\bar A = \mathcal{A}(\bar B \to F)$ and
$\phi_M=\beta$ $(-\beta_s)$ for $B_d$ $(B_s)$ mixing. Comparing the
measured $S_F$ to the one obtained from $b \to c \bar c s$ transitions
such as $B_{d(s)} \to J/\Psi K_s (\phi)$ can reveal the presence of NP
in the $b \to s$ penguin amplitude. However, to perform a precise test
of the SM we need to take into account also the doubly Cabibbo
suppressed terms in Eq.~(\ref{eq:purepenguins}). The second term then
acquires a small and calculable weak phase, leading to a small and
calculable $\Delta S = - \eta_F S_F - \sin 2 \phi_M$. Furthermore, we
must consider the contribution from the first term, \textit{i.e.} the
contribution of GIM penguins. An estimate of the latter requires some
knowledge of penguin-type hadronic matrix elements, which can be
obtained either from theory or from experimental data. Let us define
this as the ``GIM-penguin problem'': we shall come back to it in the
next Section after introducing the necessary theoretical ingredients.

Besides pure penguins, we have $b \to s$ transitions in which
emission, annihilation or emission-annihilation parameters give a
contribution to the decay amplitude. Let us call these channels
penguin-dominated. Then we can write schematically the decay amplitude
as:
\begin{equation}
  \label{eq:penguindominated}
  \mathcal{A}(B \to F) = - V^*_{ub} V_{us} \sum
  \left(
    T_i + P_i^\mathrm{GIM}
  \right) -
  V^*_{tb} V_{ts} \sum P_i \,, 
\end{equation}
where $T_i = \{ E_i, A_i, \mathit{EA}_i\}$. Also in this case,
neglecting doubly Cabibbo suppressed terms the decay amplitude has
vanishing weak phase, so that $\Delta S=0$ at this order. However,
we expect $T_i > P_j$ so that the double Cabibbo suppression can be
overcome by the enhancement in the matrix element, leading to a
sizable $\Delta S$. Once again, the evaluation of $\Delta S$ requires
some knowledge of hadronic dynamics. Let us define this as the ``tree
problem'' and return to it in the next Section.

\subsection{Evaluation of hadronic matrix elements}
\label{sec:MEs}

The last decade has witnessed remarkable progress in the theory of
nonleptonic $B$ decays. Bjorken's color transparency argument has been
put on firm grounds, and there is now a wide consensus that many $B$
two-body decay amplitudes factorize in the limit $m_b \to \infty$ and
are therefore computable in this limit in terms of few fundamental
nonperturbative quantities. Three different approaches to
factorization in $B$ decays have been put forward: the so-called QCD
factorization
\cite{hep-ph/9905312,hep-ph/0006124,hep-ph/0210085,hep-ph/0512351,hep-ph/0610322},
perturbative QCD (PQCD) \cite{hep-ph/9409313,Li:1995jr,hep-ph/9411308}
and Soft-Collinear Effective Theory (SCET)
\cite{hep-ph/0109045,hep-ph/0202088,hep-ph/0107002,hep-ph/0301055,hep-ph/0301262}.
A detailed discussion of these approaches goes beyond the scope of
this review; for our purpose, it suffices to quickly describe a few
aspects that are relevant for the study of $b \to s$ penguin
nonleptonic decays.

Unfortunately, as suggested in ref.~\cite{hep-ph/9703353} and later
confirmed in
refs.~\cite{hep-ph/9708222,hep-ph/9801420,hep-ph/0101118,hep-ph/0104126,hep-ph/0110411,hep-ph/0208048,hep-ph/0307367,hep-ph/0402290,hep-ph/0407073,hep-ph/0012152,hep-ph/0104110,hep-ph/0308039,hep-ph/0612290,hep-ph/0004173,hep-ph/0508079,hep-ph/0508041,hep-ph/0608277,hep-ph/0401188,hep-ph/0510241,hep-ph/0601214},
it turns out that subleading corrections to the infinite mass limit,
being doubly Cabibbo-enhanced in $b \to s$ penguins, are very
important (if not dominant) in these channels, so that they
reintroduce the strong model dependence that we hoped to eliminate
using factorization theorems. While different approaches to
factorization point to different sources of large corrections, no
approach is able to compute from first principles all the ingredients
needed to test the SM in $b \to s$ penguins. Therefore, it is
important to pursue, in addition to factorization studies, alternative
data-driven approaches that can in some cases lead to
model-independent predictions for CP violation in $b \to s$ penguins.

Let us now quickly review the main tools that are available for the
study of $b \to s$ penguins.

\subsubsection{QCD factorization}
\label{sec:qcdf}

The first step towards a factorization theorem was given by Bjorken's
color transparency argument \cite{Bjorken:1988kk}. Let us consider a decay
of the $B$ meson in two light pseudoscalars, where two light quarks
are emitted from the weak interaction vertex as a fast-traveling
small-size color-singlet object. In the heavy-quark limit, soft
gluons cannot resolve this color dipole and therefore soft gluon
exchange between the two light mesons decouples at lowest order in
$\Lambda/m_b$ (here and in the following $\Lambda$ denotes a typical
hadronic scale of order $\Lambda_{QCD}$).

Assuming that in $B$ decays to two light pseudoscalars perturbative
Sudakov suppression is not sufficient to guarantee the dominance of
hard spectator interactions, QCD Factorization (QCDF) states that all
soft spectator interactions can be absorbed in the heavy-to-light form
factor \cite{hep-ph/9905312}. Considering for example $B \to \pi\pi$ decays, the
following factorization formula holds at lowest order in
$\Lambda/m_b$:
\begin{eqnarray}
  \label{eq:bbns}
 \langle \pi(p^\prime) \pi(q) \vert Q_i\vert \bar B(p)\rangle &=& 
 f^{B \to \pi} (q^2) \int_0^1 \mathrm{d}x \, T_i^I(x)\phi_\pi(x) + \\
 && \int_0^1 \mathrm{d}\xi \, \mathrm{d}x\, \mathrm{d}y\,
 T^{II}_i(\xi,x,y) \phi_B(\xi) \phi_\pi(x) \phi_\pi(y), \nonumber
\end{eqnarray}
where $f^{B \to \pi} (q^2)$ is a $B\to \pi$ form factor, and
$\phi_\pi$ $(\phi_B)$ are leading-twist light cone distribution
amplitudes of the pion ($B$ meson). $T^{I,II}_i$ denote the hard
scattering amplitudes. Notice that $T^I$ starts at zeroth order in
$\alpha_s$ and at higher order contains hard gluon exchange not
involving the spectator, while $T^{II}$ contains the hard interactions
of the spectator and starts at order $\alpha_s$. 

The scheme and scale dependence of the scattering kernels $T^{I,II}_i$
matches the one of Wilson coefficients, and the final result is
consistently scale and scheme independent. 

Final state interaction phases appear in this formalism as imaginary
parts of the scattering kernels (at lowest order in
$\Lambda/m_b$). These phases appear in the computation of
penguin contractions and of hard gluon exchange between the two pions.
This means that in the heavy quark limit final state interactions can
be determined perturbatively.

A few remarks are important for the discussion of CP violation in $b
\to s$ penguins: 
\begin{itemize}
\item Penguin contractions (including charming and GIM penguins) are
  found to be factorizable, at least at one loop.
\item Subleading terms in the $\Lambda/m_b$ expansion are in general
  non-factorizable, so that they cannot be computed from first
  principles. They are important for phenomenology whenever they
  are chirally or Cabibbo enhanced. In particular, they cannot be
  neglected in $b \to s$ penguin modes. This introduces a strong model
  dependence in the evaluation of $b \to s$ penguin BR's and CP
  asymmetries.
\item Power suppressed terms can invalidate the perturbative
  calculation of strong phases performed in the infinite mass
  limit. Indeed, in this case subleading terms in the $\Lambda/m_b$
  expansion can dominate over the loop-suppressed perturbative phases
  arising at leading order in $\Lambda/m_b$.
\end{itemize}

\subsubsection{PQCD}
\label{sec:pqcd}

The basic idea underlying PQCD calculations is that the dominant
process is hard gluon exchange involving the spectator quark.  PQCD
adopts the three-scale factorization theorem \cite{hep-ph/9607214}
based on the perturbative QCD formalism by Brodsky and Lepage
\cite{Lepage:1980fj}, with the inclusion of the transverse
momentum carried by partons inside the meson.  The three different
scales are the electroweak scale $M_W$, the scale of hard gluon
exchange $t \sim O( \sqrt{\Lambda m_b})$, and the factorization scale
$1/b$, where $b$ is the conjugate variable of parton transverse
momenta.  The nonperturbative physics at scales below $1/b$ is encoded
in process-independent meson wave functions.  The inclusion of
transverse momentum leads to a Sudakov form factor which suppresses
the long distance contributions in the large $b$ region, and vanishes
as $b > 1/\Lambda$. This suppression renders the transverse momentum
flowing into the hard amplitudes of order $\Lambda m_b$.  The
off-shellness of internal particles then remains of $O(\Lambda m_b)$
even in the end-point region, and the singularities are removed.

Notice that:
\begin{itemize}
\item Contrary to QCD factorization, in PQCD all contributions are
  assumed to be calculable in perturbation theory due to the Sudakov
  suppression. This item remains controversial (see
  refs.~\cite{hep-ph/0109260} and \cite{hep-ph/0201103}).
\item The dominant strong phases in this approach come from factorized
  annihilation diagrams.
\item Also in this case, there is no control over subleading
  contributions in the $\Lambda/m_b$ expansion.
\end{itemize}

\subsubsection{SCET}
\label{sec:scet}

Soft-collinear effective theory is a powerful tool to study
factorization in multi-scale problems. The idea is to perform a
two-step matching procedure at the hard ($\mathcal{O}(m_b)$) and
hard-collinear ($\mathcal{O}(\sqrt{m_b \Lambda})$) scales. The
final expression is given in terms of perturbative hard kernels,
light-cone wave functions and jet functions. For phenomenology, it is
convenient to fit directly the nonperturbative parameters on data
using the following expression for the decay amplitude, valid at
leading order in $\alpha_s$~\cite{hep-ph/0401188,hep-ph/0510241,hep-ph/0601214}:
\begin{displaymath}
  \mathcal{A}(B \to M_1 M_2) \propto f_{M_1} \zeta^{BM_2}_J \int_0^1 du
  \phi_{M_1}(u) T_{1J}(u) + f_{M_1} \zeta^{BM_2} T_{1\zeta} + 1
  \leftrightarrow 2 + A_{cc}^{M_1M_2}\,,
\end{displaymath}
where $T$'s are perturbative hard kernels, $\zeta$'s are
nonperturbative parameters and $A_{cc}$ denotes the ``charming
penguin'' contribution.

We notice that:
\begin{itemize}
\item Charming penguins are not factorized in the infinite mass limit
  in this approach, contrary to what obtained in QCD factorization;
\item Phenomenological analyses are carried out at leading order in
  $\alpha_s$ and at leading power in $\Lambda/m_b$;
\item No control is possible on power corrections to factorization.
\end{itemize}

\subsubsection{$SU(3)$ flavor symmetry}
\label{sec:su3}

An alternative approach that has been pursued extensively in the
literature is to use $SU(3)$ flavor symmetry to extract hadronic
matrix elements from experimental data and then use them to predict
$SU(3)$-related
channels~\cite{Zeppenfeld:1980ex,Savage:1989ub,Chau:1990ay,hep-ph/9404283,hep-ph/9504327,hep-ph/9509325,hep-ph/9602218,hep-ph/0307395,hep-ph/0404073,hep-ph/0609128,hep-ph/9710331,hep-ph/9804319,hep-ph/9810260,hep-ph/9903456,hep-ph/0003323,hep-ph/0204101,hep-ph/0309012,hep-ph/0312259,hep-ph/0410407,hep-ph/0512032,hep-ph/0702275,hep-ph/0505194,hep-ph/0508046,hep-ph/0509125}. In
principle, in this way it is possible to eliminate all the
uncertainties connected to factorization and the infinite mass
limit. On the other hand, $SU(3)$-breaking must be evaluated to obtain
reliable predictions.

A few comments are in order:
\begin{itemize}
\item In some fortunate cases, such as the contribution of electroweak
  penguins $Q_{9,10}$ to $B \to K \pi$ decays, $SU(3)$ predicts some
  matrix elements to vanish, so that they can be assumed to be
  suppressed even in the presence of $SU(3)$
  breaking~\cite{hep-ph/9809311,hep-ph/9810482,hep-ph/9812396}.
\item Explicit nonperturbative calculations of two-body nonleptonic
  $B$ decays indicate that $SU(3)$-breaking corrections to $B$ decay
  amplitudes can be up to
  $80\%$, thus invalidating $SU(3)$ analyses of these
  processes~\cite{hep-ph/0308297}.
\item To take partially into account the effects of $SU(3)$ breaking,
  several authors assume that symmetry breaking follows the pattern of
  factorized matrix elements. While this is certainly an interesting
  idea, its validity for $b \to s$ penguins is questionable, given the
  importance of nonfactorizable contributions in these channels.
\end{itemize}

\subsubsection{General parameterizations}
\label{sec:charming}

The idea developed in Refs.~\cite{hep-ph/0104126,charmingnew} is to write down
the RGI parameters as the sum of their expression in the infinite mass
limit, for example using QCD factorization, plus an arbitrary
contribution corresponding to subleading terms in the power expansion.
These additional contributions are then determined by a fit to the
experimental data. In $b \to s$ penguins, the dominant
power-suppressed correction is given by charming penguins, and the
corresponding parameter can be determined with high precision from
data and is found to be compatible with a $\Lambda/m_b$ correction to
factorization \cite{hep-ph/0104126}. However, non-dominant corrections, for
example GIM penguin parameters in $b \to s$ decays, can be extracted
from data only in a few cases (for example in $B \to K \pi$ decays)
\cite{charmingnew}.  However, predictions for $\Delta S$ depend
crucially on these corrections, so that one needs external input to
constrain them. One interesting avenue is to extract the support of
GIM penguins from $SU(3)$-related channels ($b \to d$ penguins) in
which they are not Cabibbo-suppressed, and to use this support,
including a possible $SU(3)$ breaking of $100\%$, in the fit of $b \to
s$ penguin decays. Alternatively, one can omit the calculation in
factorization and fit directly the RGI parameters from the
experimental data, instead of fitting the power-suppressed
corrections~\cite{hep-ph/0507290,hep-ph/0703137}. 

We remark that:
\begin{itemize}
\item Compared to factorization approaches, general parameterizations
  have less predictive power but are more general and thus best suited
  to search for NP in a conservative way.
\item This method has the advantage that for several channels, to be
  discussed below, the predicted $\Delta S$ decreases with the
  experimental uncertainty in $BR$'s and CP asymmetries of $b \to s$
  and $SU(3)$-related $b \to d$ penguins.
\end{itemize}

We conclude this Section by remarking once again that neither the
``GIM-penguin problem'' nor the ``tree problem'' can be solved from
first principles and we must cope with model-dependent estimates. It
then becomes very important to be able to study a variety of channels
in several different approaches. In this way, we can hope to be able
to make solid predictions and to test them with high accuracy. In the
following, we quickly review the present theoretical and experimental
results, keeping in mind the goal of testing the SM and looking for
NP. 

\section{$BR$'S AND CP ASYMMETRIES WITHIN THE SM}
\label{sec:SM}

The aim of this Section is to collect pre- and post-dictions for
$BR$'s and CP asymmetries of $b \to s$ penguin decays obtained in the
approaches briefly discussed in the previous Section. The main focus
will be on $\Delta S$, but $BR$'s and rate CP asymmetries will play a
key role in assessing the reliability and the theoretical uncertainty
of the different approaches.

\subsection{$BR$'s and rate CP asymmetries}
\label{sec:BRs}

In Tables \ref{tab:PP}-\ref{tab:PV} we report some of the results
obtained in the literature for $B$ decay $BR$'s and CP asymmetries.
For QCD Factorization (QCDF) results, the first error corresponds to
variations of CKM parameters, the second to variations of the
renormalization scale, quark masses, decay constants (except for
transverse ones), form factors, and the $\eta-\eta^\prime$ mixing
angle. The third error corresponds to the uncertainty due to the
Gegenbauer moments in the expansion of the light-cone distribution
amplitudes, and also includes the scale-dependent transverse decay
constants for vector mesons.  Finally, the last error corresponds to
an estimate of the effect of the dominant power corrections. For PQCD
results from refs.~\cite{hep-ph/0508041,hep-ph/0608277}, the error only
includes the variation of Gegenbauer moments, of $\vert V_{ub}\vert$
and of the CKM phase. For PQCD results from
ref.~\cite{hep-ph/0703162}, the errors correspond to
input hadronic parameters, to scale dependence, and to CKM parameters
respectively. For SCET results, the analysis is carried out at leading
order in $\alpha_s$ and $\Lambda/m_b$ assuming exact $SU(3)$. The
errors are estimates of $SU(3)$ breaking, of $\Lambda/m_b$ corrections
and of the uncertainty due to SCET parameters respectively. SCET I and
SCET II denote two possible solutions for SCET parameters in the
fit~\cite{hep-ph/0601214}. For General Parametrization (GP) results,
the errors include the uncertainty on CKM parameters, on form factors,
quark masses and meson decay constants, and a variation of
$\Lambda/m_b$ corrections up to $50\%$ of the leading power emission
amplitude. The values in boldface correspond to predictions
(\textit{i.e.} the experimental value has not been used in the fit).

\begin{table}[t!]
  \caption{Results for CP-averaged $BR$'s (in
    units of $10^{-6}$) and
    CP asymmetries (in $\%$) in several
    approaches for $B \to PP$ decays. Experimental averages from the
    Heavy Flavor Averaging Group (HFAG)
    are also shown.}\label{tab:PP}
\vskip 0.3cm
\begin{tabular}{@{}lccccc@{}}%
\toprule
 & QCDF \cite{hep-ph/0308039} & PQCD \cite{hep-ph/0508041,hep-ph/0608277} & SCET
\cite{hep-ph/0601214} & GP \cite{charmingnew} &
exp\\
\colrule
$BR(\pi^-\bar K^0)$
 & $19.3_{\,-1.9\,-\phantom{1}7.8\,-2.1\,-\phantom{1}5.6}^{\,+1.9\,+11.3\,+1.9\,+13.2}$
& $24.5^{+13.6}_{-\ 8.1}$
& $20.8\pm7.9\pm0.6\pm0.7$
& $24.1 \pm 0.7$
& $23.1 \pm 1.0$\\
$A_\mathrm{CP}(\pi^-\bar K^0)$
 &$0.9_{\,-0.3\,-0.3\,-0.1\,-0.5}^{\,+0.2\,+0.3\,+0.1\,+0.6}$
& $\phantom{-}0\pm 0$
&$<5$
& $1.2 \pm 2.4$
& $0.9 \pm 2.5$\\
$BR(\pi^0 K^-)$
 & $11.1_{\,-1.7\,-4.0\,-1.0\,-3.0}^{\,+1.8\,+5.8\,+0.9\,+6.9}$
& $13.9^{+10.0}_{-\ 5.6}$
& $11.3\pm4.1\pm1.0\pm0.3$
& $12.6 \pm 0.5$
& $12.8 \pm 0.6$  \\
$A_\mathrm{CP}(\pi^0 K^-)$
& $7.1_{\,-1.8\,-2.0\,-0.6\,-9.7}^{\,+1.7\,+2.0\,+0.8\,+9.0}$
& $-1^{+3}_{-5}$
& $-11\pm9\pm11\pm2$
& $3.4 \pm 2.4$
& $4.7 \pm 2.6$ \\
$BR(\pi^+ K^-)$
 & $16.3_{\,-2.3\,-6.5\,-1.4\,-\phantom{1}4.8}^{\,+2.6\,+9.6\,+1.4\,+11.4}$
&  $20.9^{+15.6}_{-\ 8.3}$
& $20.1\pm7.4\pm1.3\pm0.6$
& $19.6 \pm 0.5$
& $19.4 \pm 0.6$\\
$A_\mathrm{CP}(\pi^+ K^-)$
&  $4.5_{\,-1.1\,-2.5\,-0.6\,-9.5}^{\,+1.1\,+2.2\,+0.5\,+8.7}$
& $-9^{+6}_{-8}$
&$-6\pm5\pm6\pm2$
&$-8.9 \pm 1.6$
& $-9.5 \pm 1.3$\\
$BR(\pi^0\bar K^0)$
 & $7.0_{\,-0.7\,-3.2\,-0.7\,-2.3}^{\,+0.7\,+4.7\,+0.7\,+5.4}$
&  $\phantom{0}9.1^{+\ 5.6}_{-\ 3.3}$
&$9.4\pm3.6\pm0.2\pm0.3$
& $9.5  \pm 0.4$
& $10.0 \pm 0.6$\\
$A_\mathrm{CP}(\pi^0\bar K^0)$
&$-3.3_{\,-0.8\,-1.6\,-1.0\,-3.3}^{\,+1.0\,+1.3\,+0.5\,+3.4}$
&$-7^{+3}_{-3}$
&$5\pm4\pm4\pm1$
&$\mathbf{-9.8\pm 3.7}$
& $-12 \pm 11$\\
\botrule
\end{tabular}
\end{table}

\begin{table}[t!]
\caption{Results for two-body $b \to s$ penguin decays to $\eta$ or
  $\eta^\prime$ CP-averaged $BR$'s (in unit of $10^{-6}$) and CP
  asymmetries (in $\%$) in several
  approaches. Experimental averages from HFAG are also shown.}\label{tab:singlet}
\vskip 0.3cm
\begin{tabular}{@{}lccccc@{}}%
\toprule
 & QCDF \cite{hep-ph/0308039} & SCET
I \cite{hep-ph/0601214} & SCET
II \cite{hep-ph/0601214}  & 
exp\\
\colrule
$BR(\bar K^0\eta')$
&$46.5_{\,-4.4\,-15.4\,-6.8\,-13.5}^{\,+4.7\,+24.9\,+12.3\,+31.0}$
& $63.2\pm 24.7\pm 4.2\pm 8.1$ & $62.2\pm 23.7\pm 5.5\pm 7.2$
& $64.9 \pm 3.5$\\
$A_\mathrm{CP}(\bar K^0\eta')$
&$1.8_{\,-0.5\,-0.3\,-0.2\,-0.8}^{\,+0.4\,+0.3\,+0.1\,+0.8}$
& $1.1\pm 0.6\pm 1.2\pm 0.2$ & $-2.7\pm 0.7\pm 0.8\pm 0.5$
& $9 \pm 6$ \\
$BR(\bar K^0\eta)$
&$1.1_{\,-0.1\,-1.3\,-0.5\,-0.5}^{\,+0.1\,+2.0\,+0.4\,+1.3}$
&$2.4\pm 4.4\pm 0.2\pm 0.3$ & $2.3\pm 4.4\pm 0.2\pm 0.5$
&{$<1.9$}  \\
$A_\mathrm{CP}(\bar K^0\eta)$
&$-9.0_{\,-2.1\,-12.6\,-6.2\,-7.8}^{\,+2.8\,+\phantom{1}5.4\,+2.8\,+8.2}$
& $21\pm 20\pm 4\pm 3$ & $-18\pm 22\pm 6\pm 4$
& \\
$BR(K^-\eta')$
& $49.1_{\,-4.9\,-16.3\,-7.4\,-14.6}^{\,+5.1\,+26.5\,+13.6\,+33.6}$
&$69.5\pm 27.0\pm 4.3\pm 7.7$ & $69.3\pm 26.0\pm 7.1\pm 6.3$
& $\cerr{69.7}{2.8}{2.7}$\\
$A_\mathrm{CP}(K^-\eta')$
& $2.4_{\,-0.7\,-0.8\,-0.4\,-3.5}^{\,+0.6\,+0.6\,+0.3\,+3.4}$
& $-1\pm0.6\pm0.7\pm0.5$& $0.7\pm0.5\pm0.2\pm0.9$
&$3.1 \pm 2.1$ \\
$BR(K^-\eta)$
&$1.9_{\,-0.5\,-1.6\,-0.6\,-0.7}^{\,+0.5\,+2.4\,+0.5\,+1.6}$
& $2.7\pm 4.8\pm 0.4\pm 0.3$ & $2.3\pm 4.5\pm 0.4\pm 0.3$
&$2.2 \pm 0.3$  \\
$A_\mathrm{CP}(K^-\eta)$
&  $-18.9_{\,-6.9\,-17.5\,-8.5\,-21.8}^{\,+6.4\,+11.7\,+4.8\,+25.3}$
& $33\pm 30\pm 7\pm 3$ & $-33\pm 39\pm 10\pm 4$
& $29 \pm 11$\\
\botrule
\end{tabular}
\end{table}

\begin{table}[t!]
  \caption{Results for CP-averaged $BR$'s (in
    units of $10^{-6}$) and
    CP asymmetries (in $\%$) in several
    approaches for $B \to PV$ decays. Experimental averages from HFAG
    are also shown.}\label{tab:PV}
\vskip 0.3cm
\begin{tabular}{@{}lcccc@{}}%
\toprule
& QCDF \cite{hep-ph/0308039} & PQCD \cite{hep-ph/0508041,hep-ph/0608277} & GP
\cite{charmingnew} &
exp\\
\colrule
$BR(\pi^-\bar K^{*0})$
 & $3.6_{\,-0.3\,-1.4\,-1.2\,-2.3}^{\,+0.4\,+1.5\,+1.2\,+7.7}$
&$6.0^{+2.8}_{-1.5}$
&$11.3 \pm 0.9$
& $10.7 \pm 0.8$\\
$A_\mathrm{CP}(\pi^-\bar K^{*0})$
& $1.6_{\,-0.5\,-0.5\,-0.4\,-1.0}^{\,+0.4\,+0.6\,+0.5\,+2.5}$
& $-1^{+1}_{-0}$
&$-7 \pm 6$
&$-8.5 \pm 5.7$  \\
$BR(\pi^0 K^{*-})$
 & $3.3_{\,-1.0\,-0.9\,-0.6\,-1.4}^{\,+1.1\,+1.0\,+0.6\,+4.4}$
&$4.3^{+5.0}_{-2.2}$
&$7.3 \pm 0.6$
& $6.9 \pm 2.3$\\
$A_\mathrm{CP}(\pi^0 K^{*-})$
& $8.7_{\,-2.6\,-4.3\,-3.4\,-44.2}^{\,+2.1\,+5.0\,+2.9\,+41.7}$
& $-32^{+21}_{-28}$
&$-2 \pm 13$
& $4 \pm 29$\\
$BR(\pi^+ K^{*-})$
 & $3.3_{\,-1.2\,-1.2\,-0.8\,-1.6}^{\,+1.4\,+1.3\,+0.8\,+6.2}$
& $6.0^{+6.8}_{-2.6}$
&$8.5 \pm 0.8$
& $9.8 \pm 1.1$ \\
$A_\mathrm{CP}(\pi^+ K^{*-})$
&  $2.1_{\,-0.7\,-7.9\,-5.8\,-64.2}^{\,+0.6\,+8.2\,+5.1\,+62.5}$
& $-60^{+32}_{-19}$
&$-4 \pm 13$
& $-5 \pm 14$\\
$BR(\pi^0\bar K^{*0})$
 & $0.7_{\,-0.1\,-0.4\,-0.3\,-0.5}^{\,+0.1\,+0.5\,+0.3\,+2.6}$
&$2.0^{+1.2}_{-0.6}$
&$3.1 \pm 0.4$
& $\cerr{0.0}{1.3}{0.1}$\\
$A_\mathrm{CP}(\pi^0\bar K^{*0})$
& $-12.8_{\,-3.2\,-7.0\,-4.0\,-35.3}^{\,+4.0\,+4.7\,+2.7\,+31.7}$
&$-11^{+7}_{-5}$
&$\mathbf{-11 \pm 15}$
& $-1 \pm 27$\\
$BR(\bar K^0\rho^-)$
& $5.8_{\,-0.6\,-3.3\,-1.3\,-\phantom{1}3.2}^{\,+0.6\,+7.0\,+1.5\,+10.3}$
&$8.7^{+6.8}_{-4.4}$
&$7.8 \pm 1.1$
& $\cerr{8.0}{1.5}{1.4}$  \\
$A_\mathrm{CP}(\bar K^0\rho^-)$
&$0.3_{\,-0.1\,-0.4\,-0.1\,-1.3}^{\,+0.1\,+0.3\,+0.2\,+1.6}$
&$1\pm 1$
&$0.02 \pm 0.17$
& $12 \pm 17$ \\
$BR(K^-\rho^0)$
 & $2.6_{\,-0.9\,-1.4\,-0.6\,-1.2}^{\,+0.9\,+3.1\,+0.8\,+4.3}$
&$5.1^{+4.1}_{-2.8}$
&$4.15 \pm 0.50$
& $\cerr{4.25}{0.55}{0.56}$ \\
$A_\mathrm{CP}(K^-\rho^0)$
& $-13.6_{\,-5.7\,-4.4\,-3.1\,-55.4}^{\,+4.5\,+6.9\,+3.7\,+62.7}$
& $71^{+25}_{-35}$
&$29 \pm 10$
&$\cerr{31}{11}{10}$ \\
$BR(K^-\rho^+)$
 & $7.4_{\,-1.9\,-3.6\,-1.1\,-\phantom{1}3.5}^{\,+1.8\,+7.1\,+1.2\,+10.7}$
&$8.8^{+6.8}_{-4.5}$
&$10.2 \pm 1.0$
& $\cerr{15.3}{3.7}{3.5}$ \\
$A_\mathrm{CP}(K^-\rho^+)$
& $-3.8_{\,-1.4\,-2.7\,-1.6\,-32.7}^{\,+1.3\,+4.4\,+1.9\,+34.5}$
&$64^{+24}_{-30}$
&$21 \pm 10$
& $22 \pm 23$\\
$BR(\bar K^0\rho^0)$
 & $4.6_{\,-0.5\,-2.1\,-0.7\,-2.1}^{\,+0.5\,+4.0\,+0.7\,+6.1}$
&$4.8^{+4.3}_{-2.3}$
&$5.2 \pm 0.7$
& $\cerr{5.4}{0.9}{1.0}$ \\
$A_\mathrm{CP}(\bar K^0\rho^0)$
& $7.5_{\,-2.1\,-2.0\,-0.4\,-8.7}^{\,+1.7\,+2.3\,+0.7\,+8.8}$
&$7^{+8}_{-5}$
& $\mathbf{1\pm 15}$
& $-64 \pm 46$\\
$BR(K^-\omega)$
 & $3.5_{\,-1.0\,-1.6\,-0.9\,-1.6}^{\,+1.0\,+3.3\,+1.4\,+4.7}$
&$10.6^{+10.4}_{-5.8}$
&$6.9 \pm 0.5$
& $6.8 \pm 0.5$\\
$A_\mathrm{CP}(K^-\omega)$
 &  $-7.8_{\,-3.0\,-3.6\,-1.9\,-38.0}^{\,+2.6\,+5.9\,+2.4\,+39.8}$
&$32^{+15}_{-17}$
&$5\pm6$
& $5 \pm 6$\\
$BR(\bar K^0\omega)$
 & $2.3_{\,-0.3\,-1.3\,-0.8\,-1.3}^{\,+0.3\,+2.8\,+1.3\,+4.3}$
&$9.8^{+8.6}_{-4.9}$
&$4.6 \pm 0.5$
& $5.2 \pm 0.7$\\
$A_\mathrm{CP}(\bar K^0\omega)$
 &   $-8.1_{\,-2.0\,-3.3\,-1.4\,-12.9}^{\,+2.5\,+3.0\,+1.7\,+11.8}$
&$-3^{+2}_{-4}$
&$-5\pm 11$
& $21 \pm 19 $\\
$BR(K^-\phi)$
 & $4.5_{\,-0.4\,-1.7\,-2.1\,-\phantom{1}3.3}^{\,+0.5\,+1.8\,+1.9\,+11.8}$
& $7.8^{+5.9}_{-1.8}$
&$8.39 \pm 0.59$
&$8.30 \pm 0.65$  \\
$A_\mathrm{CP}(K^-\phi)$
 &  $1.6_{\,-0.5\,-0.5\,-0.3\,-1.2}^{\,+0.4\,+0.6\,+0.5\,+3.0}$
&  $1^{+0}_{-1}$
&$ 3.0 \pm 4.5$
& $3.4 \pm 4.4$\\
$BR(\bar K^0\phi)$
 & $4.1_{\,-0.4\,-1.6\,-1.9\,-\phantom{1}3.0}^{\,+0.4\,+1.7\,+1.8\,+10.6}$
&$7.3^{+5.4}_{-1.6}$
&$7.8 \pm 0.9$
&$\cerr{8.3}{1.2}{1.0}$    \\
$A_\mathrm{CP}(\bar K^0\phi)$
 & $1.7_{\,-0.5\,-0.5\,-0.3\,-0.8}^{\,+0.4\,+0.6\,+0.5\,+1.4}$
&$3^{+1}_{-2}$
&$ 1 \pm 6$
& $-1 \pm 13$\\
\botrule
\end{tabular}
\end{table}

\begin{table}[t!]
  \caption{Results for CP-averaged $BR$'s (in
    units of $10^{-6}$) and
    CP asymmetries (in $\%$) in several
    approaches for $B_s \to PP$ decays. The only available
    experimental result is $BR(B_s \to K^+ K^-)=(24.4\pm4.8) \cdot
    10^{-6}$ \cite{hep-ex/0612018}.}\label{tab:BSPP} 
\vskip 0.3cm
\begin{tabular}{@{}lcccc@{}}%
\toprule
& QCDF \cite{hep-ph/0308039} & PQCD \cite{hep-ph/0703162} & SCET I 
\cite{hep-ph/0601214} & SCET II 
\cite{hep-ph/0601214} \\
\colrule
$BR(K^+ K^-)$
 & $22.7_{\,-3.2\,-\phantom{1}8.4\,-2.0\,-\phantom{1}9.1}^{\,+3.5\,+12.7\,+2.0\,+24.1}$
 &  $17.0^{+5.1+8.8+0.9}_{-4.1-5.0-0.3}$
& $18.2\pm6.7\pm1.1\pm0.5$
& 
\\
$A_\mathrm{CP}(K^+ K^-)$
 &  $4.0_{\,-1.0\,-2.3\,-0.5\,-11.3}^{\,+1.0\,+2.0\,+0.5\,+10.4}$
 &   $-25.8^{+1.1+5.2+0.9}_{-0.2-4.5-1.1}$
& $-6\pm5\pm6\pm2 $
& 
\\
$BR(K^0\bar K^0)$
 & $24.7_{\,-2.4\,-\phantom{1}9.2\,-2.9\,-\phantom{1}9.8}^{\,+2.5\,+13.7\,+2.6\,+25.6}$
 &  $19.6^{+6.4+10.4+0.0}_{-4.9-5.4-0.0}$
& $17.7\pm6.6\pm0.5\pm0.6$
& 
\\
$A_\mathrm{CP}(K^0\bar K^0)$
 &  $0.9_{\,-0.2\,-0.2\,-0.1\,-0.3}^{\,+0.2\,+0.2\,+0.1\,+0.2}$
 &  $0$
& $<10$
& 
\\
$BR(\eta\eta)$
 & $15.6_{\,-1.5\,-6.8\,-2.5\,-\phantom{1}5.5}^{\,+1.6\,+9.9\,+2.2\,+13.5}$
 &  $14.6^{+4.0+8.9+0.0}_{-3.2-5.4-0.0}$
& $7.1\pm 6.4\pm 0.2\pm 0.8$ & $6.4\pm 6.3\pm 0.1\pm 0.7$ \\
$A_\mathrm{CP}(\eta\eta)$
 &  $-1.6_{\,-0.4\,-0.6\,-0.7\,-2.2}^{\,+0.5\,+0.6\,+0.4\,+2.2}$
 &  $-1.6^{+0.3+0.7+0.1}_{-0.3-0.6-0.1}$
& $7.9\pm 4.9\pm 2.7\pm 1.5$ & $-1.1\pm 5.0\pm 3.9\pm 1.0$  \\
$BR(\eta\eta')$
 & $54.0_{\,-5.2\,-22.4\,-6.4\,-16.7}^{\,+5.5\,+32.4\,+8.3\,+40.5}$
 & $39.0^{+9.0+20.4+0.0}_{-7.8-13.1-0.0}$
& $24.0\pm 13.6\pm 1.4\pm 2.7$& $23.8\pm 13.2\pm 1.6\pm 2.9$  \\
$A_\mathrm{CP}(\eta\eta')$
 &  $0.4_{\,-0.1\,-0.3\,-0.1\,-0.3}^{\,+0.1\,+0.3\,+0.1\,+0.4}$
 &  $-1.2^{+0.1+0.2+0.1}_{-0.0-0.1-0.1}$
& $0.04\pm 0.14\pm 0.39\pm 0.43$ &  $2.3\pm 0.9\pm 0.8\pm 7.6$ \\
$BR(\eta'\eta')$ 
& $41.7_{\,-4.0\,-17.2\,-\phantom{1}8.5\,-15.4}^{\,+4.2\,+26.3\,+15.2\,+36.6}$
& $29.6^{+5.2+14.0+0.0}_{-5.3-8.9-0.0}$
& $44.3\pm 19.7\pm 2.3\pm 17.1$ & 
$49.4\pm 20.6\pm 8.4\pm 16.2$  \\
$A_\mathrm{CP}(\eta'\eta')$ 
& $2.1_{\,-0.6\,-0.4\,-0.3\,-1.2}^{\,+0.5\,+0.4\,+0.2\,+1.1}$
& $2.2^{+0.4+0.2+0.2}_{-0.4-0.4-0.1}$
& $0.9\pm 0.4\pm 0.6\pm 1.9$ & 
$-3.7\pm 1.0\pm 1.2\pm 5.6$ \\
\botrule
\end{tabular}
\end{table}

\begin{table}[t!]
  \caption{Results for CP-averaged $BR$'s (in
    units of $10^{-6}$) and
    CP asymmetries (in $\%$) in several
    approaches for $B_s \to PV$ decays. No experimental data are
    available yet.}\label{tab:BSPV}
\vskip 0.3cm
\begin{tabular}{@{}lccccc@{}}%
\toprule
Channel & QCDF \cite{hep-ph/0308039} & PQCD \cite{hep-ph/0703162} \\
\colrule
$BR(K^+ K^{*-})$
 & $4.1_{\,-1.5\,-1.3\,-0.9\,-2.3}^{\,+1.7\,+1.5\,+1.0\,+9.2}$
 & $7.4^{+2.1+1.9+0.9}_{-1.8-1.4-0.4}$ \\
$A_\mathrm{CP}(K^+ K^{*-})$
 & $2.2_{\,-0.7\,-8.0\,-5.9\,-71.0}^{\,+0.6\,+8.4\,+5.1\,+68.6}$
 & $-40.6^{+2.9+2.2+1.8}_{-2.4-3.0-1.3}$
\\
$BR(K^0\bar K^{*0})$
 & $3.9_{\,-0.4\,-1.4\,-1.4\,-\phantom{1}2.8}^{\,+0.4\,+1.5\,+1.3\,+10.4}$
 & $9.1^{+3.2+2.6+0.0}_{-2.2-1.5-0.0}$
\\
$A_\mathrm{CP}(K^0\bar K^{*0})$
 & $1.7_{\,-0.5\,-0.5\,-0.4\,-0.8}^{\,+0.4\,+0.6\,+0.5\,+1.4}$
 &  $0$
\\
$BR( K^- K^{*+})$
 & $5.5_{\,-1.4\,-2.6\,-0.7\,-\phantom{1}3.6}^{\,+1.3\,+5.0\,+0.8\,+14.2}$
 & $6.5^{+1.2+3.3+0.0}_{-1.2-1.8-0.1}$
\\
$A_\mathrm{CP}( K^- K^{*+})$
 & $-3.1_{\,-1.1\,-2.6\,-1.3\,-45.0}^{\,+1.0\,+3.8\,+1.6\,+47.5}$
 & $63.2^{+5.2+8.0+5.1}_{-5.8-10.2-2.6}$
\\
$BR(\bar K^0 K^{*0})$
 & $4.2_{\,-0.4\,-2.2\,-0.9\,-\phantom{1}3.2}^{\,+0.4\,+4.6\,+1.1\,+13.2}$
 & $5.9^{+0.9+2.8+0.0}_{-1.1-1.8-0.0}$
\\
$A_\mathrm{CP}(\bar K^0 K^{*0})$
 & $0.2_{\,-0.1\,-0.3\,-0.1\,-0.1}^{\,+0.0\,+0.2\,+0.1\,+0.2}$
 &  $0$
\\
$BR(\eta\omega)$
 & $0.012_{\,-0.004\,-0.003\,-0.006\,-0.006}^{\,+0.005\,+0.010\,+0.028\,+0.025}$
 & $0.10^{+0.02+0.03+0.00}_{-0.02-0.01-0.00}$
\\
$A_\mathrm{CP}(\eta\omega)$
 & 
 & $3.2^{+6.1+15.2+0.3}_{-3.9-11.2-0.1}$
\\
$BR(\eta'\omega)$
 & $0.024_{\,-0.009\,-0.006\,-0.010\,-0.015}^{\,+0.011\,+0.028\,+0.077\,+0.042}$
 & $0.66^{+0.23+0.22+0.01}_{-0.18-0.21-0.03}$
\\
$A_\mathrm{CP}(\eta'\omega)$
 & 
 & $-0.1^{+0.7+3.9+0.0}_{-0.8-4.2-0.0}$
\\
$BR(\eta\phi)$
 & $0.12_{\,-0.02\,-0.14\,-0.12\,-0.13}^{\,+0.02\,+0.95\,+0.54\,+0.32}$
 &  $1.8^{+0.5+0.1+0.0}_{-0.5-0.2-0.0}$
\\
$A_\mathrm{CP}(\eta\phi)$
 & $-8.4_{\,-2.1\,-71.2\,-44.7\,-59.7}^{\,+2.0\,+30.1\,+14.6\,+36.3}$
 & $-0.1^{+0.2+2.3+0.0}_{-0.4-1.4-0.0}$
\\
$BR(\eta'\phi)$
 & $0.05_{\,-0.01\,-0.17\,-0.08\,-0.04}^{\,+0.01\,+1.10\,+0.18\,+0.40}$
 &  $3.6^{+1.2+0.4+0.0}_{-0.9-0.4-0.0}$
\\
$A_\mathrm{CP}(\eta'\phi)$
 &  $-62.2_{\,-10.2\,-\phantom{1}84.2\,-46.8\,-\phantom{1}49.9}^{\,+15.9\,+132.3\,+80.8\,+122.4}$
 &  $1.2^{+0.1+0.4+0.1}_{-0.0-0.6-0.1}$
\\
\botrule
\end{tabular}
\end{table}
 
First of all, we notice that all approaches are able to reproduce the
experimental $BR$'s of $B \to PP$ penguins, although QCDF tends to
predict lower $BR$'s for $B \to P\eta^\prime$, albeit with large
uncertainties. Concerning $BR$'s of $B \to PV$ penguins, QCDF is
always on the low side and reproduces experimental $BR$'s only when
the upper range of the error due to power corrections is considered.
PQCD shows similar features for $K^*$ and $\rho$ modes, while it
predicts much larger values for $BR$'s of $B \to K\omega$ decays. 

The situation for rate CP asymmetries is a bit different. Both QCDF
and SCET predict $A_\mathrm{CP}(\bar B^0\to\pi^0\bar K^0) \sim -
A_\mathrm{CP}(\bar B^0\to\pi^+K^-)$ while experimentally the two
asymmetries have the same sign. PQCD reproduces the experimental
values, although it predicts $A_\mathrm{CP}(\bar B^0\to\pi^0\bar K^0)$
on the low side of the experimental value. It is interesting to notice
that the GP approach is able to predict the correct value and sign of
$A_\mathrm{CP}(\bar B^0\to\pi^0\bar K^0)$ in spite of the complete
generality of the method. 

Notice also that $B \to K \pi$ data in Tab.~\ref{tab:PP} are perfectly
reproduced in the GP approach, thus showing on general grounds the
absence of any ``$K \pi$ puzzle'', although specific dynamical
assumptions may lead to discrepancies between theory and
experiment~\cite{hep-ph/0309012,hep-ph/0412086,hep-ph/0505060,hep-ph/0701181,hep-ph/0701217}.

\begin{figure}[t!]
\begin{center}
\includegraphics[width=0.32\textwidth]{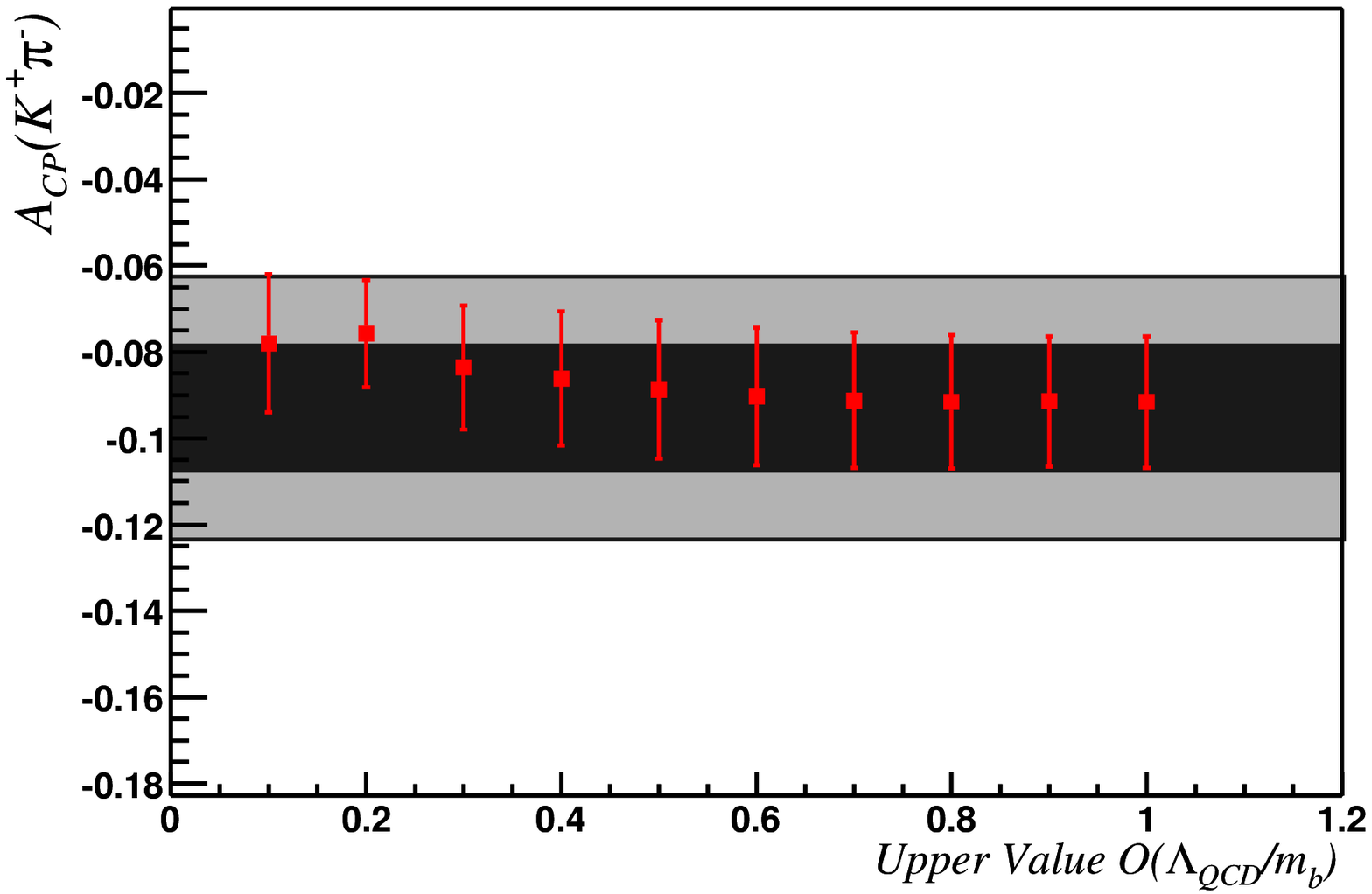}
\includegraphics[width=0.32\textwidth]{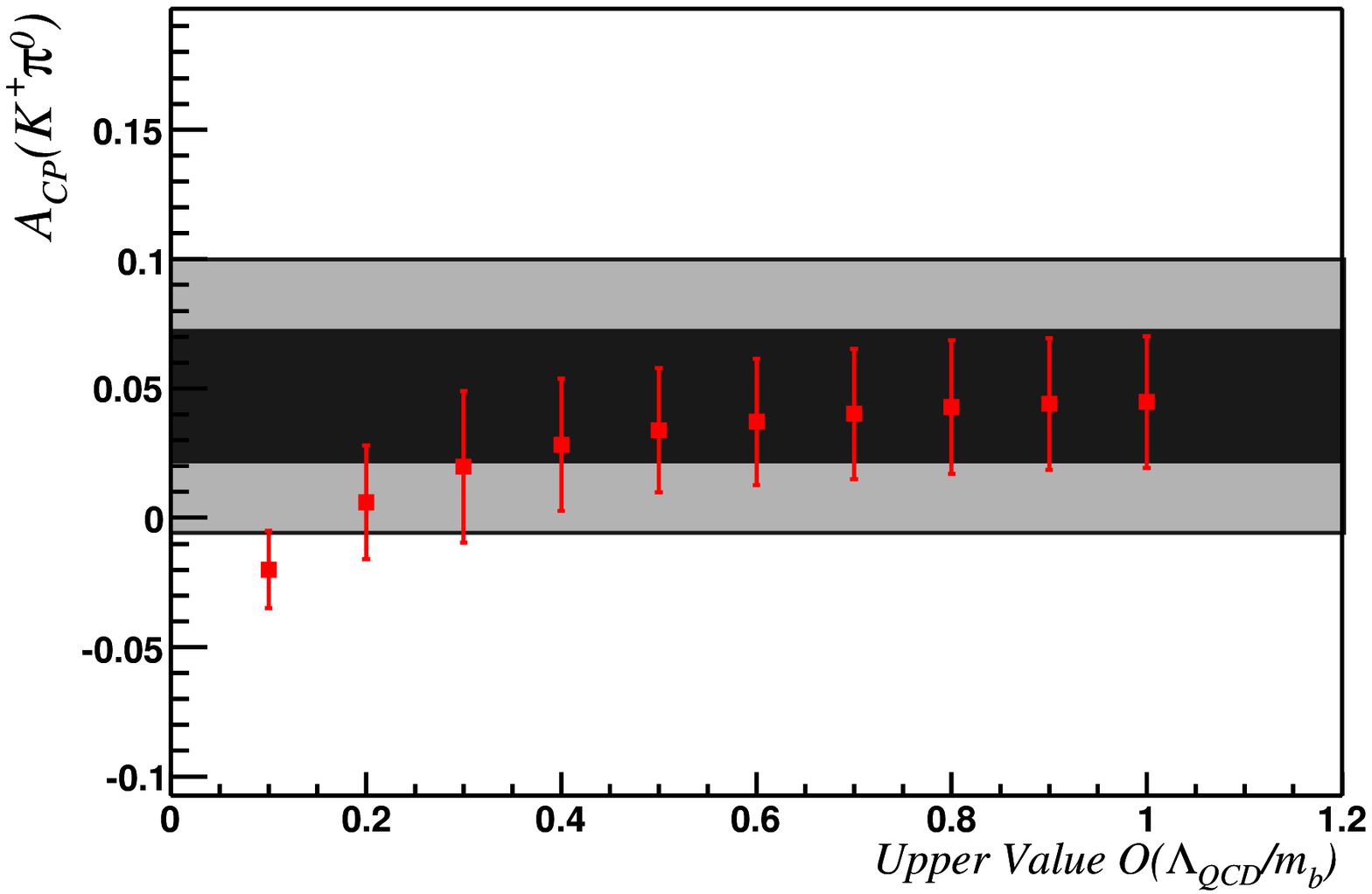} 
\includegraphics[width=0.32\textwidth]{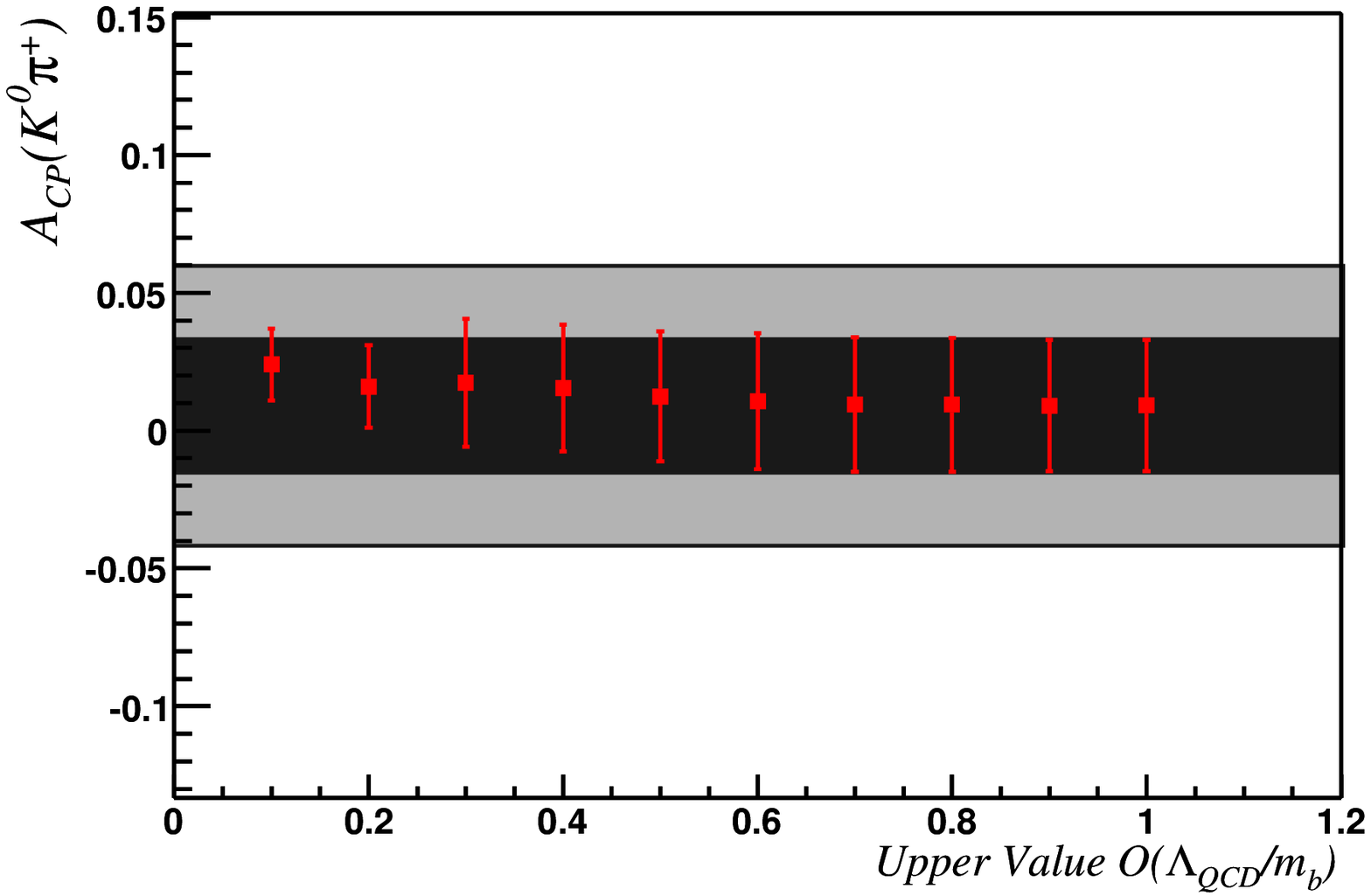}
\end{center}
\caption{$A_\mathrm{CP}$ values for $B \to K \pi$ in the GP
  approach \cite{charmingnew}, obtained varying ${\cal
    O}(\Lambda_{QCD}/m_b)$ contributions in the range [0, UV], with
  the upper value UV scanned between zero and one (in units of the
  factorized emission amplitude).  For comparison, the experimental
  $68\%$ ($95\%$) probability range is given by the dark (light)
  band.}
\label{fig:ACPKP}
\end{figure}

We conclude that factorization approaches in general show a remarkable
agreement with experimental data, but their predictions suffer from
large uncertainties. Furthermore, QCDF and SCET cannot reproduce rate
asymmetries in $B \to K \pi$; this might be a hint that some delicate
aspects of the dynamics of penguin decays, for example rescattering
and final state interaction phases, are not fully under control. It is
then reassuring that a more general approach as GP can reproduce the
experimental data with reasonable (but not too small) values of the
$\Lambda/m_b$ corrections to factorization. To quantify this
statement, we report in Fig.~\ref{fig:ACPKP} the results of the GP fit
for $A_\mathrm{CP}(B \to K \pi)$ as a function of the upper bound on
$\Lambda/m_b$ corrections \cite{charmingnew}. It is clear that
imposing a too low upper bound, of order $10\%$, would generate a
spurious tension between theory and experiment.

For the reader's convenience, we report in Tabs.~\ref{tab:BSPP} and
\ref{tab:BSPV} the predictions obtained in several approaches for
$BR$'s and CP asymmetries of $B_s$ penguin-dominated $b \to s$
decays.

\subsection{Predictions for $S$ and $\Delta S$ in $b \to s$ penguins}
\label{sec:DeltaS}

Keeping in mind the results of Sec.~\ref{sec:BRs}, we now turn to the
main topic of this review, namely our ability to test the SM using
time-dependent CP asymmetries in $b \to s$ penguin nonleptonic
decays. 

Starting from Eq.~(\ref{eq:penguindominated}), we write down the
expression for $S_F$ as follows:
\begin{equation}
  \label{eq:DeltaSgen}
  S_F = \frac{\sin (2
    (\beta_s+\phi_M))+\vert r_F \vert^2 \sin (2
    (\phi_M+\gamma))+2 \,\mathrm{Re}\,r_F \sin (\beta_s+2
    \phi_M+\gamma)}{1+\vert r_F \vert^2+2\, \mathrm{Re}\,r_F \cos
    (\beta_s-\gamma)}\,,
\end{equation}
where $r_F=\vert V_{us}V_{ub}\vert/\vert V_{ts}V_{tb}\vert \times \sum
(T_i + P_i^\mathrm{GIM})/\sum P_i$ with $T_i=0$ for pure penguin
channels. Since the angle $\beta_s$ is small and very well known
($\beta_s=(2.1 \pm 0.1)^\circ$), the problem is then reduced to the
evaluation of $\kappa_F=\sum (T_i + P_i^\mathrm{GIM})/\sum P_i$ for
each channel (notice that $T_i=0$ for pure penguin channels).
Factorization methods have been used to provide estimates of
$\kappa_F$, $S_F$ and $\Delta S_F$ for $b \to s$ channels. The latter
are reported in Tables~\ref{tab:S} and \ref{tab:DeltaS}.  A few
remarks are important.  First of all, the evaluation of
$P^\mathrm{GIM}_i$ relies on the factorization of penguin contractions
of charm and up quarks, which is debatable even in the infinite mass
limit. In addition to that, in factorization $P^\mathrm{GIM}$ has a
perturbative loop suppression so that it is likely to be dominated by
power corrections. Furthermore, the contribution of $T_i$ and
$P_i^\mathrm{GIM}$ is particularly difficult to estimate for $\eta$
and $\eta^\prime$ channels. Last but not least, the determination of
the sign of $\Delta S_F$ relies heavily on the determination of the sign
of Re $\kappa_F$. If $P_i^\mathrm{GIM}$ is dominated by power
corrections, there is no guarantee that the sign given by the
perturbative calculation is correct. 

\begin{table}[t!]
  \caption{Predictions for $S$ parameters in $\%$ for
    $B$ decays. Experimental averages from HFAG 
     are also shown.}\label{tab:S} 
\vskip 0.3cm
\begin{tabular}{@{}lcccccc@{}}%
\toprule
& PQCD \cite{hep-ph/0508041,hep-ph/0608277} & SCET I
\cite{hep-ph/0601214} & SCET II
\cite{hep-ph/0601214} & GP \cite{charmingnew} &
exp\\
\colrule
$ S_{\pi^0 K_S}$ 
& $74^{+2}_{-3}$
& $80\pm2\pm2\pm1$ 
&
& $74.3 \pm 4.4$
& $33 \pm 21$
\\
$S_{\eta^\prime K_S}$ 
& 
&$70.6\pm 0.5\pm 0.6\pm 0.3$
&$71.5\pm 0.5\pm 0.8\pm 0.2$
& $70.9 \pm 3.9$
& $61 \pm 7$
\\
$S_{\eta K_S}$ 
&
&$69\pm 15\pm 5\pm 1$
&$79\pm 14\pm 4\pm 1$
&
& 
\\
$ S_{\phi K_S}$ 
& $71^{+1}_{-1}$
& 
&
& $71.5 \pm 8.7$
& $39 \pm 18$
\\
$ S_{\rho^0 K_S}$ 
& $50^{+10}_{-\phantom{1}6}$
& 
&
& $64 \pm 11$
& $20 \pm 57$
\\
$ S_{\omega K_S}$ 
& $84^{+3}_{-7}$
& 
&
& $75.7 \pm 10.3$
& $48 \pm 24$
\\
\botrule
\end{tabular}
\end{table}

\begin{table}[t!]
  \caption{Predictions for $\Delta S$ parameters in $\%$ for
    $B$ decays. Experimental averages from HFAG 
     are also shown.}\label{tab:DeltaS} 
\vskip 0.3cm
\begin{tabular}{@{}lcccccc@{}}%
\toprule
& QCDF \cite{hep-ph/0505075} & SCET I
\cite{hep-ph/0601214} & SCET II
\cite{hep-ph/0601214} & GP \cite{charmingnew}& exp\\
\colrule
$\Delta S_{\pi^0 K_S}$ 
& $7^{+5}_{-4}$
& $7.7\pm2.2\pm1.8\pm1$ 
&
& $2.4 \pm 5.9$
& $-35 \pm 21$
\\
$\Delta S_{\eta^\prime K_S}$ 
& $1^{+1}_{-1}$
&$-1.9\pm 0.5\pm 0.6\pm 0.3$
&$-1.0\pm 0.5\pm 0.8\pm 0.2$
& $-0.7 \pm 5.4$
&$-7 \pm 7$
\\
$\Delta S_{\eta K_S}$ 
&$10^{+11}_{-\phantom{1}7}$
&$-3.4\pm 15.5 \pm 5.4\pm 1.4$
&$7.0\pm 13.6\pm 4.2\pm 1.1$
&
\\
$\Delta S_{\phi K_S}$ 
& $2^{+1}_{-1}$
& 
&
& $0.4 \pm 9.2$
& $-29 \pm 18$
\\
$\Delta S_{\rho^0 K_S}$ 
& $-8^{+\phantom{1}8}_{-12}$
& 
&
& $-6.2 \pm 8.4$
& $-48 \pm 57$
\\
$\Delta S_{\omega K_S}$ 
& $13^{+8}_{-8}$
& 
&
& $5.6 \pm 10.7$
& $-20 \pm 24$
\\
\botrule
\end{tabular}
\end{table}

With the above \textit{caveat} in mind, from Tables~\ref{tab:S} and
\ref{tab:DeltaS} we learn that:
\begin{itemize}
\item Experimentally there is a systematic trend for negative $\Delta
  S$.  This might be a hint of the presence of new sources of CP
  violation in the $b \to s$ penguin amplitude.
\item The experimental uncertainty is dominant in all channels. In
  addition to that, the GP estimate of the theoretical uncertainty,
  which is certainly conservative, can be reduced with experimental
  improvements on $BR$'s and CP asymmetries.
\item As discussed in Sec.~\ref{sec:basic}, the theoretical
  uncertainty estimated from first principles is much smaller for pure
  penguin decays such as $B \to \phi K_s$ than for penguin-dominated
  channels.
\item In the model-independent GP approach, the theoretical uncertainty
  is smaller for $B \to \pi^0 K_s$ because the number of observables
  in the $B \to K \pi$ system is sufficient to constrain efficiently
  the hadronic parameters. This means that the theoretical error can
  be kept under control by improving the experimental data in these
  channels. On the other hand, the information on $B \to \phi K_s$ is
  not sufficient to bound the subleading terms and this results in a
  relatively large theoretical uncertainty that cannot be decreased
  without additional input on hadronic parameters. Furthermore, using
  $SU(3)$ to constrain $\Delta S_{\phi K_s}$ is difficult because the
  number of amplitudes involved is very large~\cite{Zeppenfeld:1980ex,hep-ph/0505194,hep-ph/0508046,hep-ph/0509125}.
\end{itemize}

The ideal situation would be represented by a pure penguin decay for
which the information on $P_i^\mathrm{GIM}$ is available with minimal
theoretical input. Such situation is realized by the pure penguin
decays $B_s \to K^{0(*)} \bar K^{0(*)}$. An upper bound for the
$P_i^\mathrm{GIM}$ entering this amplitude can be obtained from the
$SU(3)$-related channels $B_d \to K^{0(*)} \bar K^{0(*)}$. Then, even
adding a generous $100 \%$ $SU(3)$ breaking and an arbitrary strong
phase, it is possible to have full control over the theoretical error
in $\Delta S$ \cite{hep-ph/0703137}.

For the reader's convenience, we report in Tab.~\ref{tab:SBs} the
predictions for the $S$ coefficient of the time-dependent CP asymmetry
for several $B_s$ penguin-dominated decays.

\begin{table}[t!]
  \caption{Predictions for $S$ parameters for
    $B_s$ decays.}\label{tab:SBs} 
\vskip 0.3cm
\begin{tabular}{@{}lcccccc@{}}%
\toprule
 & PQCD \cite{hep-ph/0703162} & SCET I
\cite{hep-ph/0601214} & SCET II
\cite{hep-ph/0601214} \\
\colrule
$\bar B^0_s\to K_S\pi^0$ 	& $-0.46^{+0.14+0.19+0.02}_{-0.13-0.20-0.04}$ 	& $-0.16\pm0.41\pm0.33\pm0.17$	& \\
$\bar B^0_s\to K_S\eta$ 	& $-0.31^{+0.05+0.16+0.02}_{-0.05-0.17-0.03}$	& $0.82\pm 0.32\pm 0.11\pm 0.04$ & $0.63\pm 0.61\pm 0.16\pm 0.08$\\
$\bar B^0_s\to K_S\eta'$ 	& $-0.72^{+0.02+0.04+0.00}_{-0.02-0.03-0.00}$ 	& $0.38\pm 0.08\pm 0.10\pm 0.04$ & $0.24\pm 0.09\pm 0.15\pm 0.05$\\
$\bar B^0_s\to K^-K^+$ 		& $0.28^{+0.04+0.04+0.02}_{-0.04-0.03-0.01}$	& $0.19\pm0.04\pm0.04\pm0.01$	& \\
$\bar B^0_s\to \pi^0\eta$ 	& $0.00^{+0.03+0.09+0.00}_{-0.02-0.10-0.01}$ 	& $0.45\pm 0.14\pm 0.42\pm 0.30$ & $0.38\pm 0.20\pm 0.42\pm 0.37$\\
$\bar B^0_s\to \eta\eta$ 	& $0.03^{+0.00+0.01+0.00}_{-0.00-0.01-0.00}$ 	& $-0.026\pm 0.040\pm 0.030\pm 0.014$ & $-0.077\pm 0.061\pm 0.022\pm \
0.026$\\
$\bar B^0_s\to \eta\eta'$ 	& $0.04^{+0.00+0.00+0.00}_{-0.00-0.00-0.00}$ 	& $0.041\pm 0.004\pm 0.002\pm 0.051$ & $0.015\pm 0.010\pm 0.008\pm \
0.069$\\
$\bar B^0_s\to \eta'\eta'$ 	& $0.04^{+0.00+0.00+0.00}_{-0.00-0.00-0.00}$ 	& $0.049\pm 0.005\pm 0.005\pm 0.031$ & $0.051\pm 0.009\pm 0.017\pm \
0.039$\\
 $\bar B^0_s\to\omega\eta$                     &
 $0.07^{+0.00+0.04+0.00}_{-0.01-0.11-0.00}$\\
 $\bar B^0_s\to\omega\eta^\prime$              &  $-0.19^{+0.01+0.04+0.01}_{-0.01-0.04-0.03}$\\
 $\bar B^0_s\to\phi\eta$               &
 $0.10^{+0.01+0.04+0.01}_{-0.01-0.03-0.00}$\\
 $\bar B^0_s\to\phi\eta^\prime$   &
 $0.00^{+0.00+0.02+0.00}_{-0.00-0.02-0.00}$\\
 $\bar B^0_s\to K_S\phi$    &  $-0.72 $\\
\botrule
\end{tabular}
\end{table}

Before closing this Section, let us mention non-resonant three-body
$B$ decays such as $B \to K_s \pi^0 \pi^0$, $B \to K_s K_s K_s$ or $B
\to K^+ K^- K_s$. In this case, a theoretical estimate of $\kappa_F$
is extremely challenging, and using $SU(3)$ to constrain $\kappa_F$ is
difficult because of the large number of channels
involved~\cite{hep-ph/0505194}. Nevertheless, they are certainly
helpful in completing the picture of CP violation in $b \to s$
penguins. 

To summarize the status of $b \to s$ penguins in the SM, we can say
that additional experimental data will allow us to establish whether
the trend of negative $\Delta S$ shown by present data really signals
the presence of NP in $b \to s$ penguins. Theoretical errors are not
an issue in this respect, because the estimates based on factorization
can in most cases be checked using the GP approach based purely on
experimental data. $B_s$ decays will provide additional useful
channels and will help considerably in assessing the presence of NP in
$b \to s$ penguins.

\section{CP VIOLATION IN $b \to s$ PENGUINS BEYOND THE SM }
\label{sec:NP}

We have seen that there is a hint of NP in CP-violating $b \to s$ hadronic
penguins. In this Section, we would like to answer two basic questions
that arise when considering NP contributions to these decays:
\begin{enumerate}
\item What are the constraints from other processes on new sources of
  CP violation in $b \to s$ transitions?
\item Are NP contributions to $b \to s$ transitions well motivated
  from the theoretical point of view?
\end{enumerate}
We consider here only model-independent aspects of these two
questions, and postpone model-dependent analyses to
Section \ref{sec:SUSY}. 

\subsection{Model-independent constraints on $b \to s$ transitions }
\label{sec:miconstr}

The last year has witnessed enormous progress in the experimental
study of $b \to s$ transitions. In particular, the TeVatron
experiments have provided us with the first information on the $B_s -
\bar B_s$ mixing amplitude \cite{hep-ex/0609040}, which can be
translated into constraints on the $\Delta B=\Delta S=2$ effective
Hamiltonian. In any given model, as we shall see for example in
Sec.~\ref{sec:SUSY}, these constraints can be combined with the ones
from $b \to s \gamma$ and $b \to s \ell^+\ell^-$ decays to provide
strong bounds on NP effects in $b \to s$ hadronic penguins.

Let us now summarize the presently available bounds on the $B_s - \bar
B_s$ mixing amplitude, following the discussion of
ref.~\cite{UTfitDF2}. General NP contributions to the $\Delta B=\Delta
S=2$ effective Hamiltonian can be incorporated in the analysis in a
model-independent way, parametrizing the shift induced in the mixing
frequency and phase with two parameters, $C_{B_s}$ and $\phi_{B_s}$,
equal to 1 and 0 in the
SM~\cite{Soares:1992xi,Deshpande:1996yt,hep-ph/9610208,hep-ph/9610252,hep-ph/9704287}:
\begin{equation} 
C_{B_s}
e^{2 i \phi_{B_s}} = \frac{\langle
  B_s|\mathcal{H}_\mathrm{eff}^\mathrm{full}|\bar{B}_s\rangle} {\langle
  B_s|\mathcal{H}_\mathrm{eff}^\mathrm{SM}|\bar{B}_s\rangle} \, .  
\end{equation}

As for the absorptive part of the $B_s -\bar B_s$ mixing amplitude,
which is derived from the double insertion of the $\Delta B=1$
effective Hamiltonian, it can be affected by non-negligible NP effects
in $\Delta B=1$ transitions through penguin contributions. Following
refs.~\cite{hep-ph/0509219,hep-ph/0605213}, we thus introduce two additional
parameters, $C_s^\mathrm{Pen}$ and $\phi_s^\mathrm{Pen}$, which encode
NP contributions to the penguin part of the $\Delta B=1$ Hamiltonian in
analogy to what $C_{B_s}$ and $\phi_{B_s}$ do for the mixing
amplitude.

The available experimental information is the following: the
measurement of $\Delta m_s$~\cite{hep-ex/0609040}, the semileptonic
asymmetry in $B_s$ decays $A_\mathrm{SL}^{s}$ and the dimuon asymmetry
$A_\mathrm{CH}$ from D$\O$~\cite{hep-ex/0701007,hep-ex/0609014}, the measurement of the
$B_s$ lifetime from flavor-specific final
states~\cite{PHLTA.B377.205,hep-ex/9808003,hep-ex/0107077,hep-ex/9802002,hep-ex/0604046,hep-ex/0603003}, the determination of $\Delta \Gamma_s
/\Gamma_s$ from the time-integrated angular analysis of $B_s \to
J/\psi \phi$ decays by CDF~\cite{hep-ex/0412057}, the three-dimensional
constraint on $\Gamma_s$, $\Delta \Gamma_s$, and $B_s$--$\bar B_s$
mixing phase $\phi_s$ from the time-dependent angular analysis of $B_s
\to J/\psi \phi$ decays by D$\O$~\cite{hep-ex/0701012}.

\begin{figure}[t!]
\begin{center}
\includegraphics[width=0.45\textwidth]{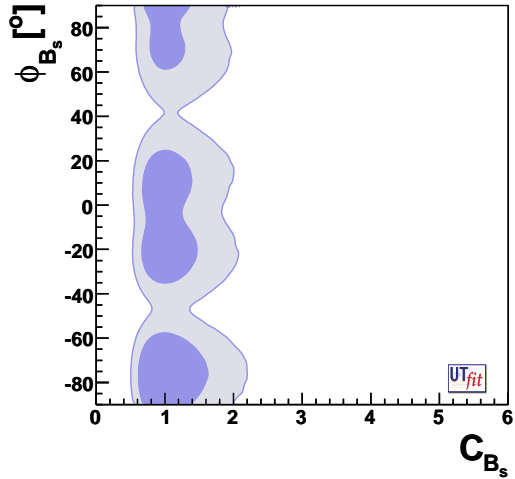}
\caption{Constraints on the $\phi_{B_s}$ vs. $C_{B_s}$ plane
  \cite{UTfitDF2}. Darker (lighter) regions correspond to $68 \%$
  ($95\%$) probability.}
\label{fig:cbsphibs}
\end{center}
\end{figure}

Making use of this experimental information it is possible to
constrain $C_{B_s}$ and
$\phi_{B_s}$~\cite{hep-ph/0605213,hep-ph/0604112,hep-ph/0604249,hep-ph/0605028,hep-ph/0612167,UTfitDF2}. The fourfold
ambiguity for $\phi_{B_s}$ inherent in the untagged analysis of
ref.~\cite{hep-ex/0701012} is somewhat reduced by the measurements of
$A_\mathrm{SL}^s$ and $A_\mathrm{SL}$~\cite{hep-ex/0603053}, which prefer negative
values of $\phi_{B_s}$. The results for $C_{B_s}$ and $\phi_{B_s}$,
obtained from the general analysis allowing for NP in all sectors, are
\cite{UTfitDF2}
\begin{equation}
C_{B_s} = 1.03 \pm 0.29\,, \quad
\phi_{B_s}= (-75 \pm 14)^\circ \cup (-19 \pm 11)^\circ \cup (9 \pm 10)^\circ
\cup (102 \pm 16)^\circ ~.
\label{eq:cbsphibs}
\end{equation}
Thus, the deviation from zero in $\phi_{B_s}$ is below the $1 \sigma$
level, although clearly there is still ample room for values of
$\phi_{B_s}$ very far from zero. The corresponding p.d.f. in the
$C_{B_s}$-$\phi_{B_s}$ plane is shown in
Fig.~\ref{fig:cbsphibs}.

The experimental information on $b \to s \gamma$ and $b \to s \ell^+
\ell^-$ decays
\cite{hep-ex/0308044,hep-ex/0403004,hep-ex/0503044,hep-ex/0603018,hep-ex/0404006,hep-ex/0508004,hep-ex/0604007}
can also be combined in a model-independent way along the lines of
refs.~\cite{hep-ph/9408213,hep-ph/0112300,hep-ph/0310219,hep-ph/0410155}.
In this way, it is possible to constrain the coefficients of the $b
\to s \gamma$, $b \to s \gamma^*$ and $b \to s Z$ vertices, which also
contribute to $b \to s$ hadronic penguins. It turns out that
order-of-magnitude enhancements of these vertices are excluded, so
that they are unlikely to give large effects in $b \to s$ nonleptonic
decays. On the other hand, the $b \to s g$ vertex is only very weakly
constrained, so that it can still give large contributions to $b \to
s$ hadronic penguins.  Finally, the information contained in
Eq.~(\ref{eq:cbsphibs}) can be used to constrain NP effects in $b \to
s$ hadronic decays only within a given model, since a connection
between $\Delta B=2$ and $\Delta B=1$ effective Hamiltonians is
possible only once the model is specified. We shall return to this
point in Sec.~\ref{sec:SUSY}.

\subsection{Theoretical motivations for NP in $b \to s$ transitions}
\label{sec:motivations}

We now turn to the second question formulated at the beginning of this
Section, namely whether on general grounds it is natural to expect NP
to show up in $b \to s$ transitions. The general picture emerging from
the generalized Unitarity Triangle analysis performed in
ref.~\cite{hep-ph/0509219,hep-ph/0605213,UTfitDF2} and from the very
recent data on $D - \bar D$
mixing~\cite{hep-ex/0703020,hep-ex/0703036,arXiv:0704.1000,hep-ph/0703204}
is that no new sources of CP violation are present in $B_d$, $K$ and
$D$ mixing amplitudes.  Conversely, large NP contributions to $s \to d
g$, $b \to d g$ and $b \to s g$ transitions are not at all excluded.
Therefore, although the idea of minimal flavor violation is
phenomenologically
appealing~\cite{hep-lat/9407029,hep-ph/9703442,hep-ph/9806308,hep-ph/0007085,hep-ph/0207036,hep-ph/0505110,hep-ph/0604057},
an equally possible alternative is that NP is contributing more to
$\Delta F=1$ transitions than to $\Delta F=2$ ones. Within the class
of $\Delta F=1$ transitions, (chromo)-magnetic vertices are peculiar
since they require a chirality flip to take place, which leads to a
down-type quark mass suppression within the SM. On the other hand, NP
models can weaken this suppression if they contain additional heavy
fermions and/or additional sources of chiral mixing. In this case,
they can lead to spectacular enhancements for the coefficients of
(chromo)-magnetic operators. Furthermore, if the relevant new
particles are colored, they can naturally give a strong enhancement of
chromomagnetic operators while magnetic operators might be only
marginally modified \cite{hep-ph/9604438}. The electric dipole moment
of the neutron puts strong constraints on new sources of CP violation
in chirality-flipping flavor-conserving operators involving light
quarks, but this does not necessarily imply the suppression of
flavor-violating operators, especially those involving $b$ quarks.
Therefore, assuming that NP is sizable in hadronic $b \to s$ penguins
is perfectly legitimate given the present information available on
flavor physics.

From a theoretical point of view, a crucial observation is the strong
breaking of the SM $SU(3)^5$ flavor symmetry by the top quark Yukawa
coupling. This breaking necessarily propagates in the NP sector, so
that in general it is very difficult to suppress NP contributions to
CP violation in $b$ decays, and these NP contributions could be
naturally larger in $b \to s$ transitions than in $b \to d$
ones. This is indeed the case in several flavor models (see for
example Ref.~\cite{hep-ph/0104101}).

Another interesting argument is the connection between quark and
lepton flavor violation in grand unified
models~\cite{hep-ph/0002141,hep-ph/0212180,hep-ph/0303071,hep-ph/0304130}. The idea is very simple: the
large flavor mixing present in the neutrino sector, if mainly
generated by Yukawa couplings, should be shared by right-handed
down-type quarks that sit in the same $SU(5)$ multiplet with
left-handed leptons. Once again, one expects in this case large NP
contributions to $b \to s$ transitions.

We conclude that the possibility of large NP effects in $b \to s$
penguin hadronic decays is theoretically well motivated on general
grounds. The arguments sketched above can of course be put on firmer
grounds in the context of specific models, and we refer the reader to
the rich literature on this subject.

\section{SUSY MODELS}
\label{sec:SUSY}

Let us now focus on SUSY and discuss the phenomenological effects of
the new sources of flavor and CP violation in $b \to s$ processes that
arise in the squark
sector~\cite{hep-ph/9604387,hep-ph/9704402,hep-ph/9803368,hep-ph/9803401,hep-ph/9806266,hep-ph/0103121,hep-ph/0105292,hep-ph/0109149,hep-ph/0207070,hep-ph/0207356,hep-ph/0212023,hep-ph/0212092,hep-ph/0301269,hep-ph/0304229,hep-ph/0306086,hep-ph/0306076,hep-ph/0303214,hep-ph/0307024,hep-ph/0404055,hep-ph/0407284,hep-ph/0407291,hep-ph/0411151,hep-ph/0505151}.
In general, in the MSSM squark masses are neither flavor-universal,
nor are they aligned to quark masses, so that they are not flavor
diagonal in the super-CKM basis, in which quark masses are diagonal
and all neutral current vertices are flavor diagonal. The ratios of
off-diagonal squark mass terms to the average squark mass define four
new sources of flavor violation in the $b \to s$ sector: the mass
insertions $(\delta^d_{23})_{AB}$, with $A,B=L,R$ referring to the
helicity of the corresponding quarks. These $\delta$'s are in general
complex, so that they also violate CP. One can think of them as
additional CKM-type mixings arising from the SUSY sector. Assuming
that the dominant SUSY contribution comes from the strong interaction
sector, \textit{i.e.} from gluino exchange, all FCNC processes can be
computed in terms of the SM parameters plus the four $\delta$'s plus
the relevant SUSY parameters: the gluino mass $m_{\tilde g}$, the
average squark mass $m_{\tilde q}$, $\tan \beta$ and the $\mu$
parameter. The impact of additional SUSY contributions such as
chargino exchange has been discussed in detail in Ref.
\cite{hep-ph/0306076}. We consider only the case of small or moderate
$\tan \beta$, since for large $\tan \beta$ the constraints from $B_s
\to \mu^+ \mu^-$ and $\Delta m_s$ preclude the possibility of having
large effects in $b \to s$ hadronic penguin decays
\cite{hep-ph/0110121,hep-ph/0210145,hep-ph/0301269,hep-ph/0510422,hep-ph/0604121,hep-ph/0605012,hep-ph/0703035}.

Barring accidental cancellations, one can consider one single $\delta$
parameter, fix the SUSY masses and study the phenomenology. The
constraints on $\delta$'s come at present from $B \to X_s \gamma$, $B
\to X_s l^+ l^-$ and from the $B_s - \bar B_s$ mixing amplitude as
given in Eq.~(\ref{eq:cbsphibs}).  We refer the reader to
refs.~\cite{hep-ph/0212397,hep-ph/0603114,noisusynew} for all the details of
this analysis.

\begin{figure}[t!]
\begin{center}
\includegraphics[width=0.45\textwidth]{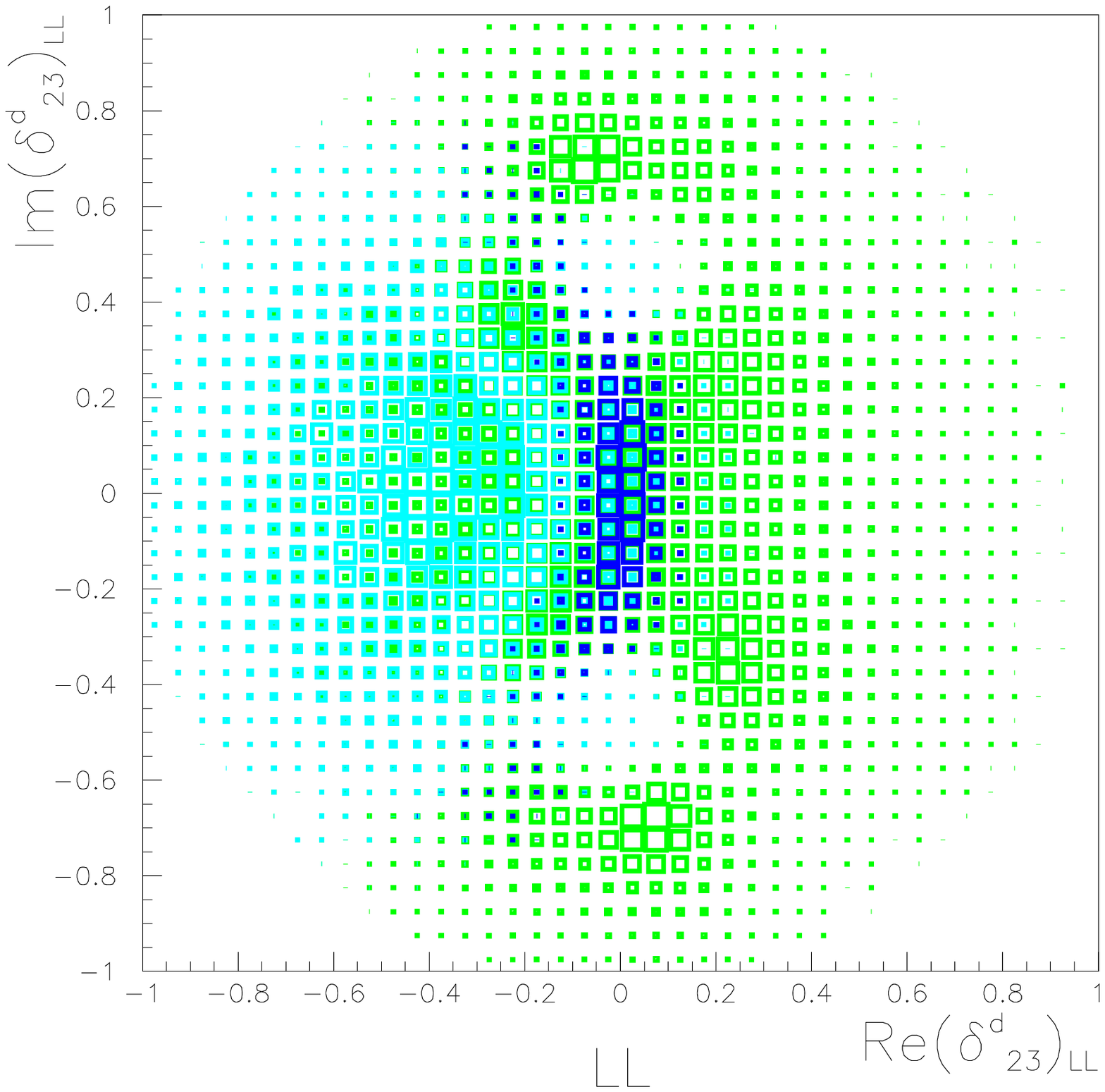} 
\includegraphics[width=0.45\textwidth]{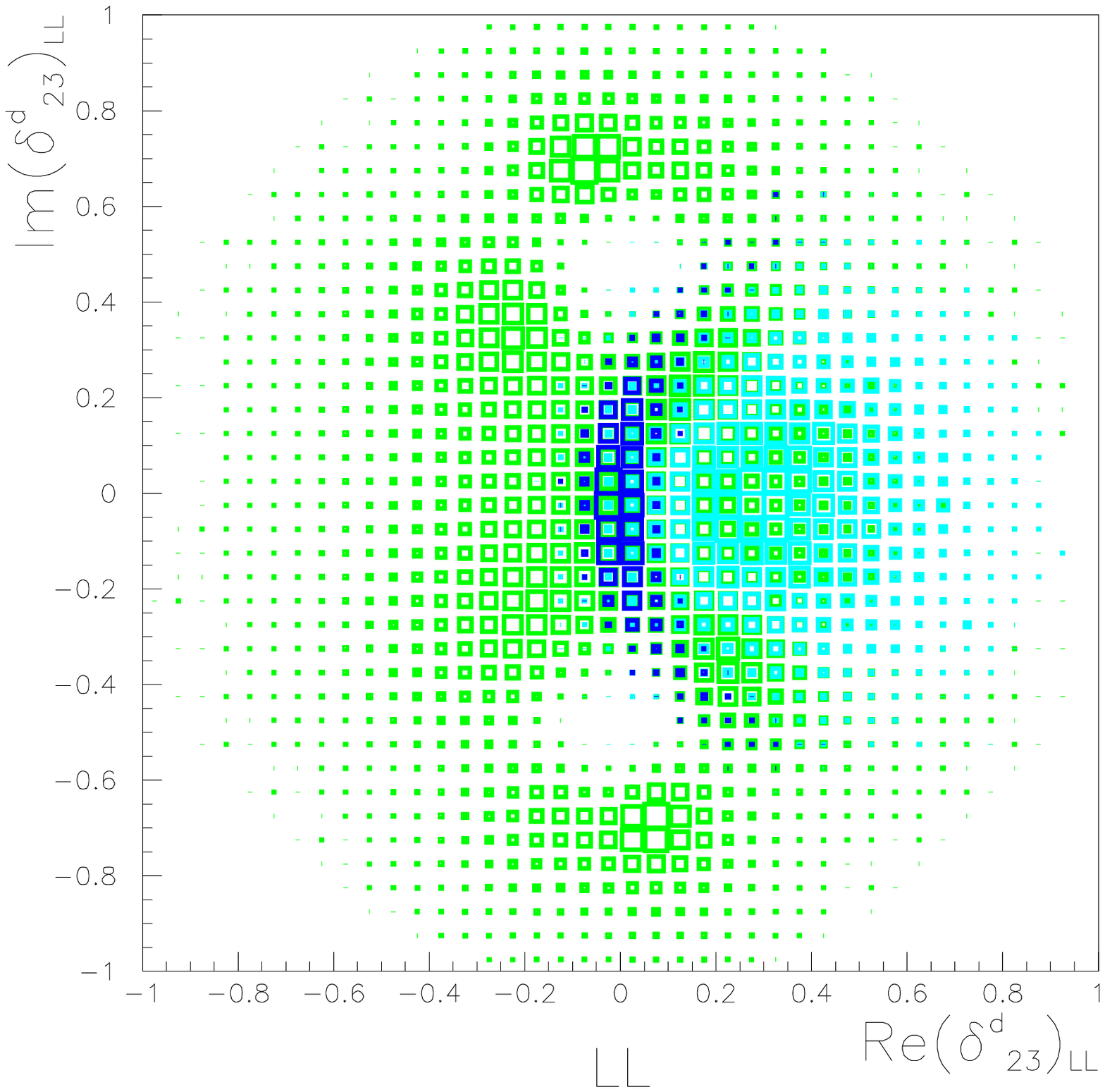}
\includegraphics[width=0.45\textwidth]{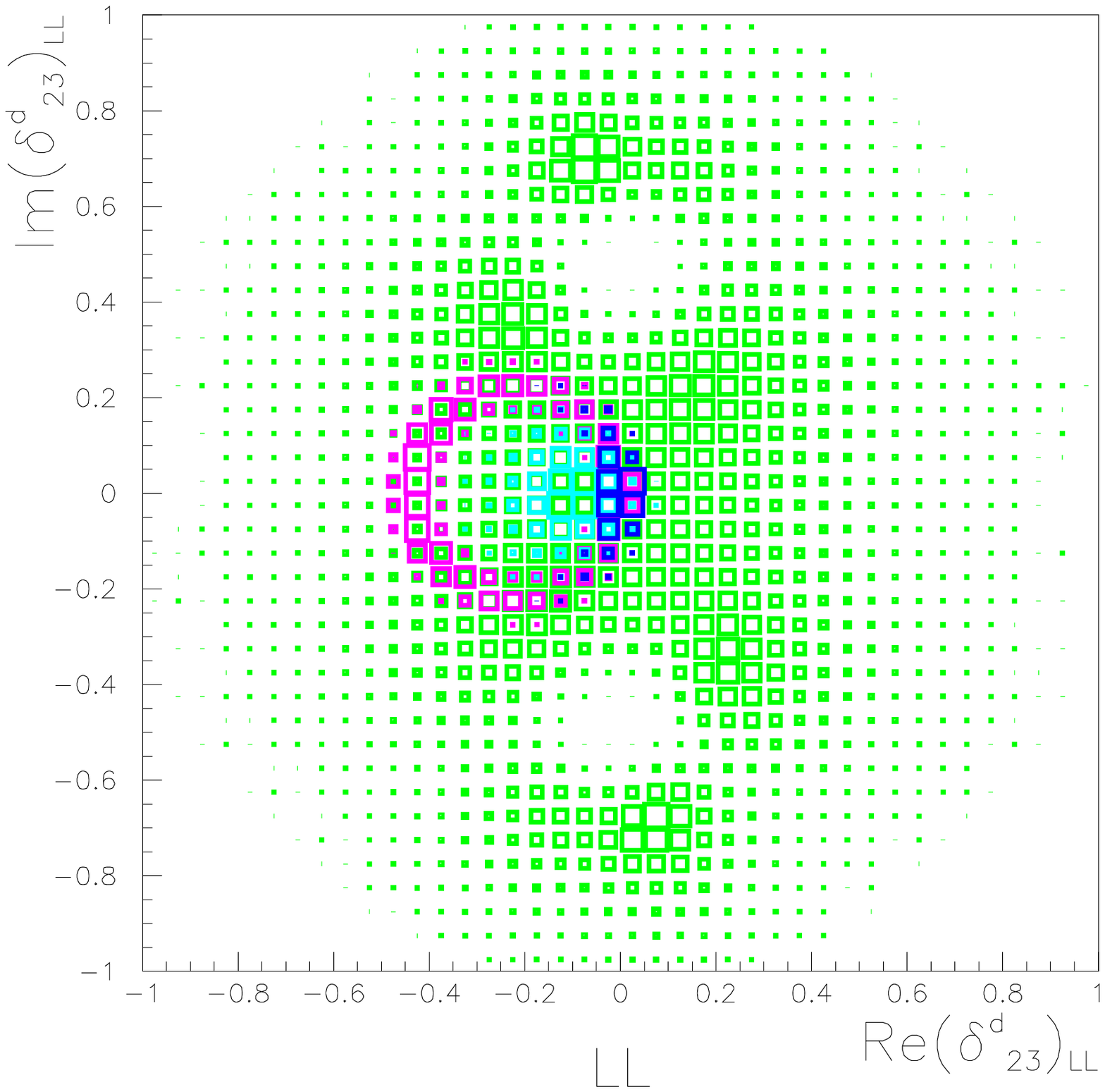} 
\includegraphics[width=0.45\textwidth]{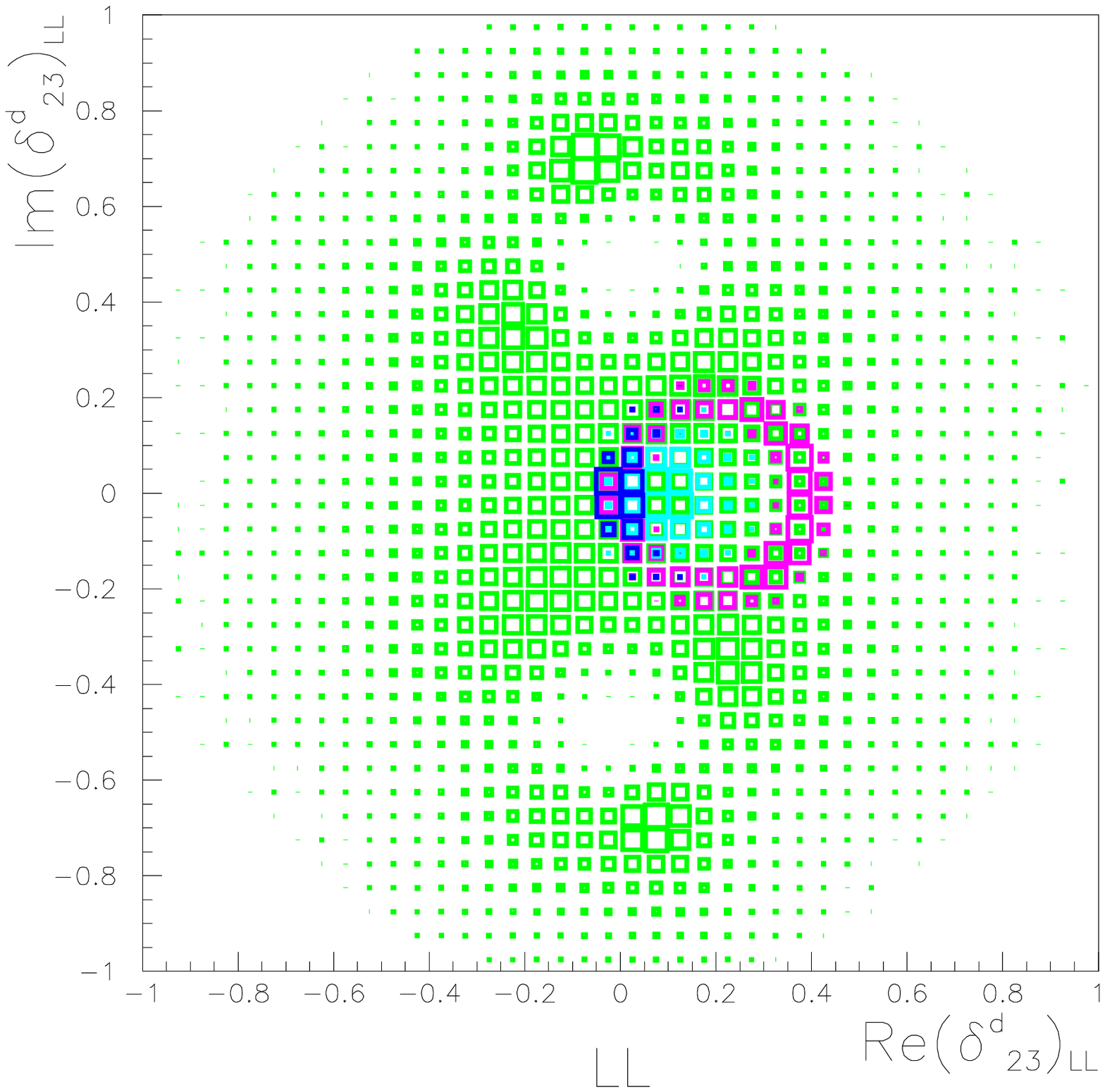}
\caption{Allowed region in the
  Re$\left(\delta^d_{23}\right)_{LL}$-Im$\left(\delta^d_{23}\right)_{LL}$
  plane. In the plots on the left (right), negative (positive) $\mu$
  is considered. Plots in the upper (lower)
  row correspond to $\tan\beta=3$ ($\tan\beta=10$). See the text for
  details.}
\label{fig:LL}
\end{center}
\end{figure}

\begin{figure}[t!]
\begin{center}
\includegraphics[width=0.45\textwidth]{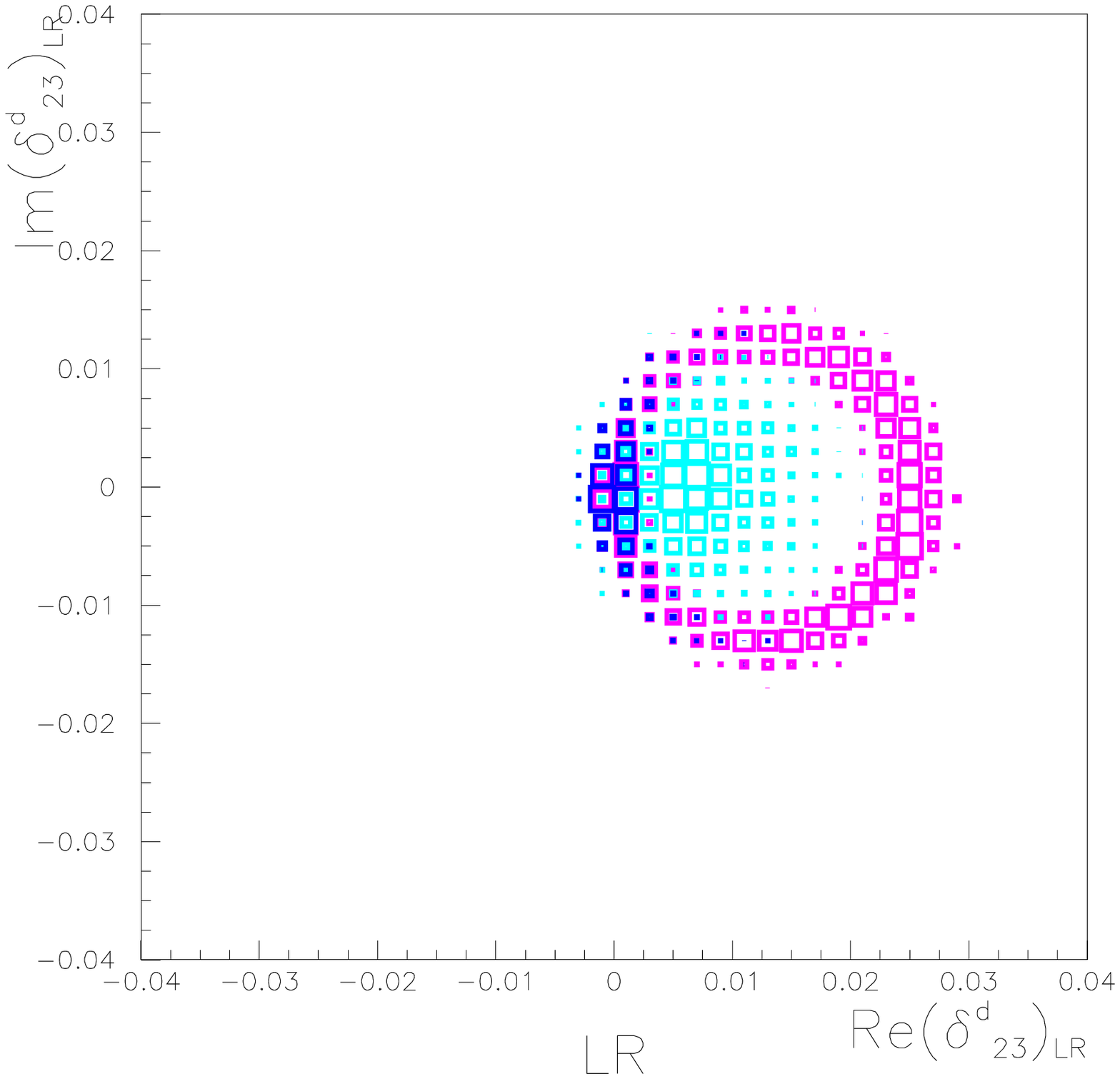} 
\includegraphics[width=0.45\textwidth]{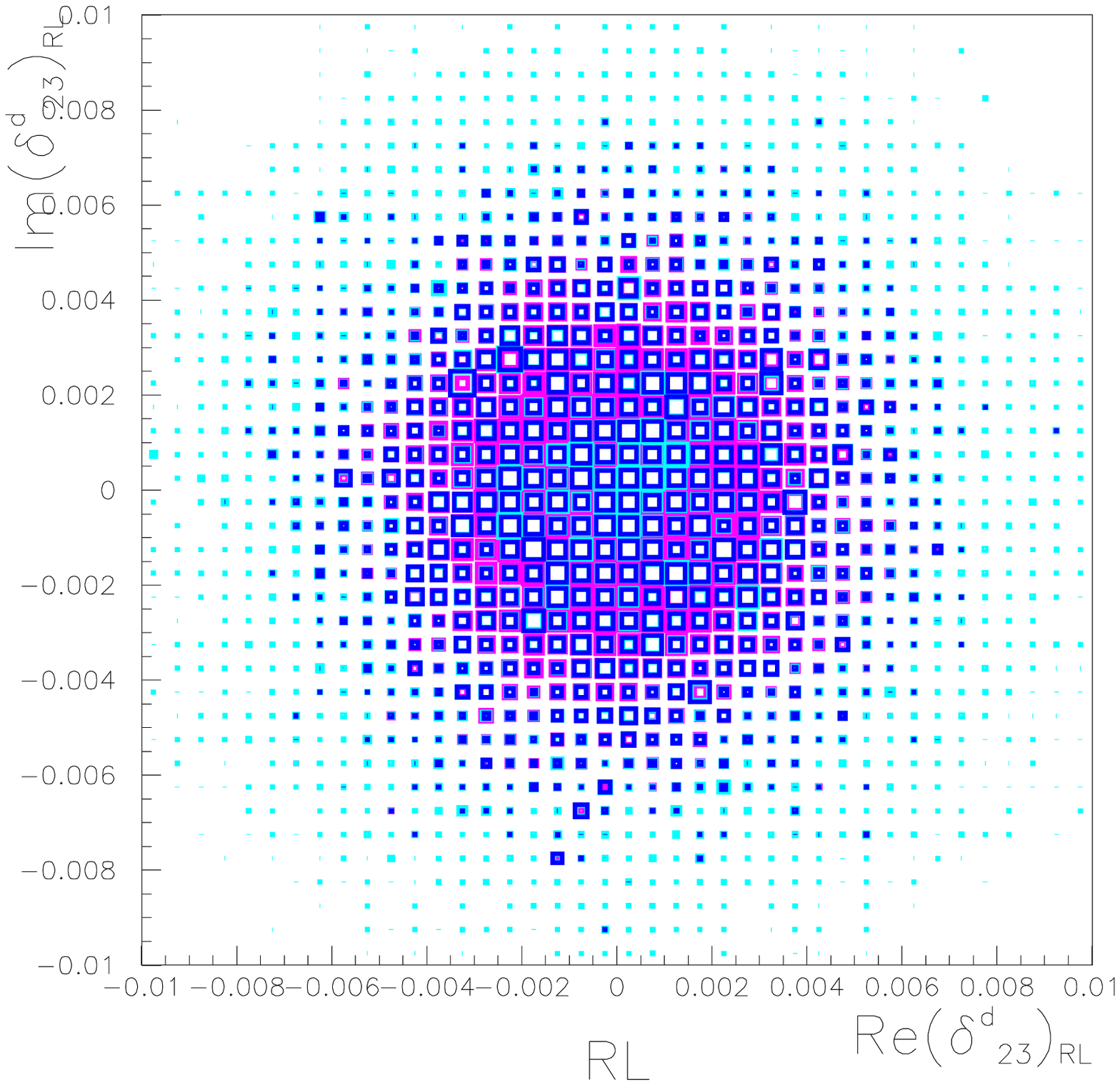}
\caption{Allowed region in the
  Re$\left(\delta^d_{23}\right)_{LR}$-Im$\left(\delta^d_{23}\right)_{LR}$
  (left) and
  Re$\left(\delta^d_{23}\right)_{RL}$-Im$\left(\delta^d_{23}\right)_{RL}$
  (right) plane. Results do not depend on the sign of $\mu$ or on the
  value of $\tan \beta$.}
\label{fig:LR}
\end{center}
\end{figure}

\begin{figure}[t!]
\begin{center}
\includegraphics[width=0.45\textwidth]{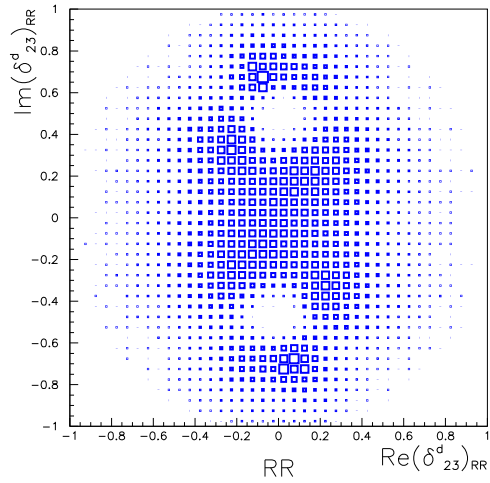} 
\caption{Allowed region in the
  Re$\left(\delta^d_{23}\right)_{RR}$-Im$\left(\delta^d_{23}\right)_{RR}$
  plane.  Results do not depend on the sign of $\mu$ or on the
  value of $\tan \beta$. See the text for
  details.}
\label{fig:RR}
\end{center}
\end{figure}

\begin{figure}[t!]
\begin{center}
\includegraphics[width=0.45\textwidth]{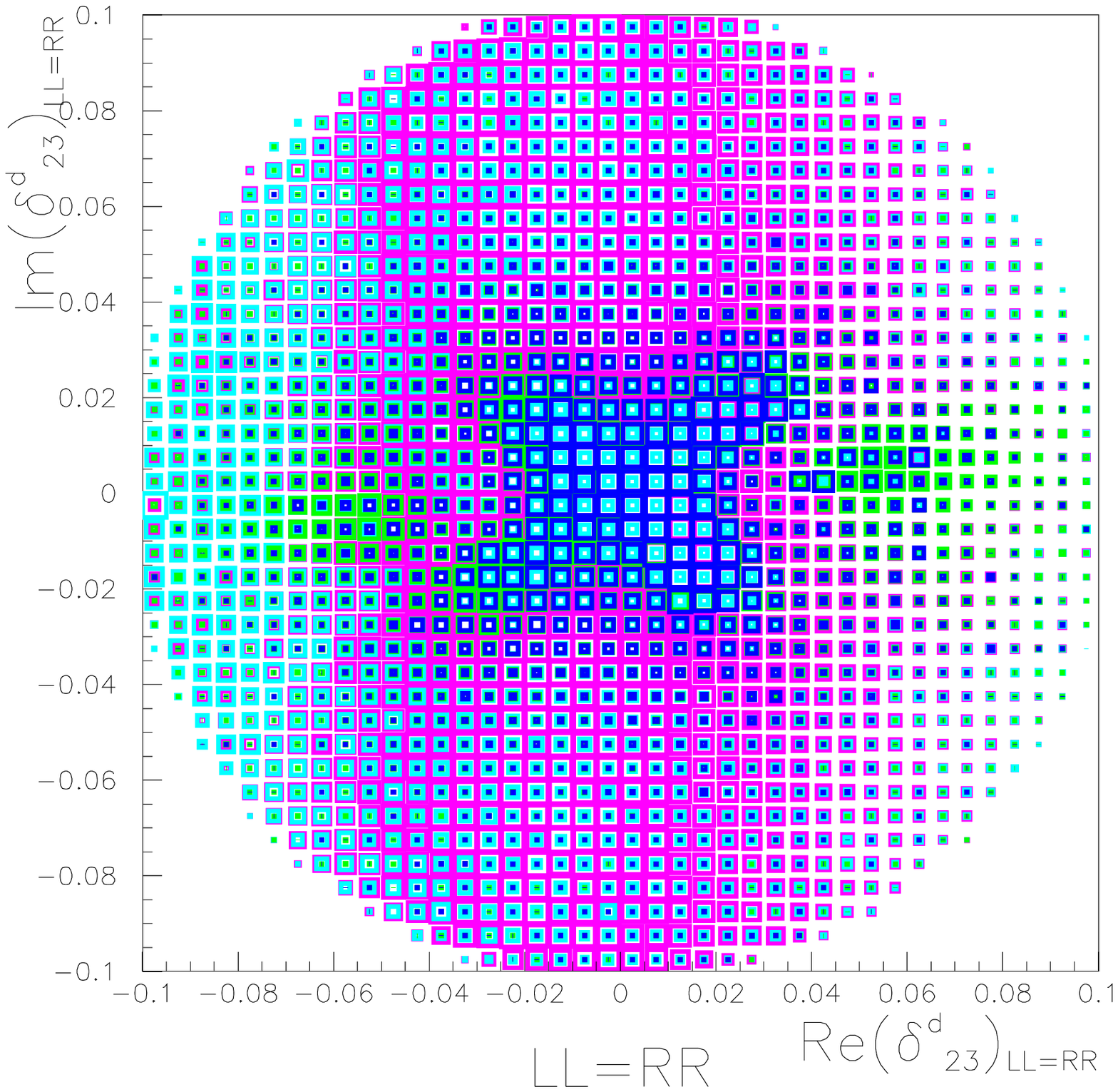} 
\includegraphics[width=0.45\textwidth]{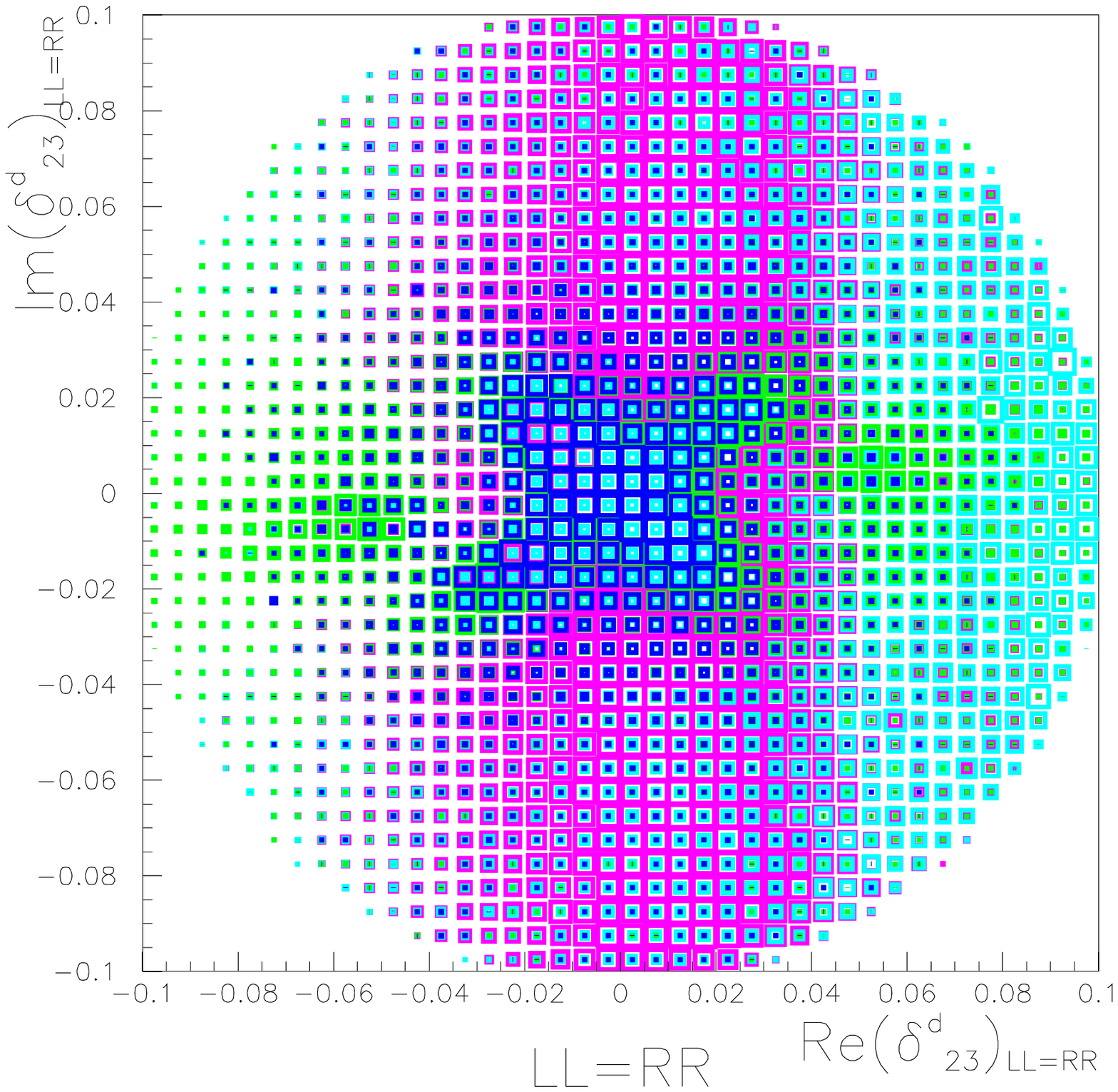} 
\includegraphics[width=0.45\textwidth]{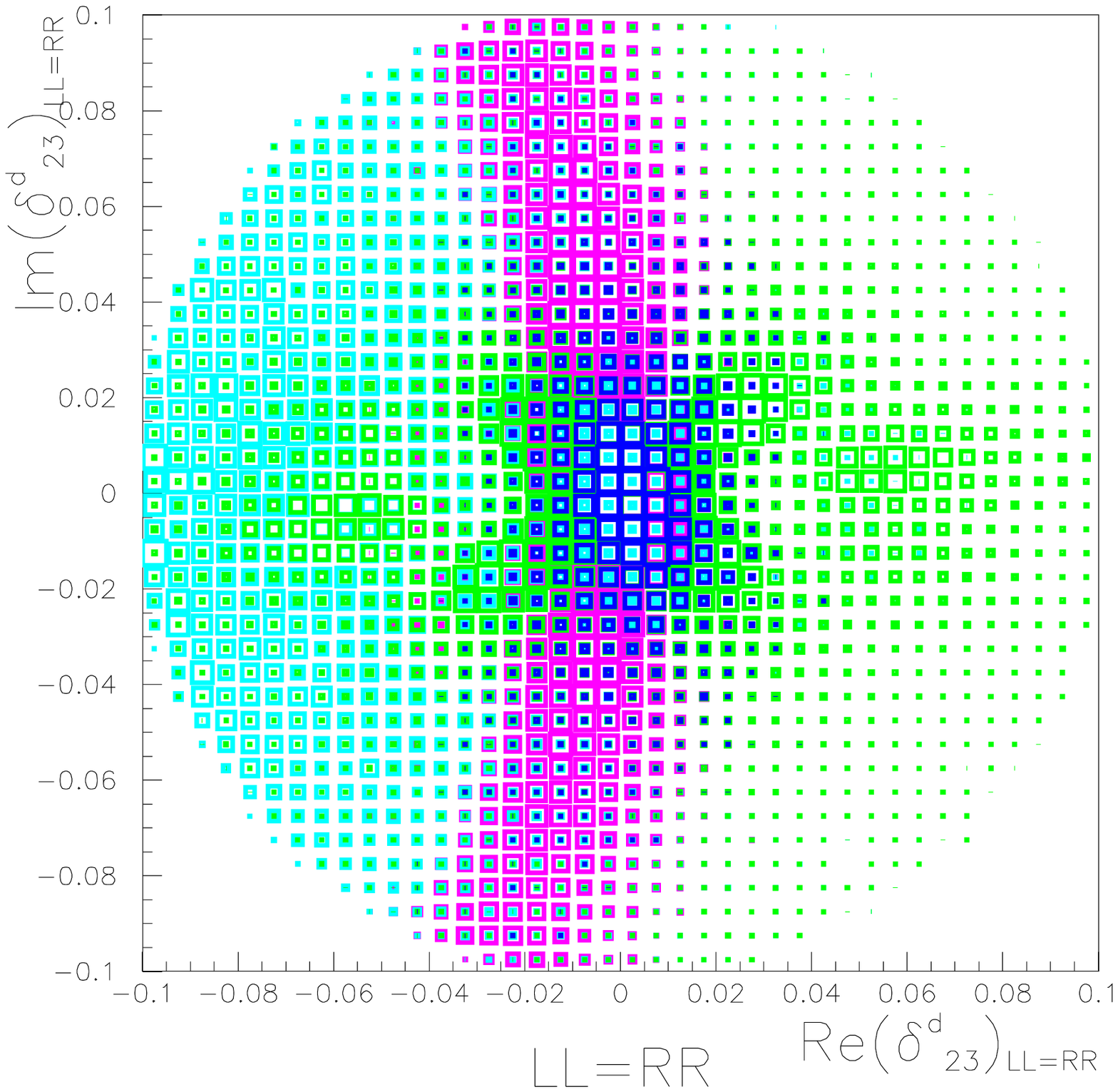} 
\includegraphics[width=0.45\textwidth]{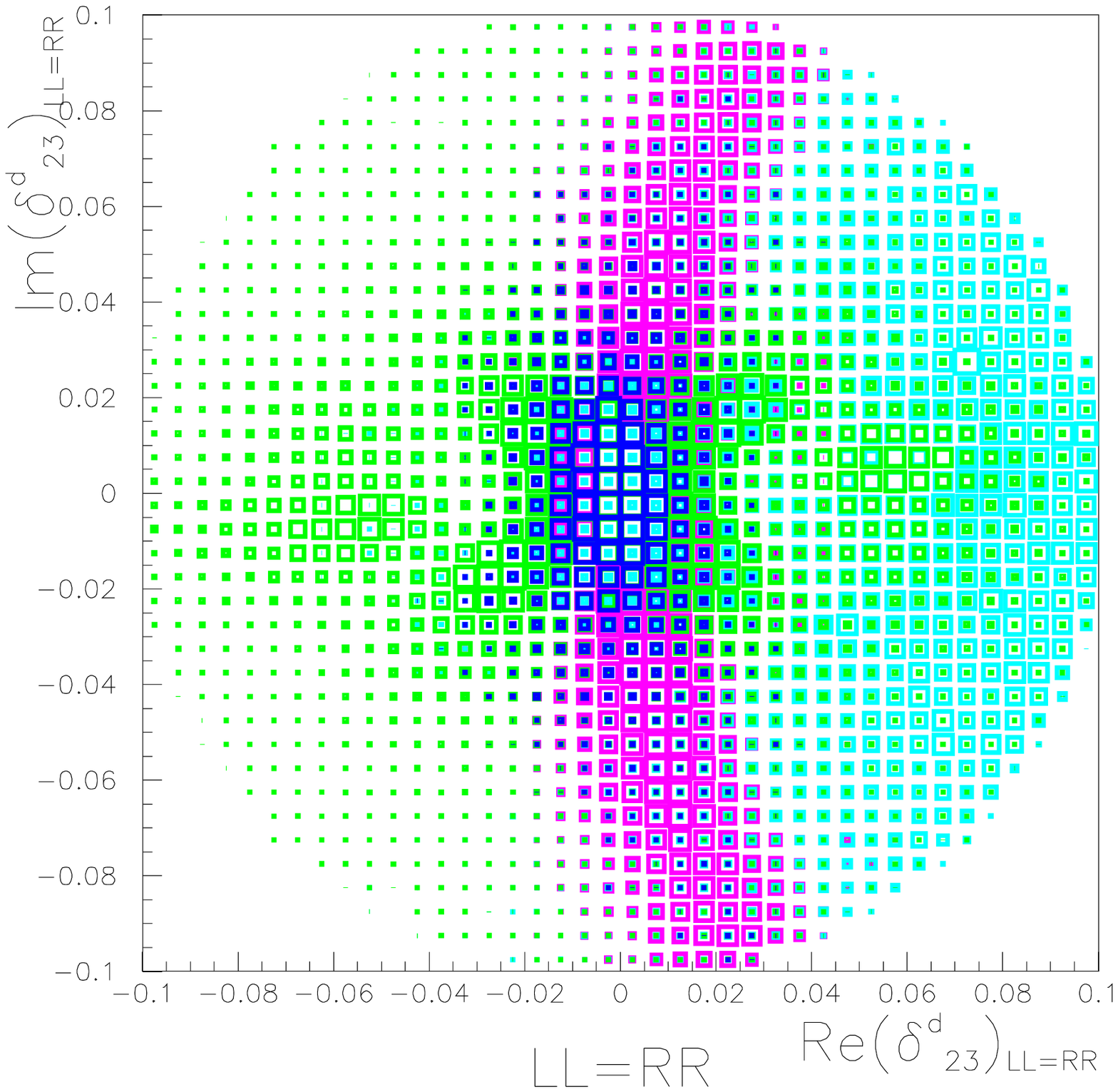} 
\caption{Allowed region in the
  Re$\left(\delta^d_{23}\right)_{LL=RR}$-Im$\left(\delta^d_{23}\right)_{LL=RR}$
  plane.  In the plots on the left (right), negative (positive) $\mu$
  is considered. Plots in the upper (lower)
  row correspond to $\tan\beta=3$ ($\tan\beta=10$). See the
  text for details.}
\label{fig:LLRR}
\end{center}
\end{figure}

Fixing as an example $m_{\tilde g}=m_{\tilde q}= \vert \mu \vert =$
350 GeV and $\tan \beta = 3$ or $10$, one obtains the constraints on
$\delta$'s reported in Figs.~\ref{fig:LL}-\ref{fig:RR}
\cite{hep-ph/0603114,noisusynew}. We plot in light green the allowed
region considering only the constraint from the $C_{B_s}$
vs. $\phi_{B_s}$ p.d.f. of Fig.~\ref{fig:cbsphibs}, in light blue the
allowed region considering only the constraint from $b \to s \ell^+
\ell^-$ and in violet the allowed region considering only the
constraint from $b \to s \gamma$. The dark blue region is the one
selected imposing all constraints simultaneously.

Several comments are in order at this point:
\begin{itemize}
\item Only $(\delta^d_{23})_{\mathrm{LL},\mathrm{LR}}$ generate
  amplitudes that interfere with the SM in rare decays. Therefore,
  the constraints from rare decays for
  $(\delta^d_{23})_{\mathrm{RL},\mathrm{RR}}$ are symmetric around
  zero, while the interference with the SM produces the circular shape
  of the $B \to X_s \gamma$ constraint on
  $(\delta^d_{23})_{\mathrm{LL},\mathrm{LR}}$.
\item We recall that
  $\mathrm{LR}$ and $\mathrm{RL}$ mass insertions generate much larger
  contributions to the (chromo)magnetic operators, since the necessary
  chirality flip can be performed on the gluino line ($\propto
  m_{\tilde g}$) rather than on the quark line ($\propto m_{
    b}$).  Therefore, the constraints from rare decays are much more
  effective on these insertions, so that the bound from $B_s - \bar
  B_s$ has no impact in this case. 
\item The $\mu \tan \beta$ 
  flavor-conserving $\mathrm{LR}$ squark mass term generates,
  together with a flavor changing $\mathrm{LL}$ mass insertion, an
  effective $(\delta^d_{23})_{\mathrm{LR}}^\mathrm{eff}$ that
  contributes to $B \to X_s \gamma$. For positive (negative) $\mu$,
  we have $(\delta^d_{23})_{\mathrm{LR}}^\mathrm{eff} \propto +(-)
  (\delta^d_{23})_{\mathrm{LL}}$ and therefore the circle determined
  by $B \to X_s \gamma$ in the $\mathrm{LL}$ and $\mathrm{LR}$ cases
  lies on the same side (on opposite sides) of the origin (see
  Figs.~\ref{fig:LL} and \ref{fig:LR}).
\item For $\tan \beta=3$, we see from the upper row of
  Fig.~\ref{fig:LL} that the bound on $(\delta^d_{23})_{\mathrm{LL}}$
  from $B_s - \bar B_s$ mixing is competitive with the one from rare
  decays, while for $\tan \beta=10$ rare decays give the strongest
  constraints (lower row of Fig.~\ref{fig:LL}). The bounds on all
  other $\delta$'s do not depend on the sign of $\mu$ and on the value
  of $\tan \beta$ for this choice of SUSY parameters.
\item For $\mathrm{LL}$ and $\mathrm{LR}$ cases, $B \to X_s \gamma$
  and $B \to X_s l^+ l^-$ produce bounds with different shapes on the
  Re $\delta$ -- Im $\delta$ plane (violet and light blue regions in
  Figs.~\ref{fig:LL} and \ref{fig:LR}), so that applying them
  simultaneously a much smaller region around the origin survives
  (dark blue regions in Figs.~\ref{fig:LL} and \ref{fig:LR}). This
  shows the key role played by rare decays in constraining new sources
  of flavor and CP violation in the squark sector.
\item For the $\mathrm{RR}$ case, the constraints
  from rare decays are very weak, so that the only significant bound
  comes from $B_s - \bar B_s$ mixing. 
\item If $(\delta^d_{23})_{\mathrm{LL}}$ and
  $(\delta^d_{23})_{\mathrm{RR}}$ insertions are simultaneously
  nonzero, they generate chirality-breaking contributions that are
  strongly enhanced over chirality-conserving ones, so that the
  product $(\delta^d_{23})_{\mathrm{LL}}(\delta^d_{23})_{\mathrm{RR}}$
  is severely bounded. In Fig.~\ref{fig:LLRR} we report the allowed
  region obtained in the case
  $(\delta^d_{23})_{\mathrm{LL}}=(\delta^d_{23})_{\mathrm{RR}}$. For
  $(\delta^d_{23})_{\mathrm{LL}}\neq (\delta^d_{23})_{\mathrm{RR}}$,
  this constraint can be interpreted as a bound on
  $\sqrt{(\delta^d_{23})_{\mathrm{LL}}(\delta^d_{23})_{\mathrm{RR}}}$.
  We observe a very interesting interplay between the constraints from
  rare decays and the one from $B_s -\bar B_s$ mixing. Increasing
  $\tan \beta$ from $3$ to $10$, the bound from rare decays becomes
  tighter, but $B_s -\bar B_s$ mixing still plays a relevant role.
\item All constraints scale approximately linearly with squark and
  gluino masses.
\end{itemize}

\begin{figure}[!ht]
\begin{center}
\includegraphics[width=0.45\textwidth]{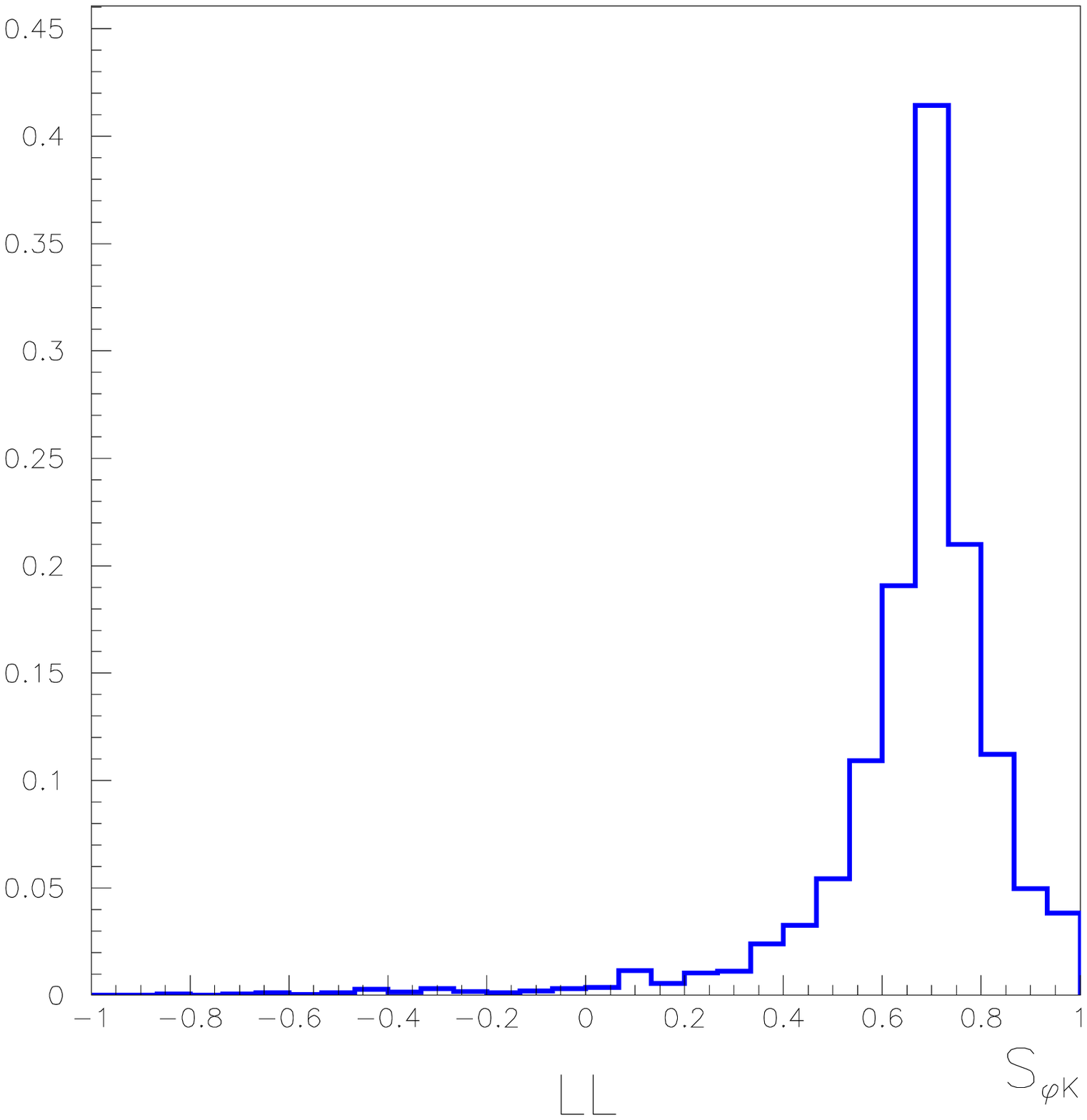} 
\includegraphics[width=0.45\textwidth]{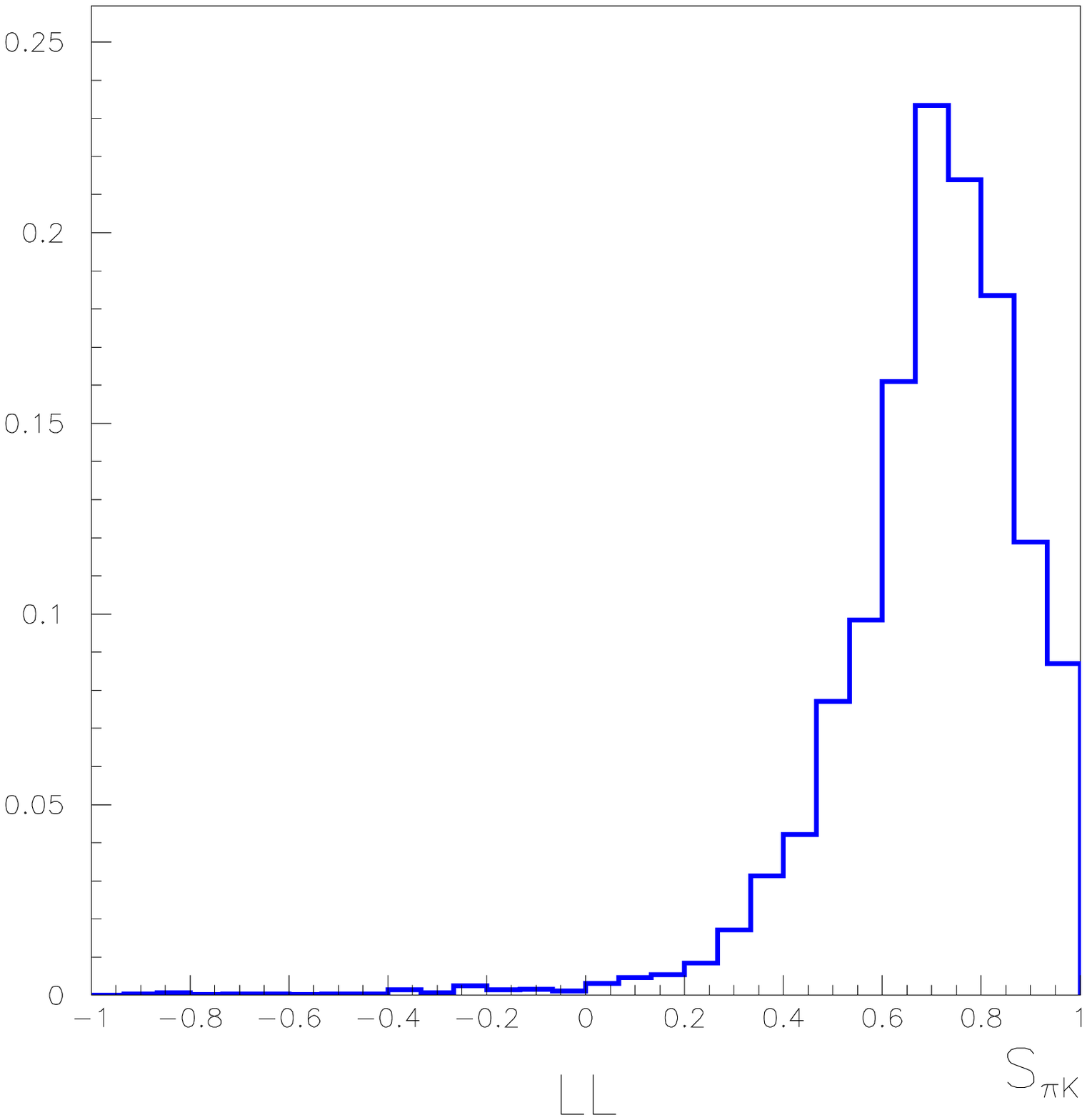} 
\includegraphics[width=0.45\textwidth]{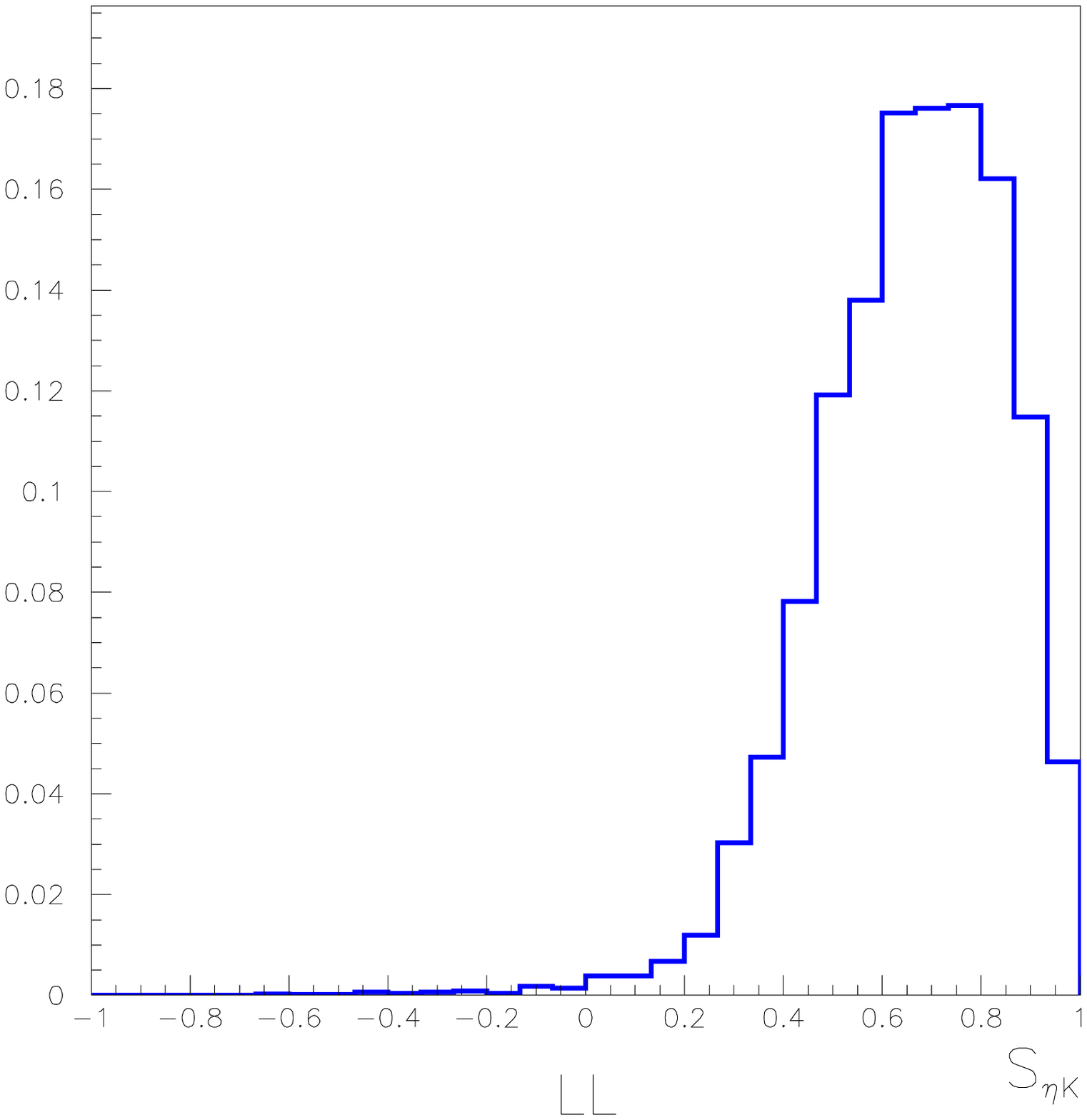} 
\includegraphics[width=0.45\textwidth]{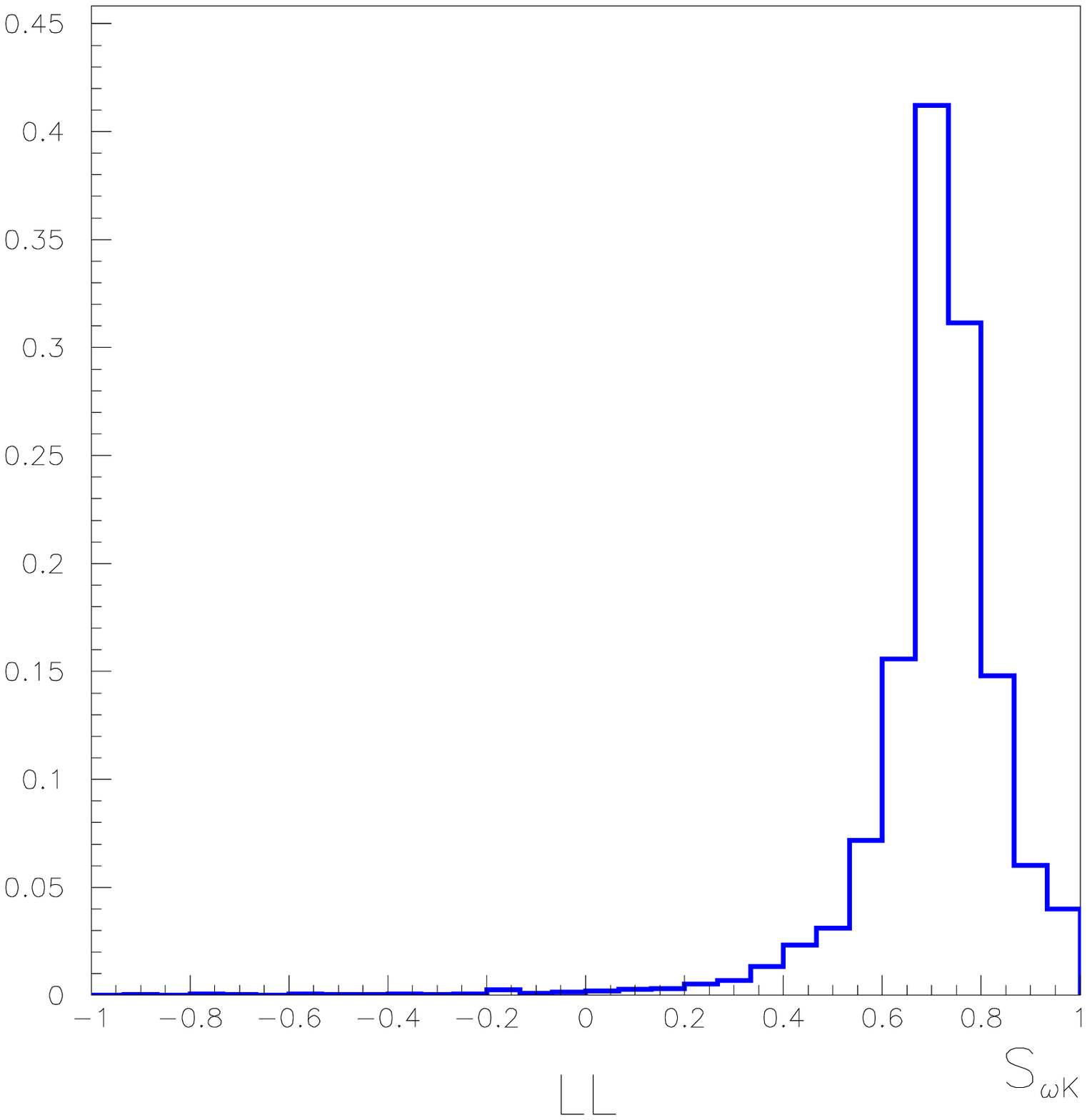} 
\caption{Probability density functions for $S_{\phi K_s}$, $S_{\pi^0
    K_s}$, $S_{\eta^\prime K_s}$ and $S_{\omega K_s}$ induced by
  $(\delta^d_{23})_{\mathrm{LL}}$.}
\label{fig:ll}
\end{center}
\end{figure}

\begin{figure}[!ht]
\begin{center}
\includegraphics[width=0.45\textwidth]{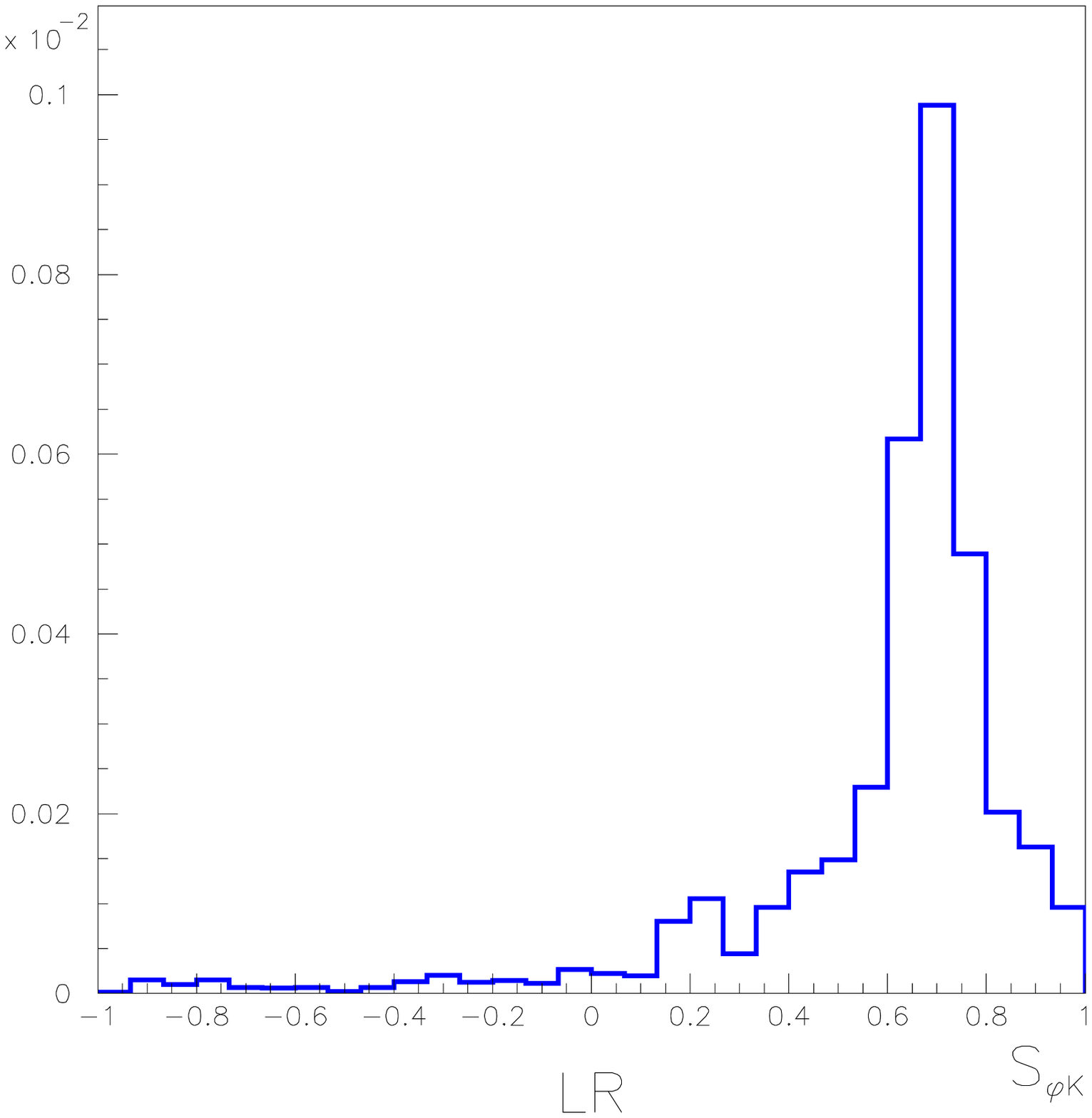} 
\includegraphics[width=0.45\textwidth]{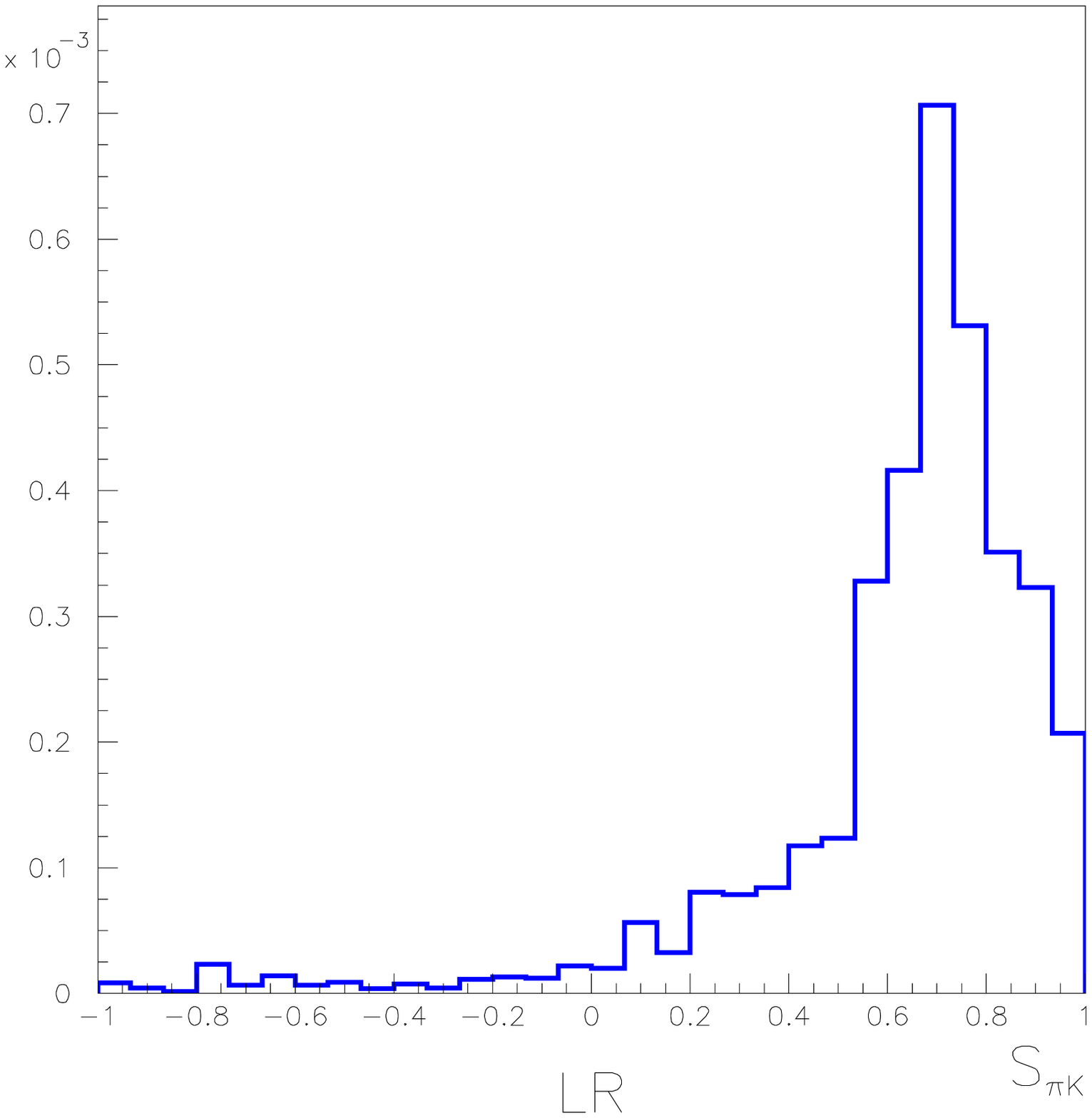} 
\includegraphics[width=0.45\textwidth]{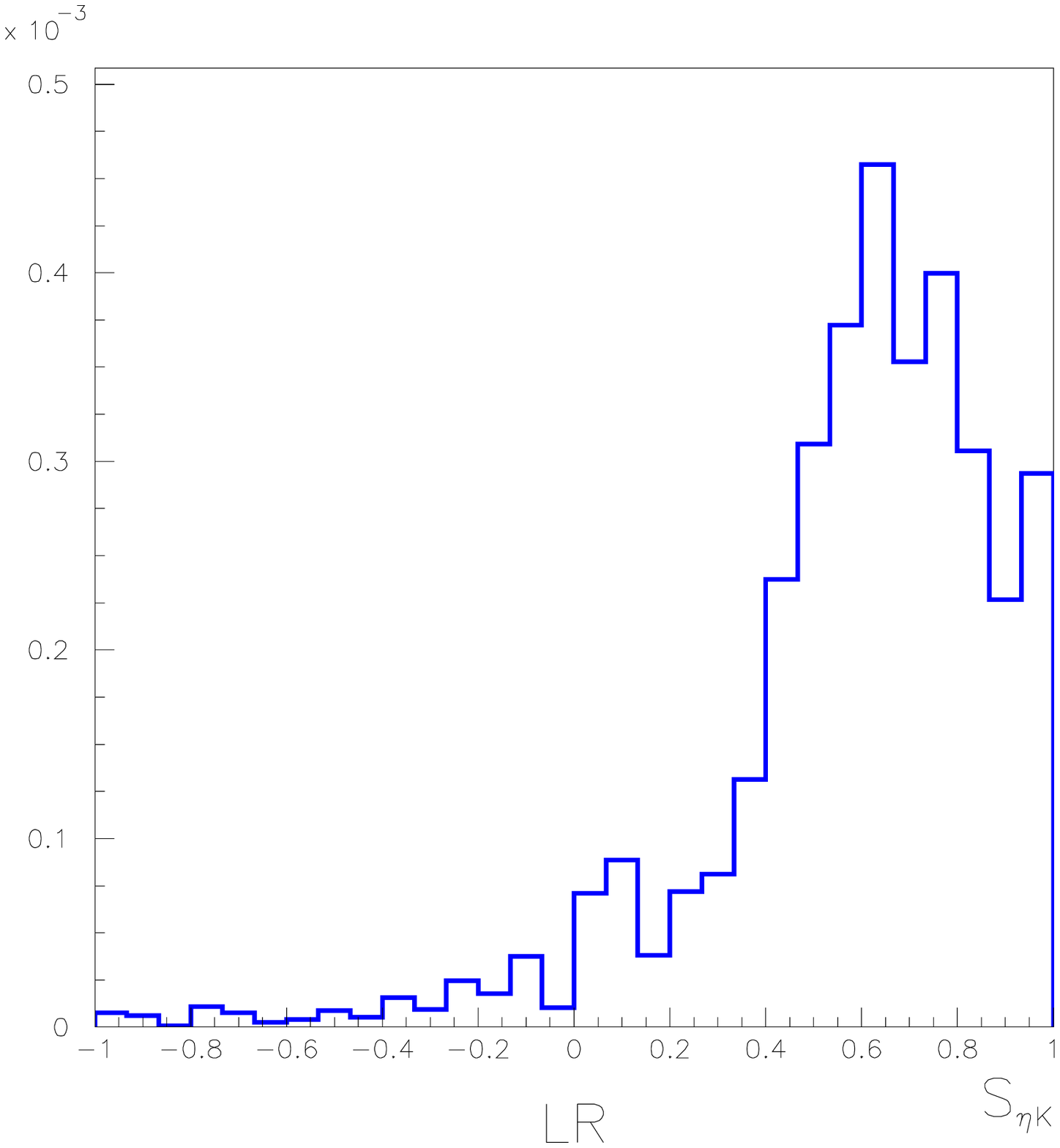} 
\includegraphics[width=0.45\textwidth]{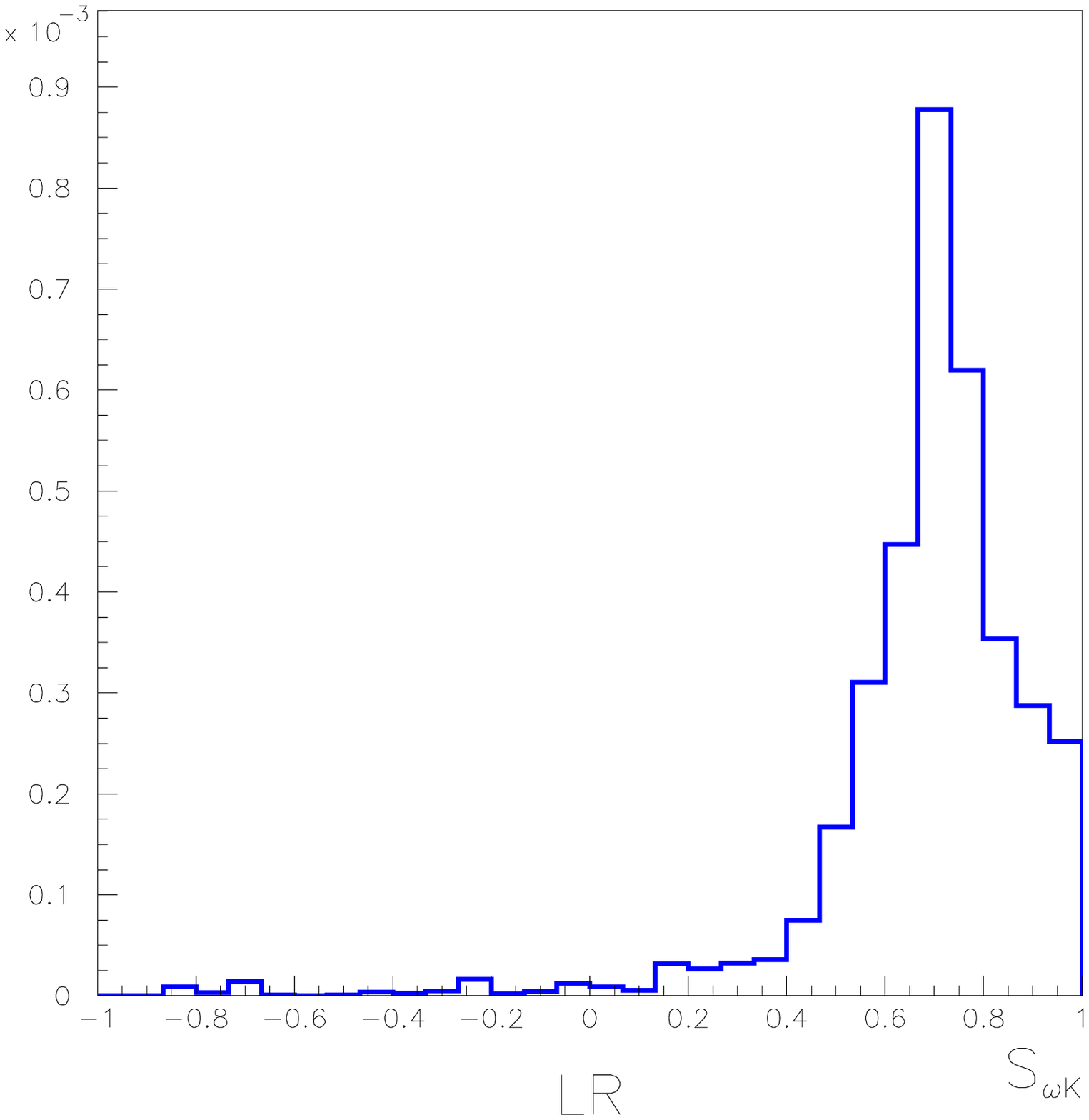} 
\caption{Probability density functions for $S_{\phi K_s}$, $S_{\pi^0
    K_s}$, $S_{\eta^\prime K_s}$ and $S_{\omega K_s}$ induced by
  $(\delta^d_{23})_{\mathrm{LR}}$.}
\label{fig:lr}
\end{center}
\end{figure}

\begin{figure}[!ht]
\begin{center}
\includegraphics[width=0.45\textwidth]{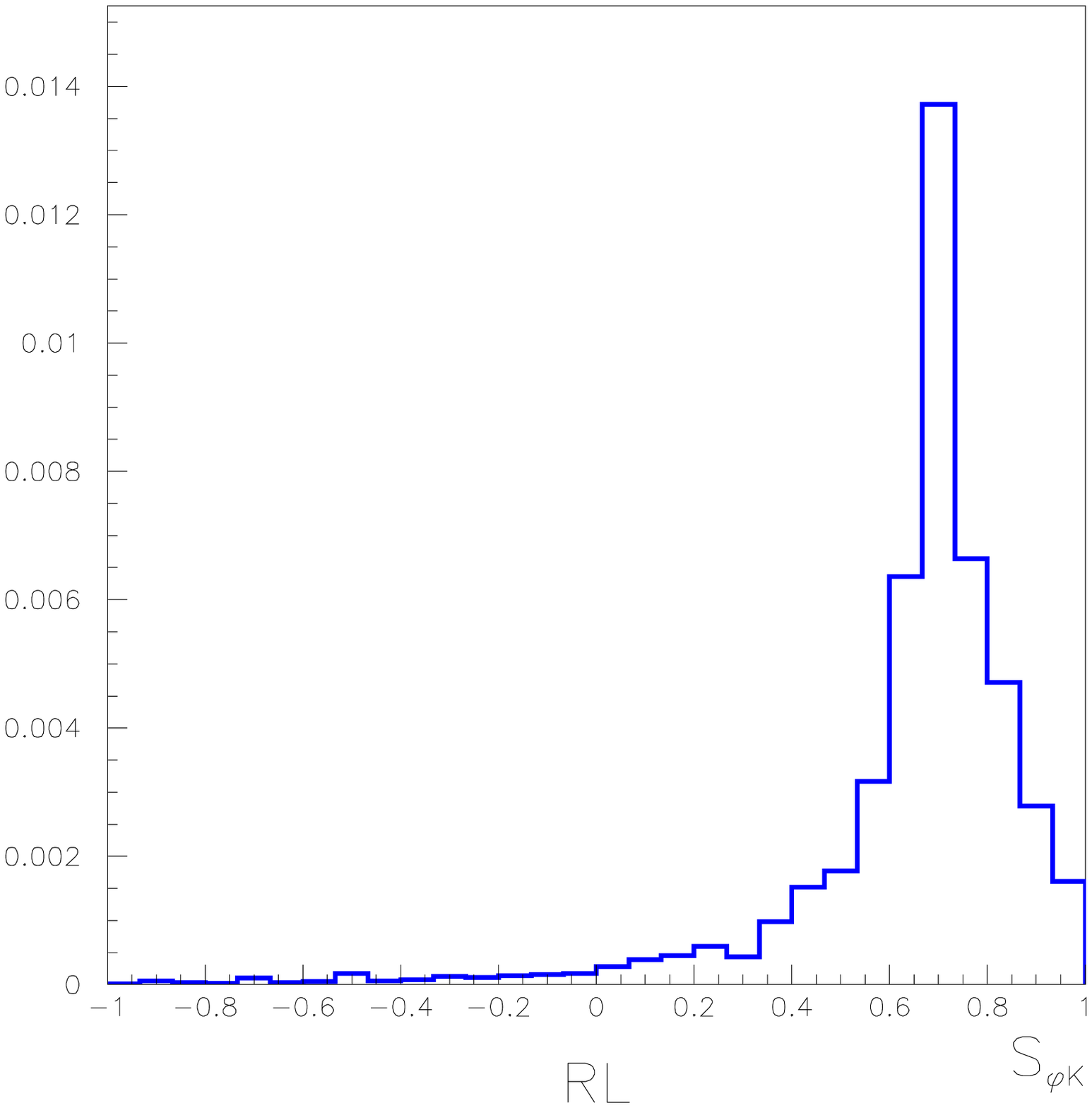} 
\includegraphics[width=0.45\textwidth]{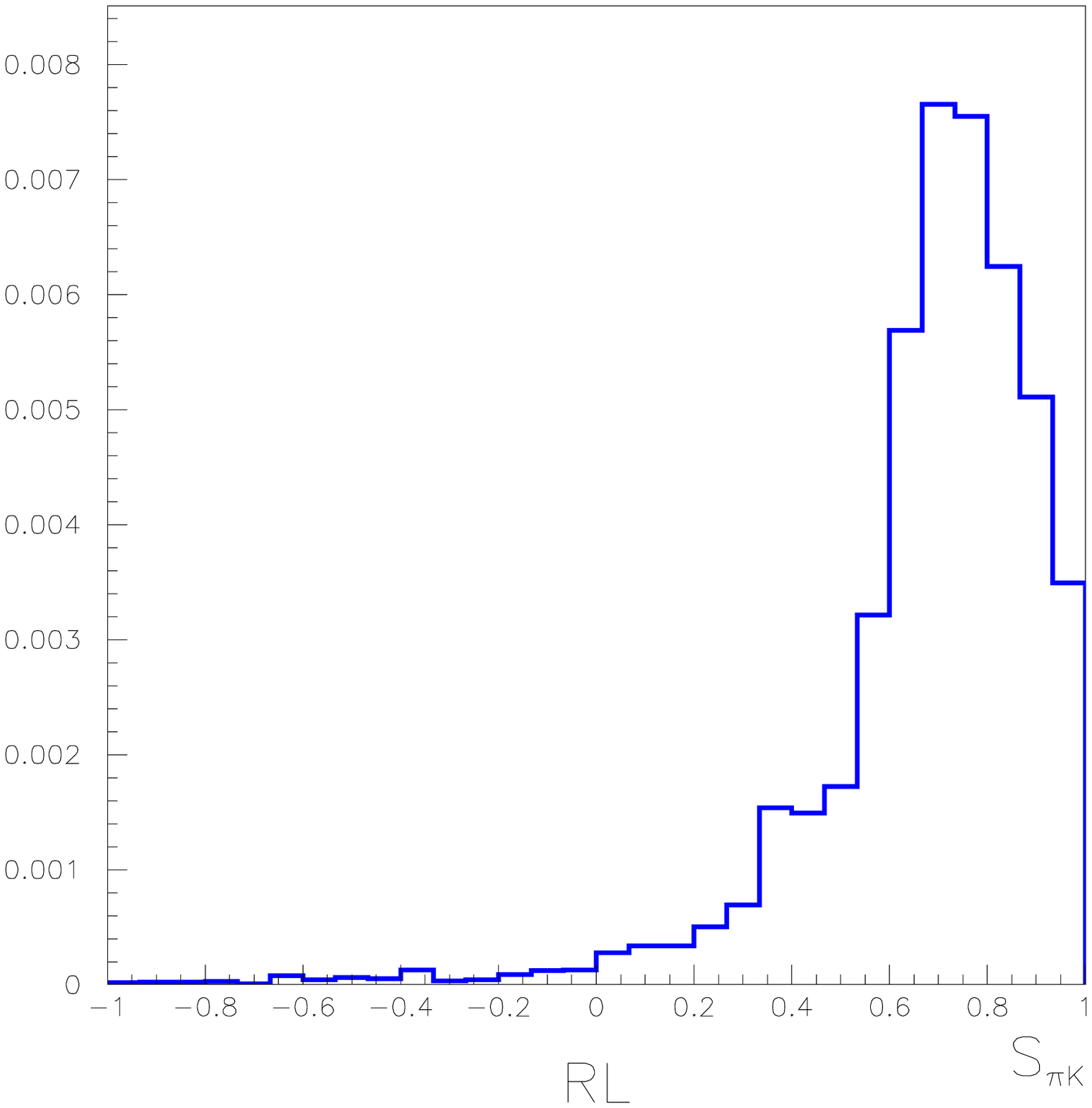} 
\includegraphics[width=0.45\textwidth]{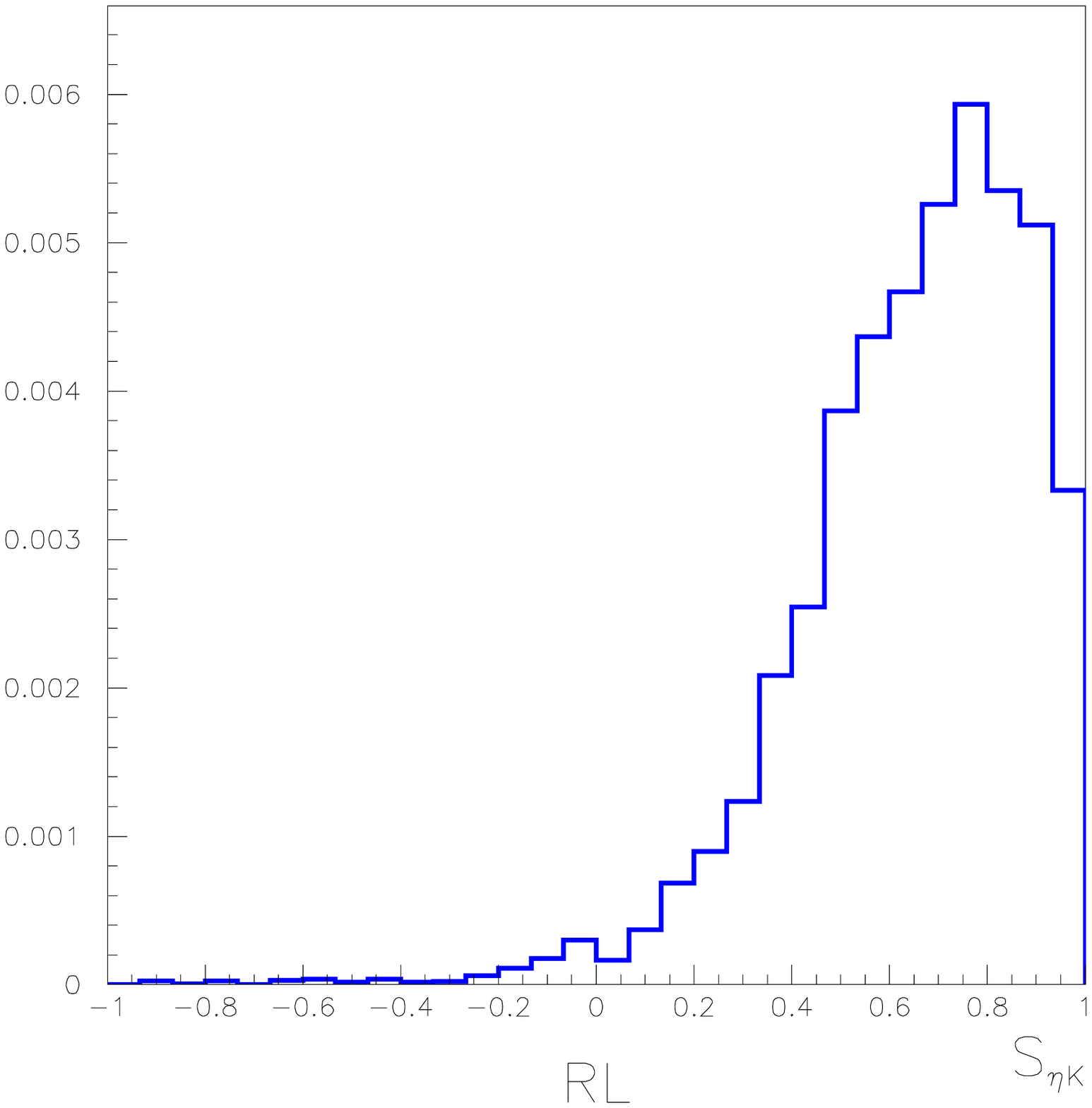} 
\includegraphics[width=0.45\textwidth]{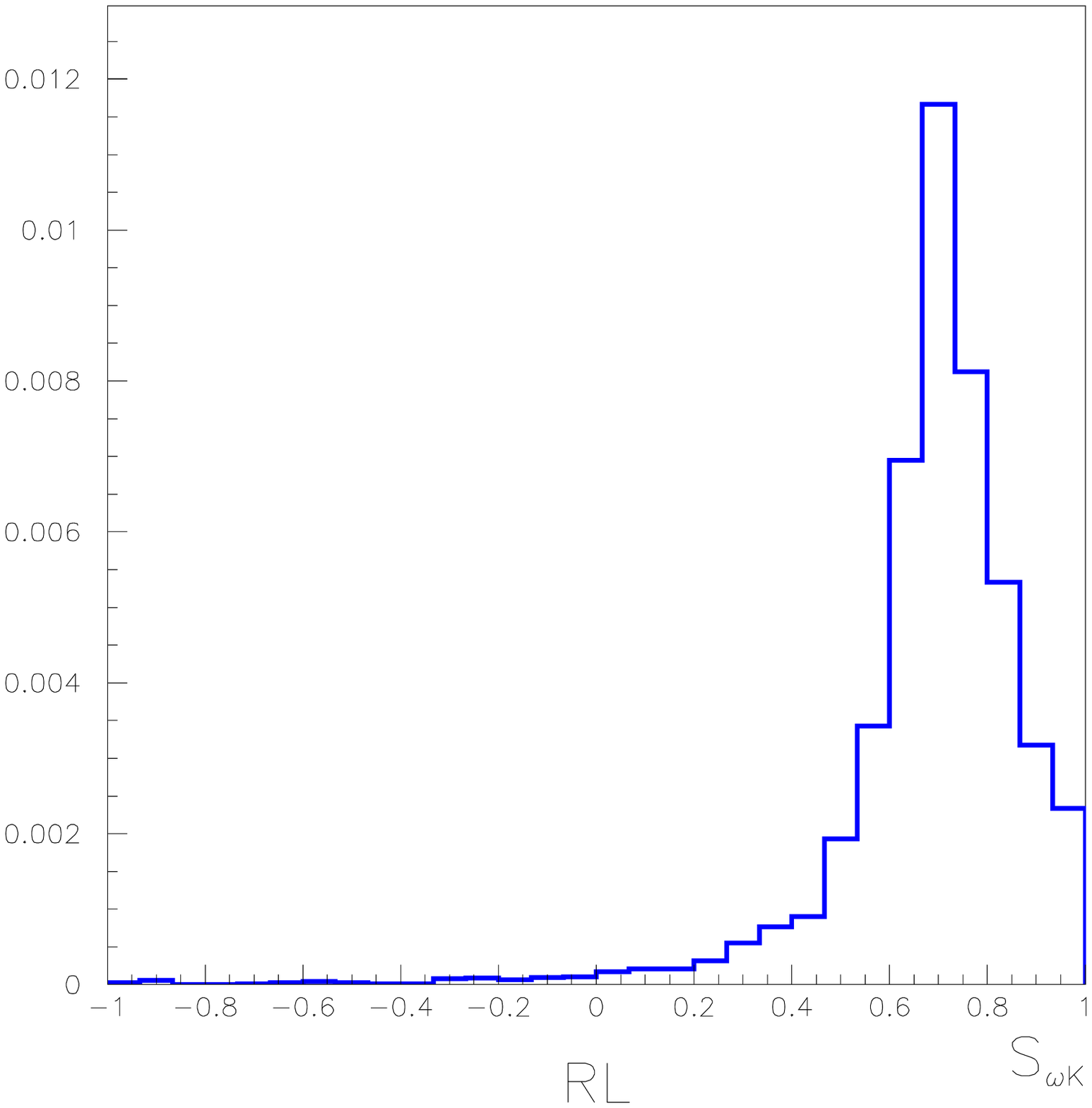} 
\caption{Probability density functions for $S_{\phi K_s}$, $S_{\pi^0
    K_s}$, $S_{\eta^\prime K_s}$ and $S_{\omega K_s}$ induced by
  $(\delta^d_{23})_{\mathrm{RL}}$.}
\label{fig:rl}
\end{center}
\end{figure}

\begin{figure}[!ht]
\begin{center}
\includegraphics[width=0.45\textwidth]{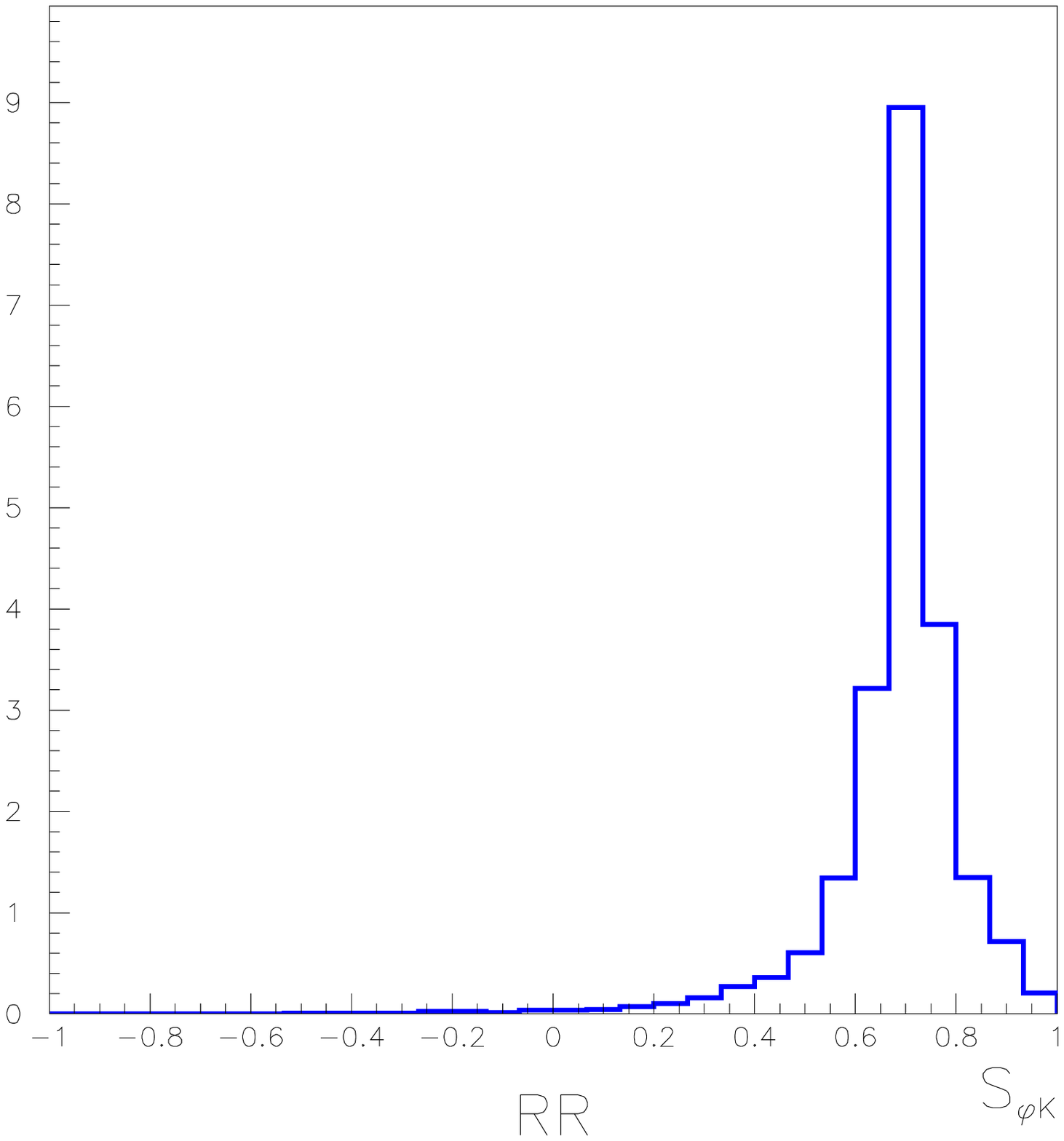} 
\includegraphics[width=0.45\textwidth]{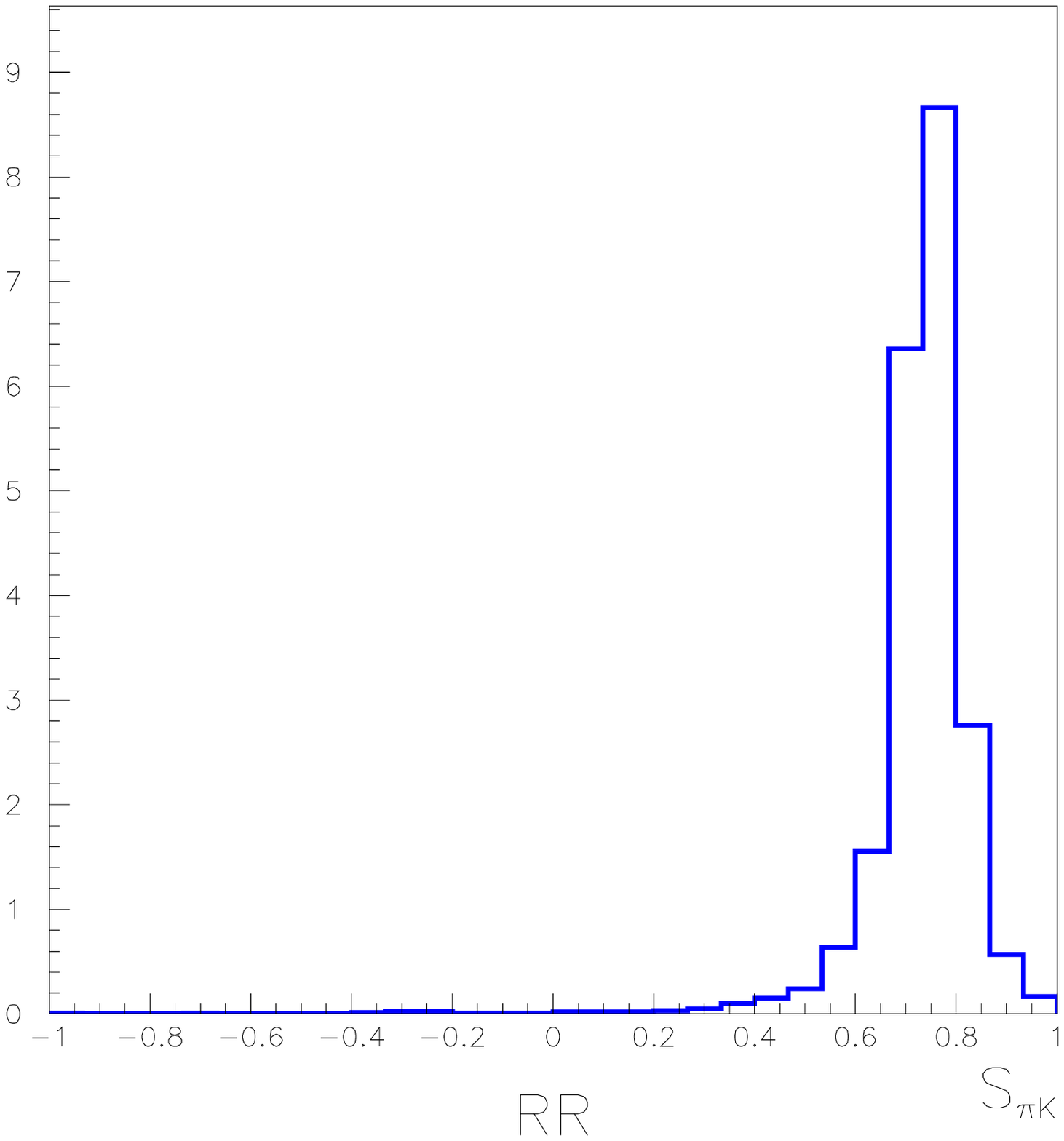} 
\includegraphics[width=0.45\textwidth]{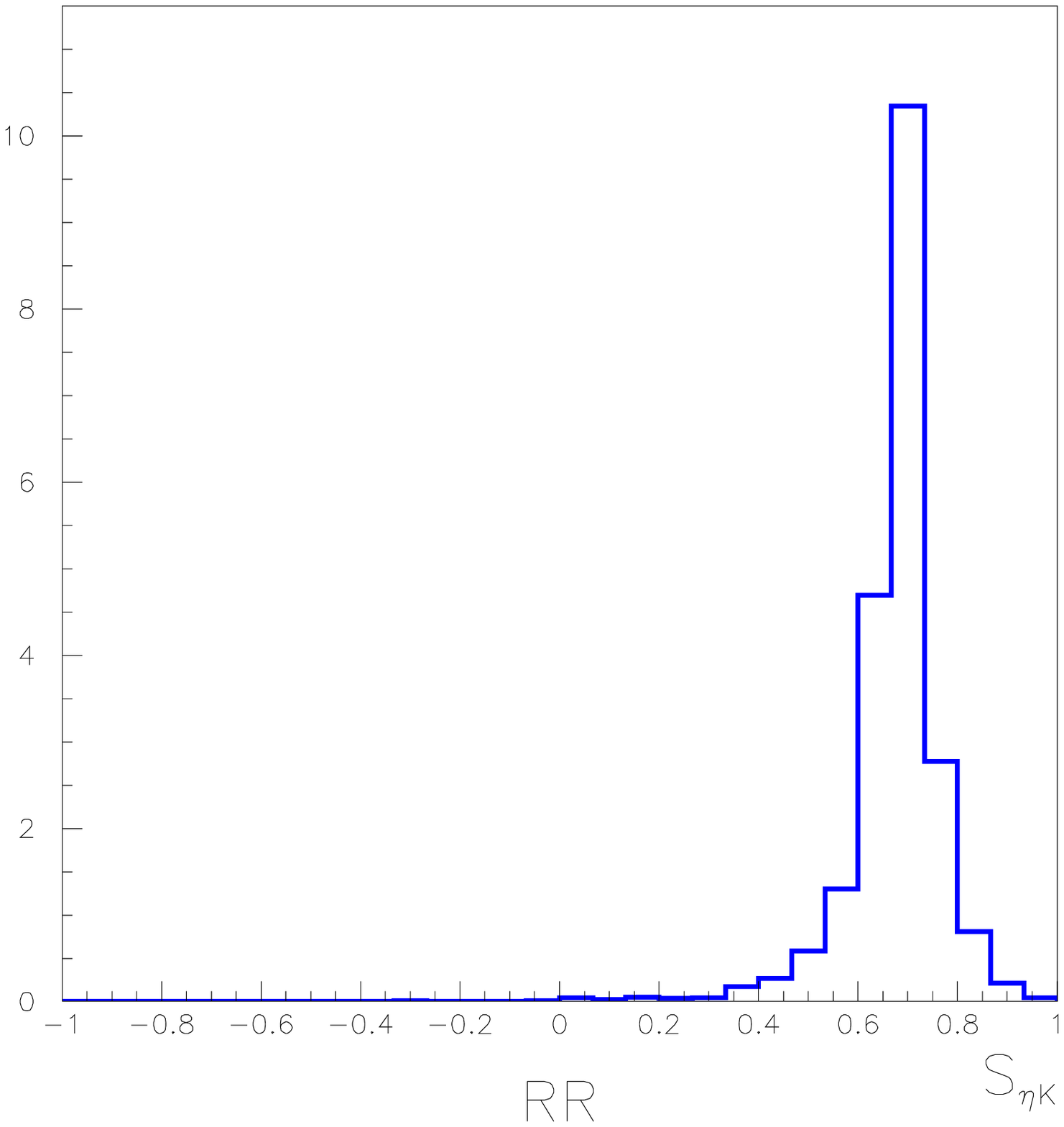} 
\includegraphics[width=0.45\textwidth]{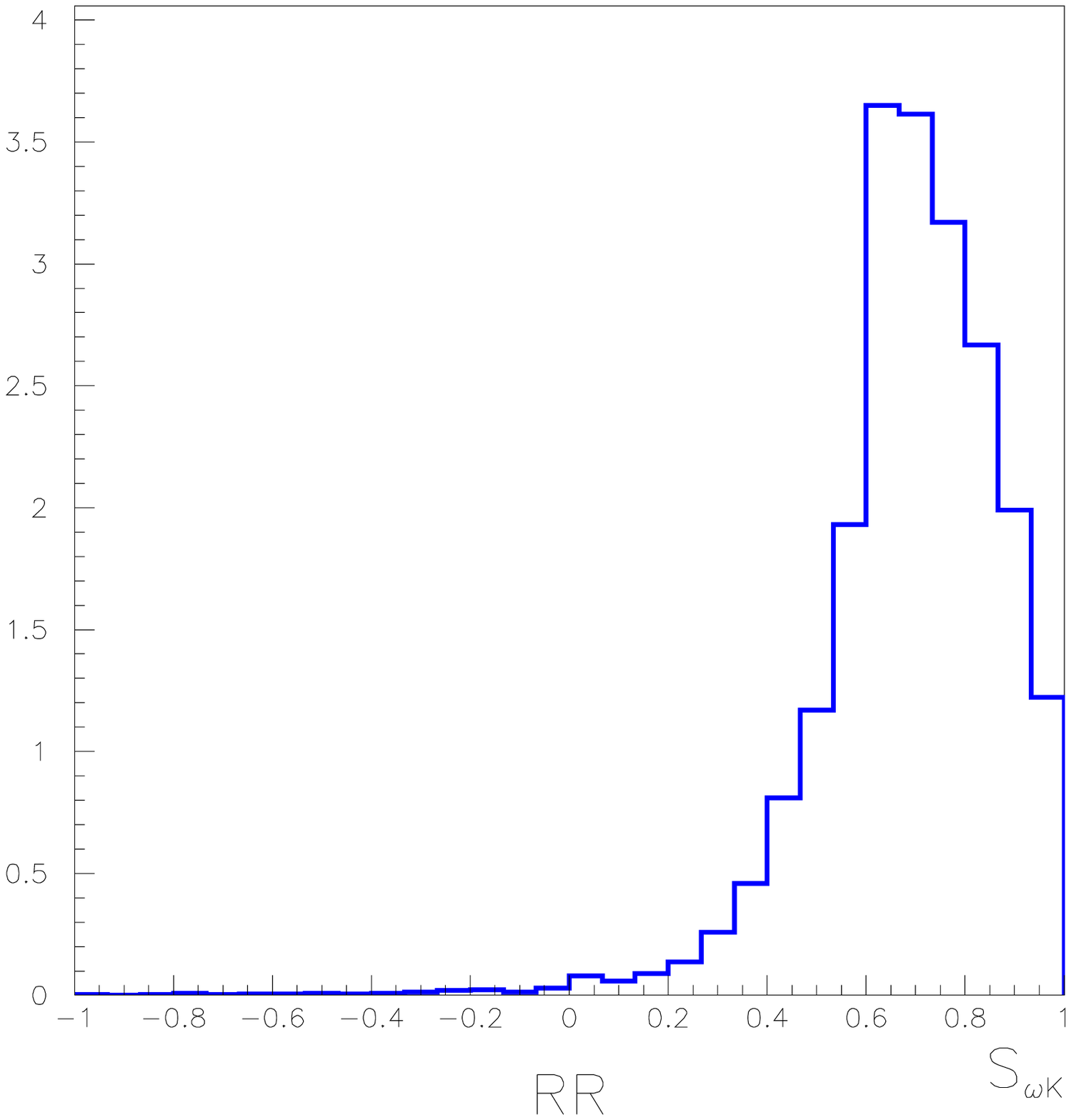} 
\caption{Probability density functions for $S_{\phi K_s}$, $S_{\pi^0
    K_s}$, $S_{\eta^\prime K_s}$ and $S_{\omega K_s}$ induced by
  $(\delta^d_{23})_{\mathrm{RR}}$.}
\label{fig:rr}
\end{center}
\end{figure}

\begin{figure}[!ht]
\begin{center}
\includegraphics[width=0.45\textwidth]{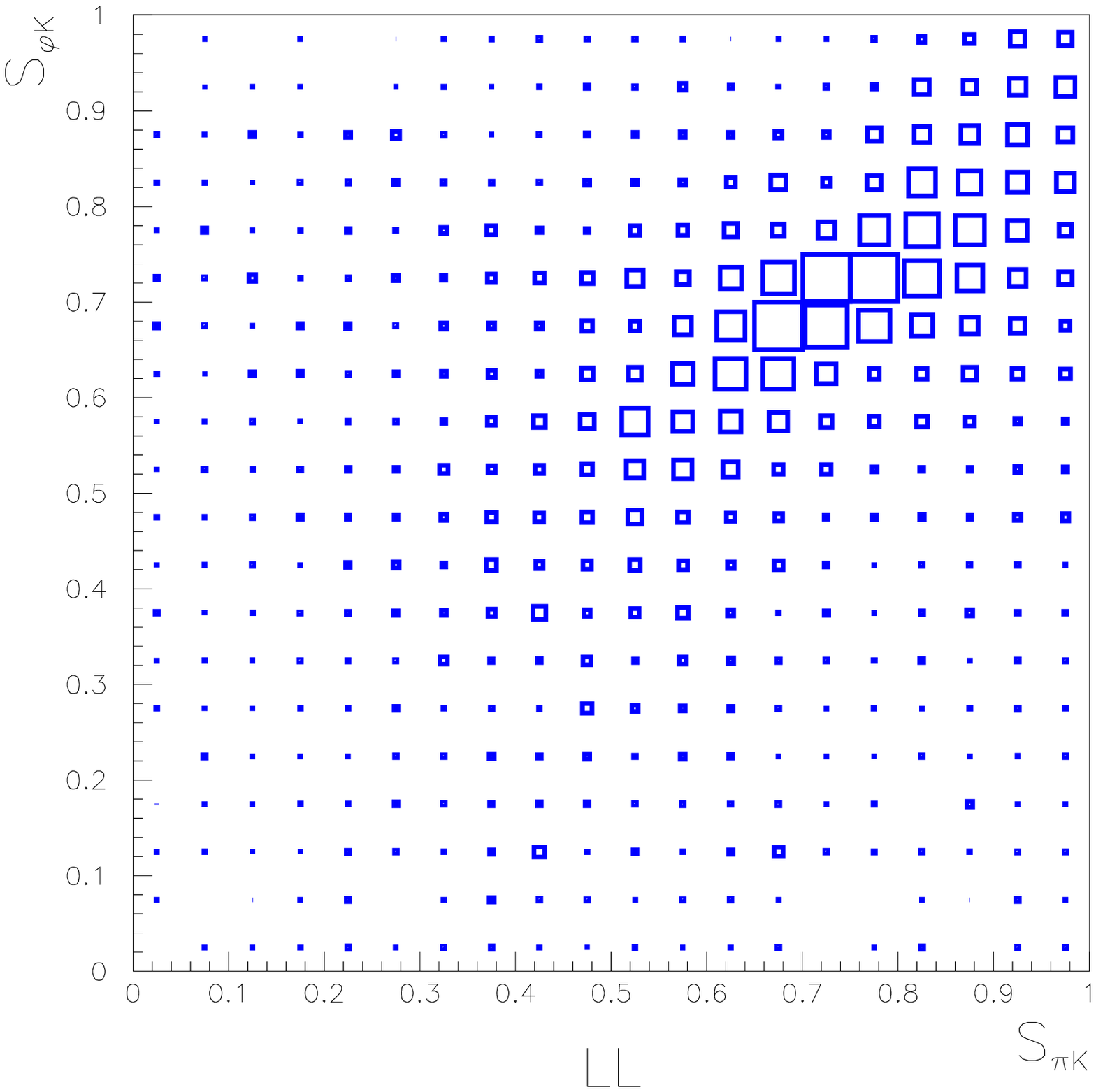} 
\includegraphics[width=0.45\textwidth]{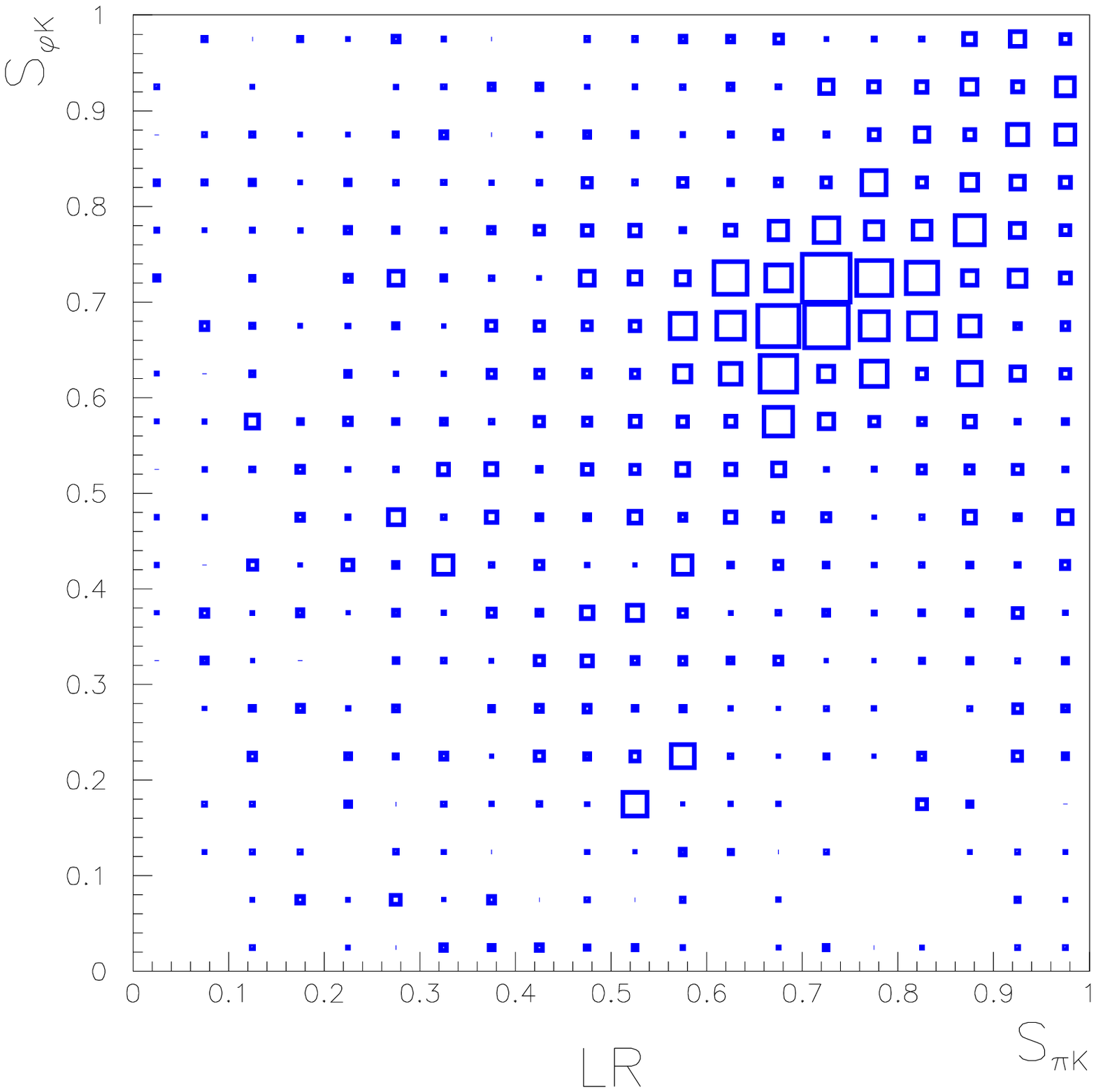} 
\includegraphics[width=0.45\textwidth]{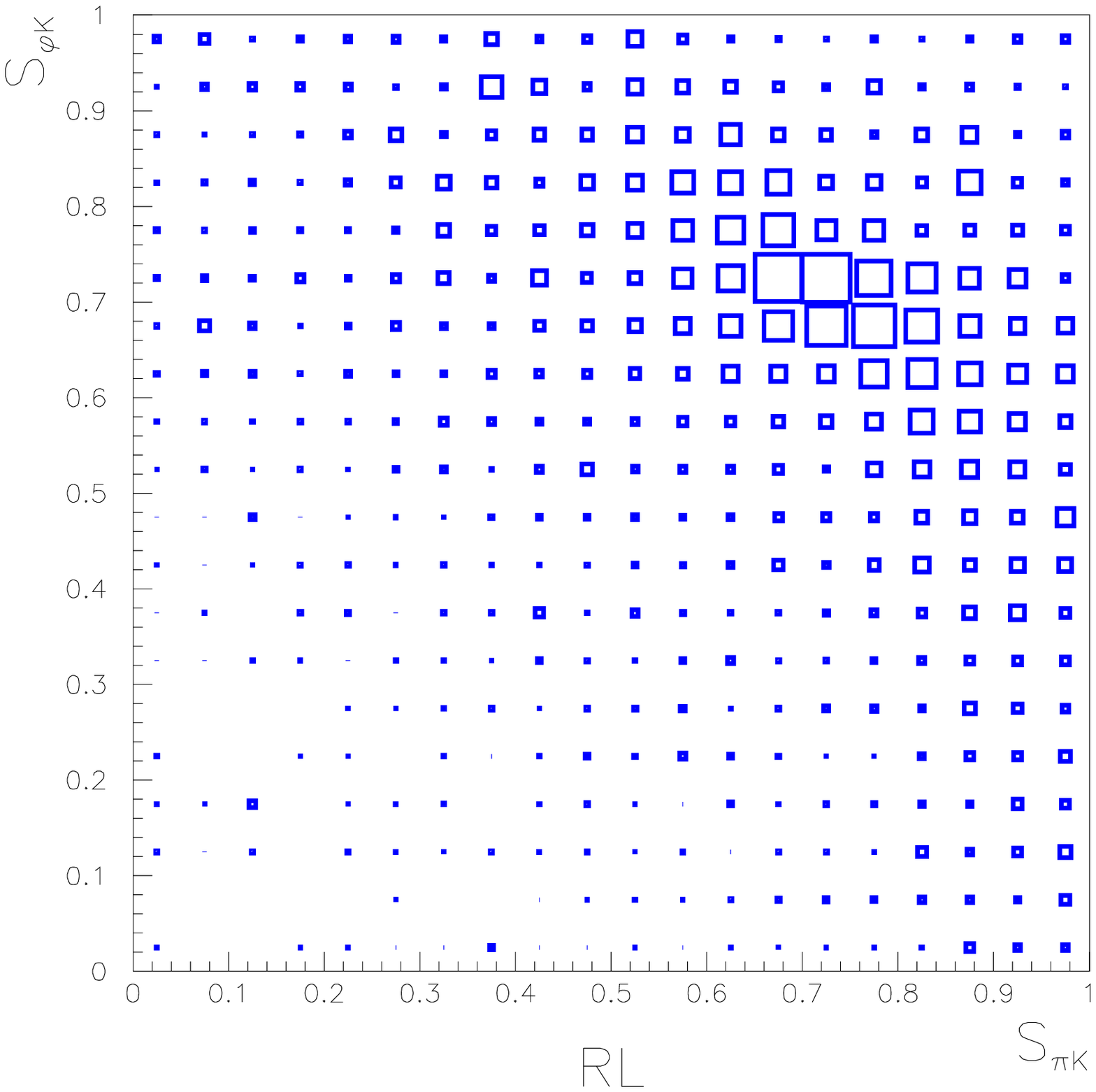} 
\includegraphics[width=0.45\textwidth]{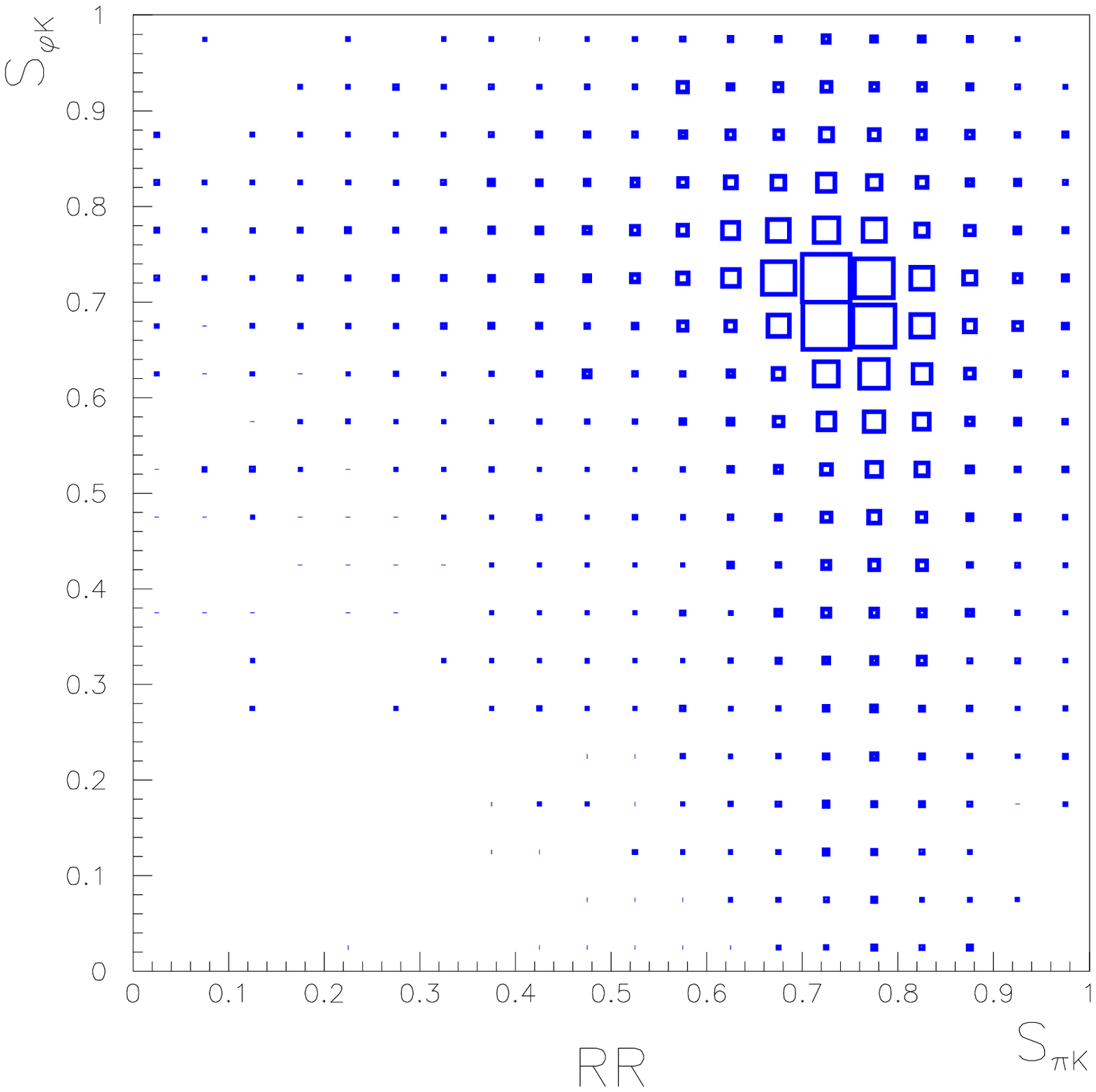} 
\caption{Correlation between $S_{\phi K_s}$ and $S_{\pi^0
    K_s}$ for $LL$, $LR$, $RL$ and $RR$ mass insertions.}
\label{fig:kphivskp0}
\end{center}
\end{figure}

Having determined the p.d.f's for the four $\delta$'s, we now turn to
the evaluation of the time-dependent CP asymmetries. As we discussed
in Sec.~\ref{sec:basic}, the uncertainty in the calculation of SUSY
effects is even larger than the SM one. Furthermore, we cannot use the
GP approach since to estimate the SUSY contribution we need to
evaluate the hadronic matrix elements explicitly.  Following
ref.~\cite{hep-ph/0212397}, we use QCDF, enlarging the range for
power-suppressed contributions to annihilation chosen in
Ref.~\cite{hep-ph/0308039} as suggested in Ref.~\cite{hep-ph/0104126}.
We warn the reader about the large theoretical uncertainties that
affect this evaluation.

In Figs.~\ref{fig:ll}-\ref{fig:rr} we present the results for $S_{\phi
  K_s}$, $S_{\pi^0 K_s}$, $S_{\eta^\prime K_s}$ and $S_{\omega K_s}$.
They do not show a sizable dependence on the sign of $\mu$ or on $\tan
\beta$ for the chosen range of SUSY parameters. We see that:
\begin{itemize}
\item deviations from the SM expectations are possible in all
  channels, and the present experimental central values can be reproduced;
\item they are more easily generated by $LR$ and $RL$ insertions, due
  to the enhancement mechanism discussed above.
\item As noticed in refs.~\cite{hep-ph/0407076,hep-ph/0409245}, the
  correlation between $\Delta S_{PP}$ and $\Delta S_{PV}$ depends on
  the chirality of the NP contributions. For example, we show in
  Fig.~\ref{fig:kphivskp0} the correlation between $\Delta S_{K_S\phi}$
  and $\Delta S_{K_s\pi^0}$ for the four possible choices for mass
  insertions. We see that the $\Delta S_{K_S\phi}$ and $\Delta
  S_{K_s\pi^0}$ are correlated for $LL$ and $LR$ mass insertions, and
  anticorrelated for $RL$ and $RR$ mass insertions.
\end{itemize}

\begin{figure}[!ht]
\begin{center}
\includegraphics[width=0.45\textwidth]{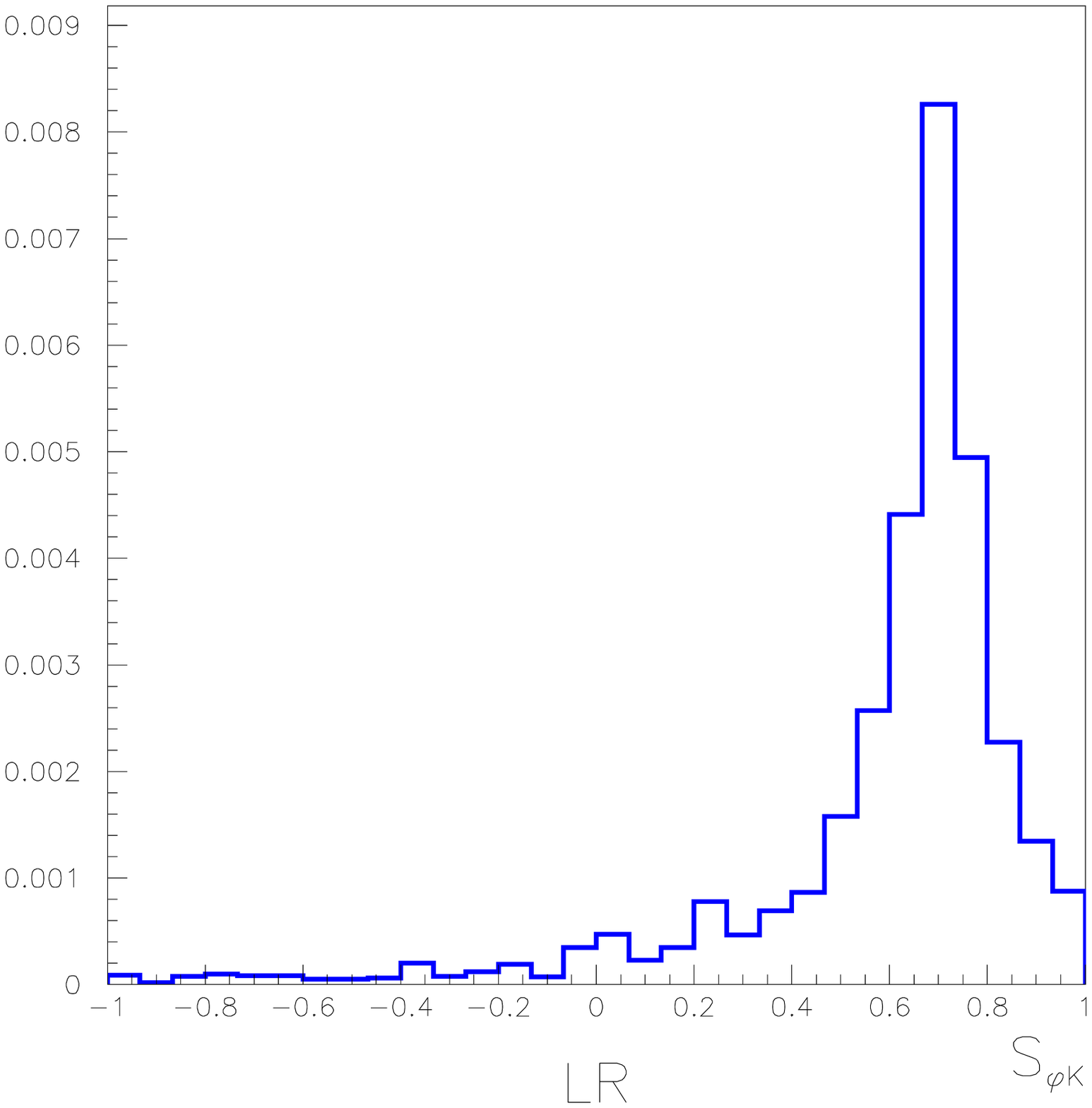} 
\includegraphics[width=0.45\textwidth]{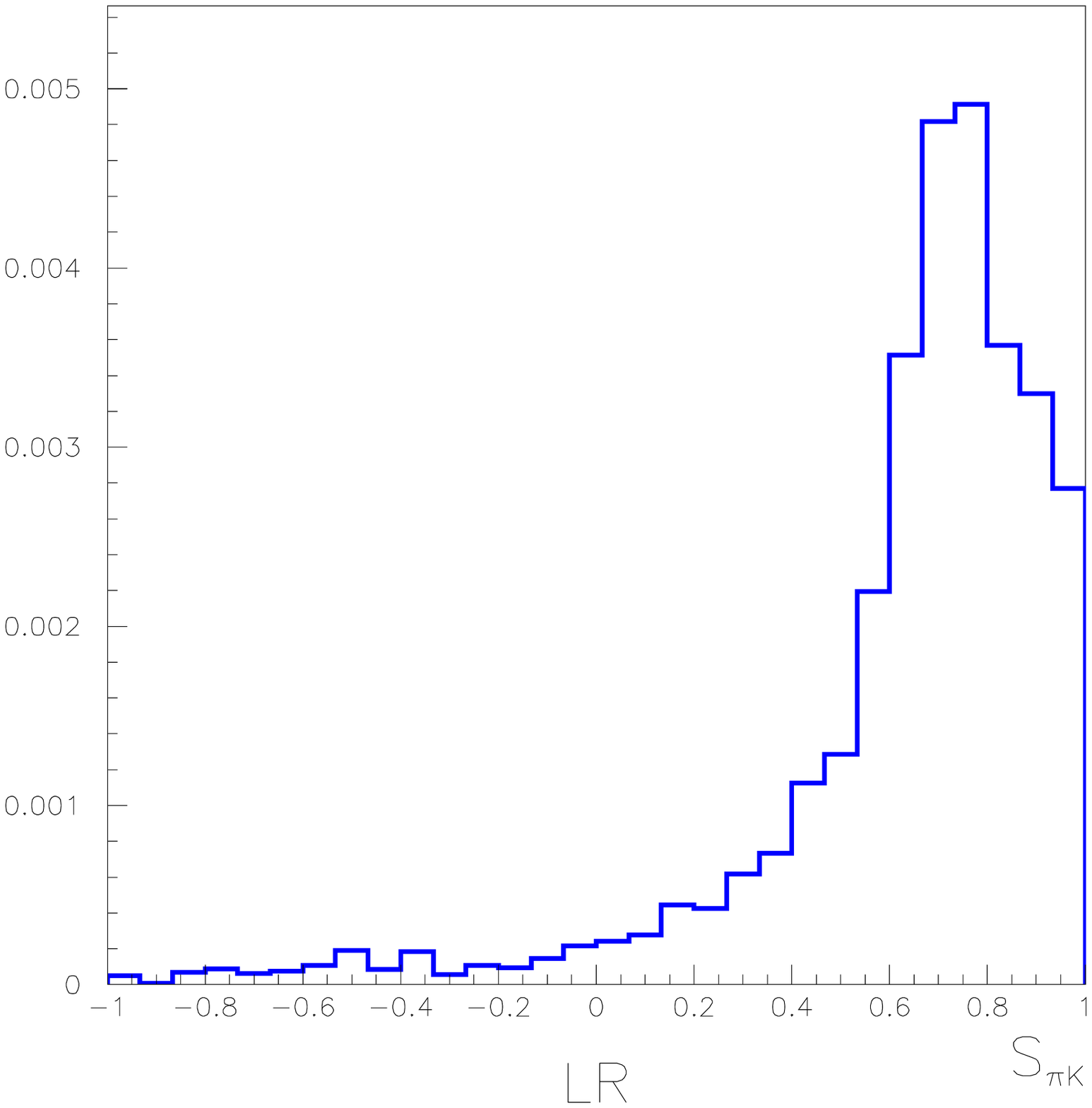} 
\includegraphics[width=0.45\textwidth]{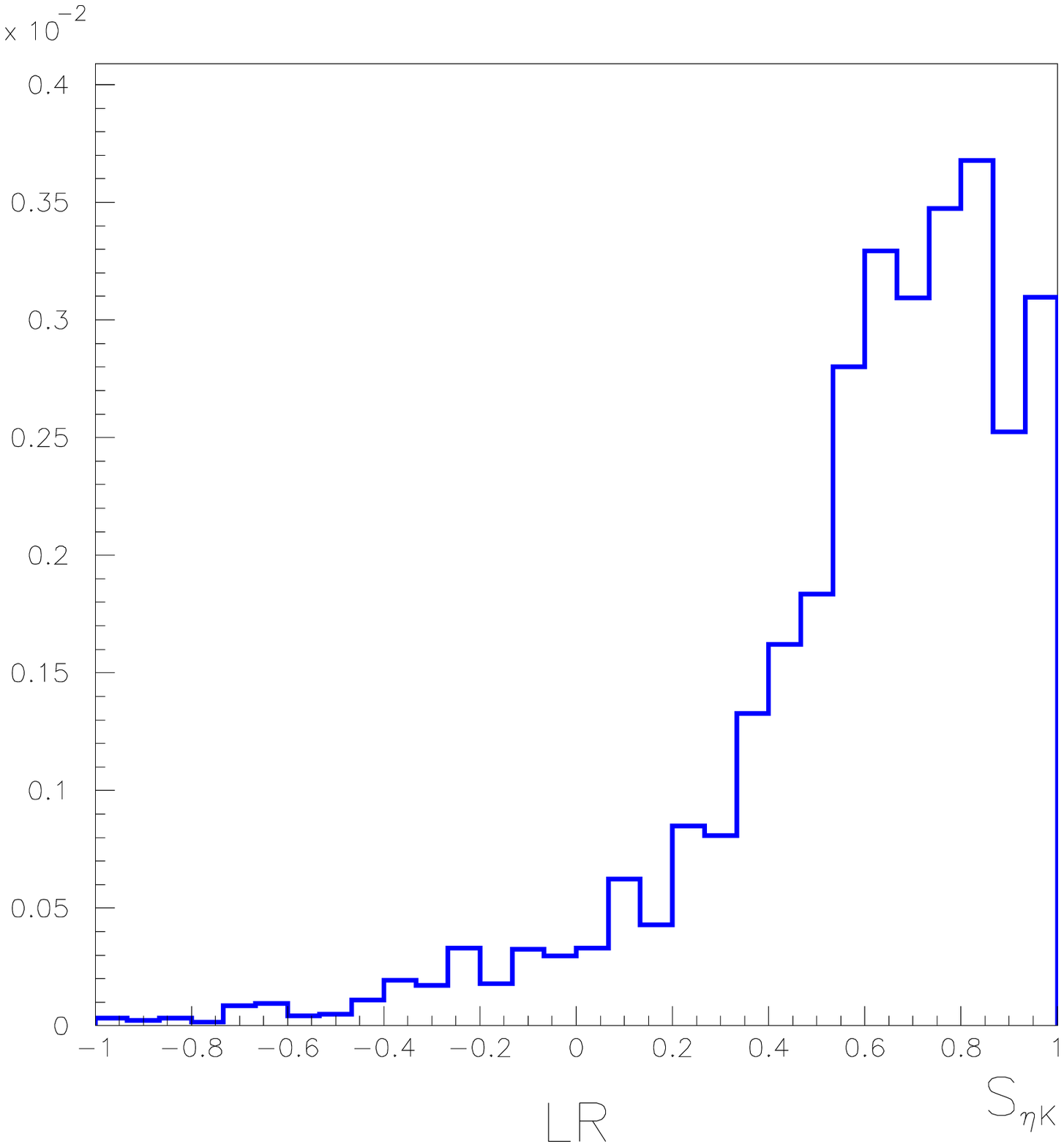} 
\includegraphics[width=0.45\textwidth]{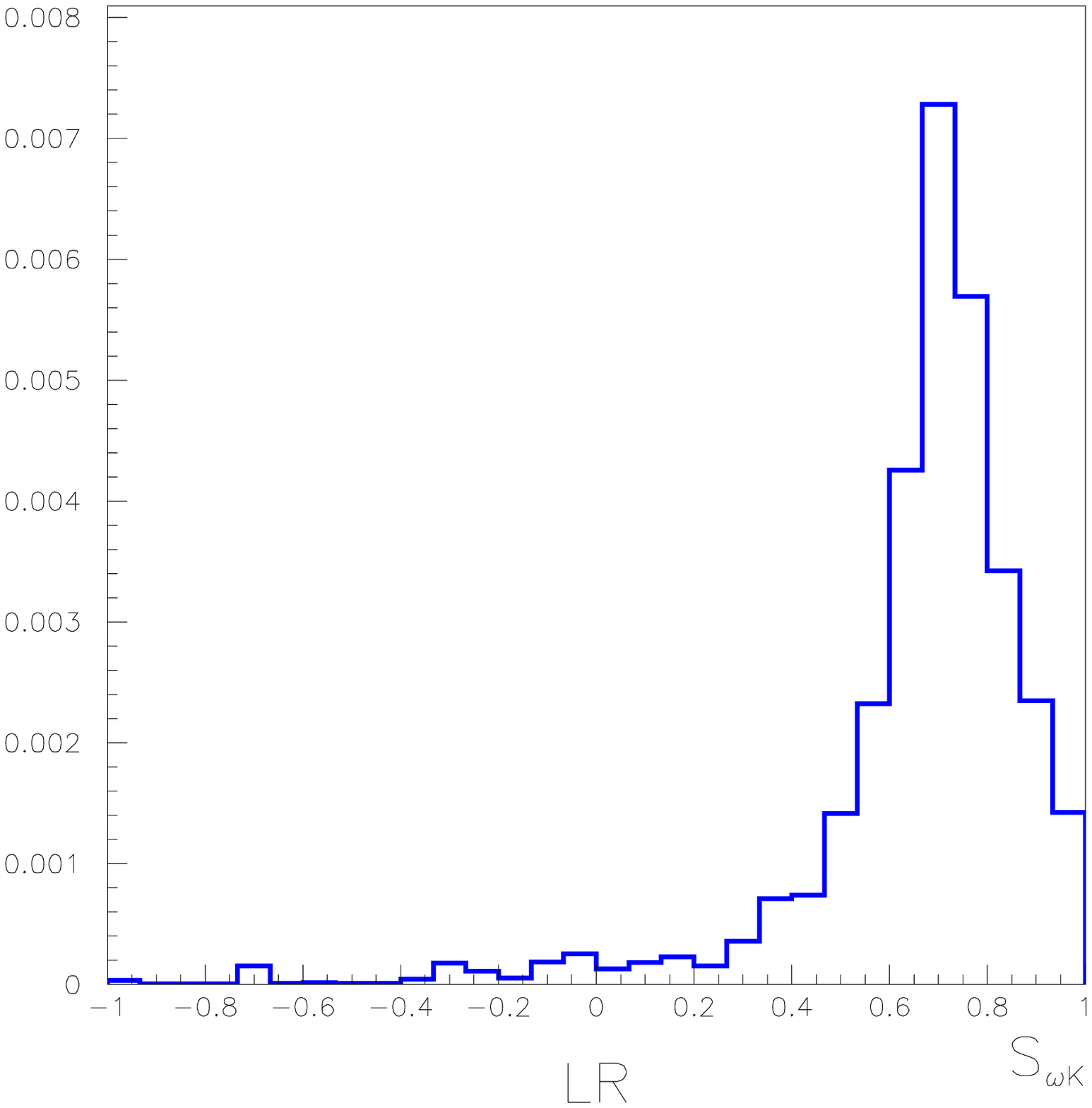} 
\caption{Probability density functions for $S_{\phi K_s}$, $S_{\pi^0
    K_s}$, $S_{\eta^\prime K_s}$ and $S_{\omega K_s}$ induced by
  $(\delta^d_{23})_{\mathrm{LR}}$ for $m_{\tilde g} = m_{\tilde q}
  =\mu=1$ TeV.}
\label{fig:1TeVLR}
\end{center}
\end{figure}

\begin{figure}[!ht]
\begin{center}
\includegraphics[width=0.45\textwidth]{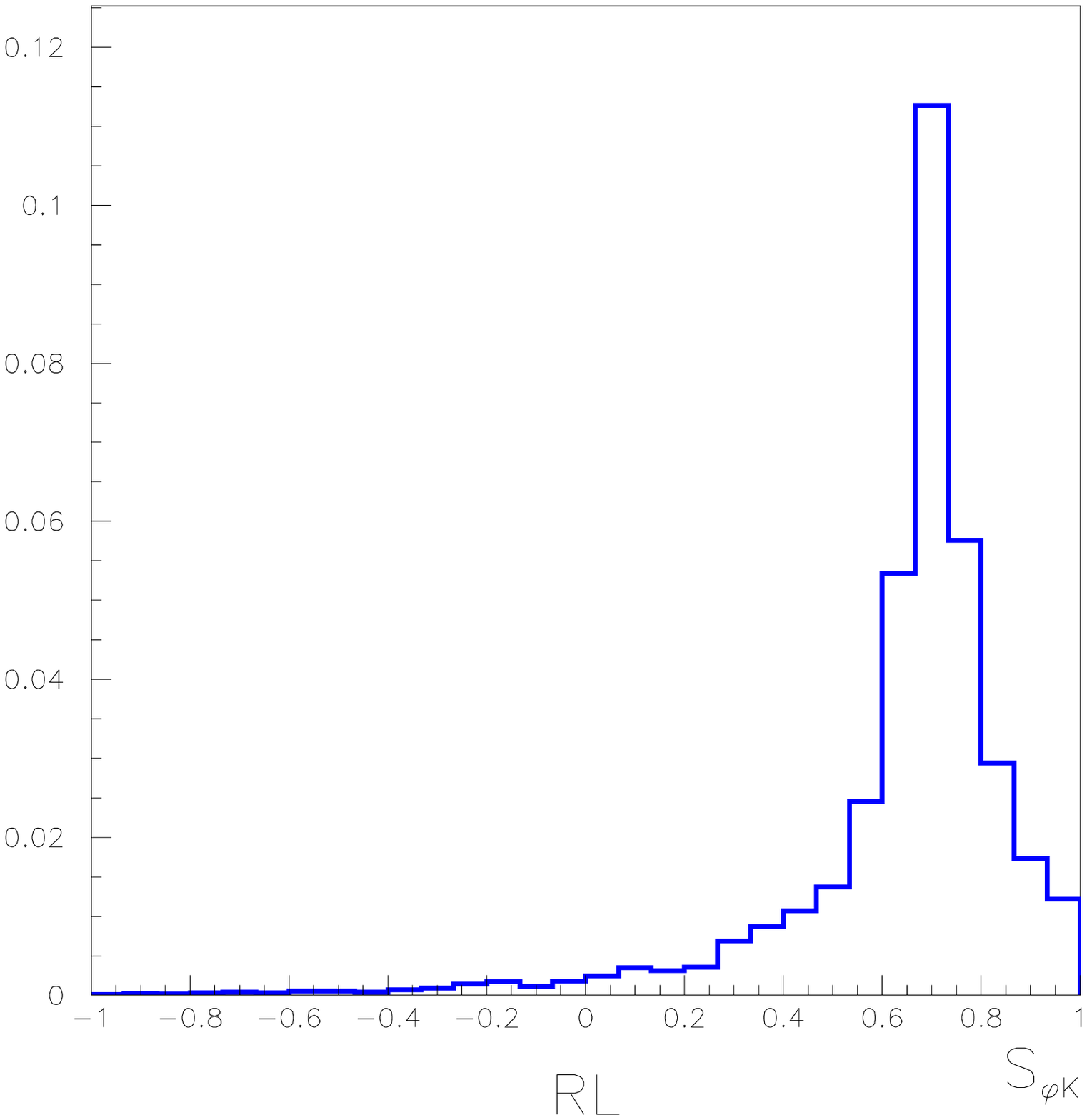} 
\includegraphics[width=0.45\textwidth]{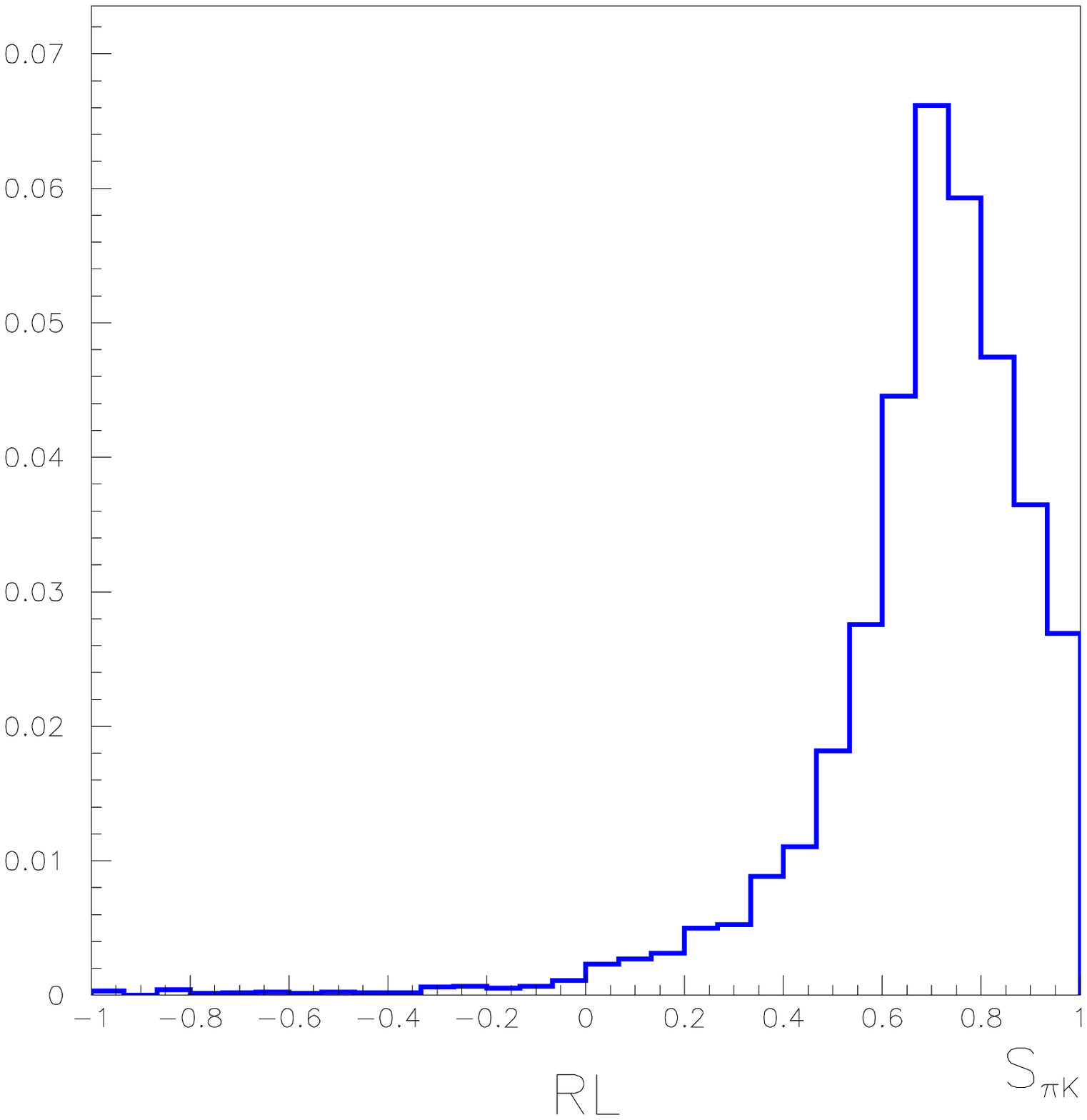} 
\includegraphics[width=0.45\textwidth]{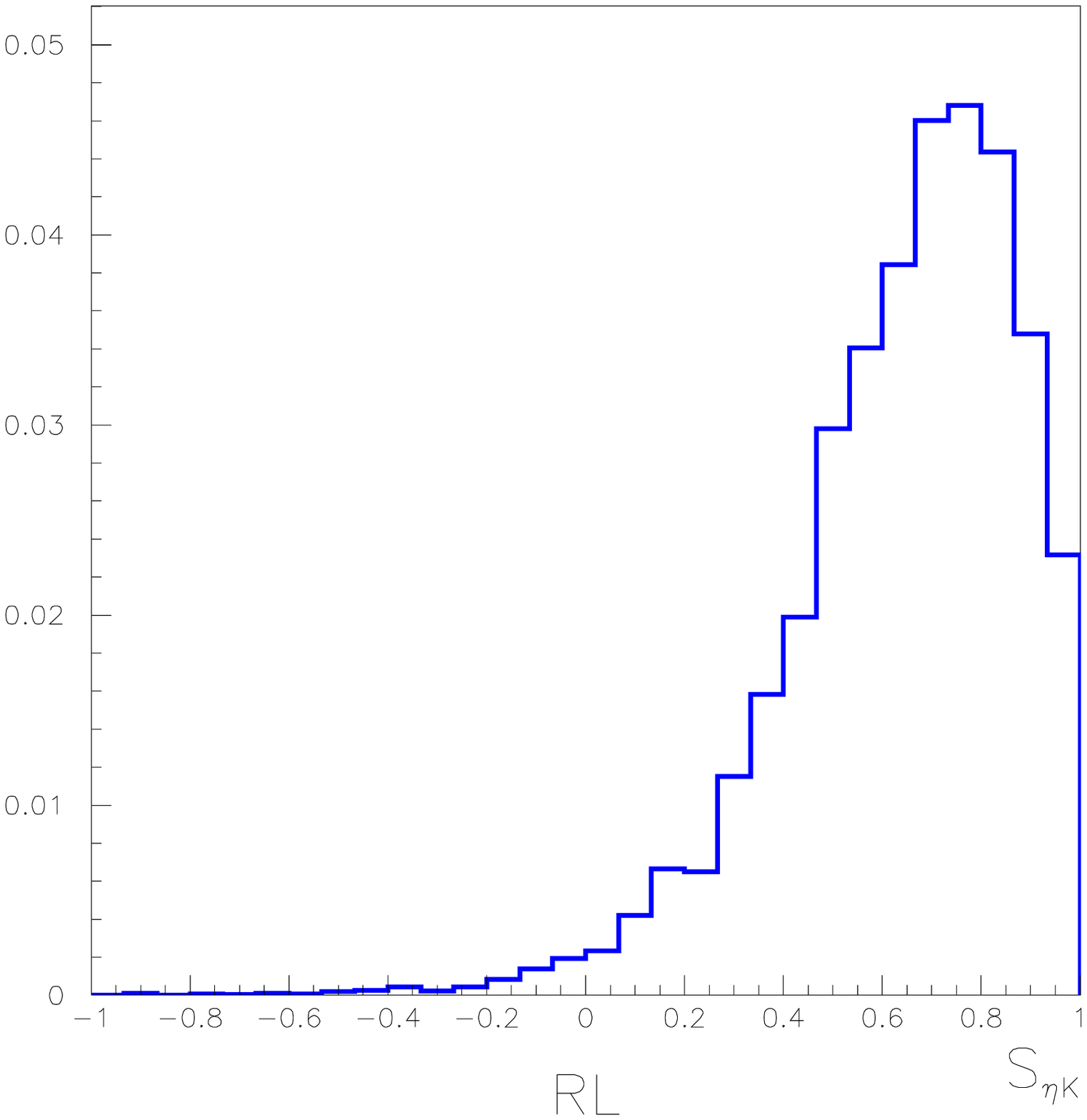} 
\includegraphics[width=0.45\textwidth]{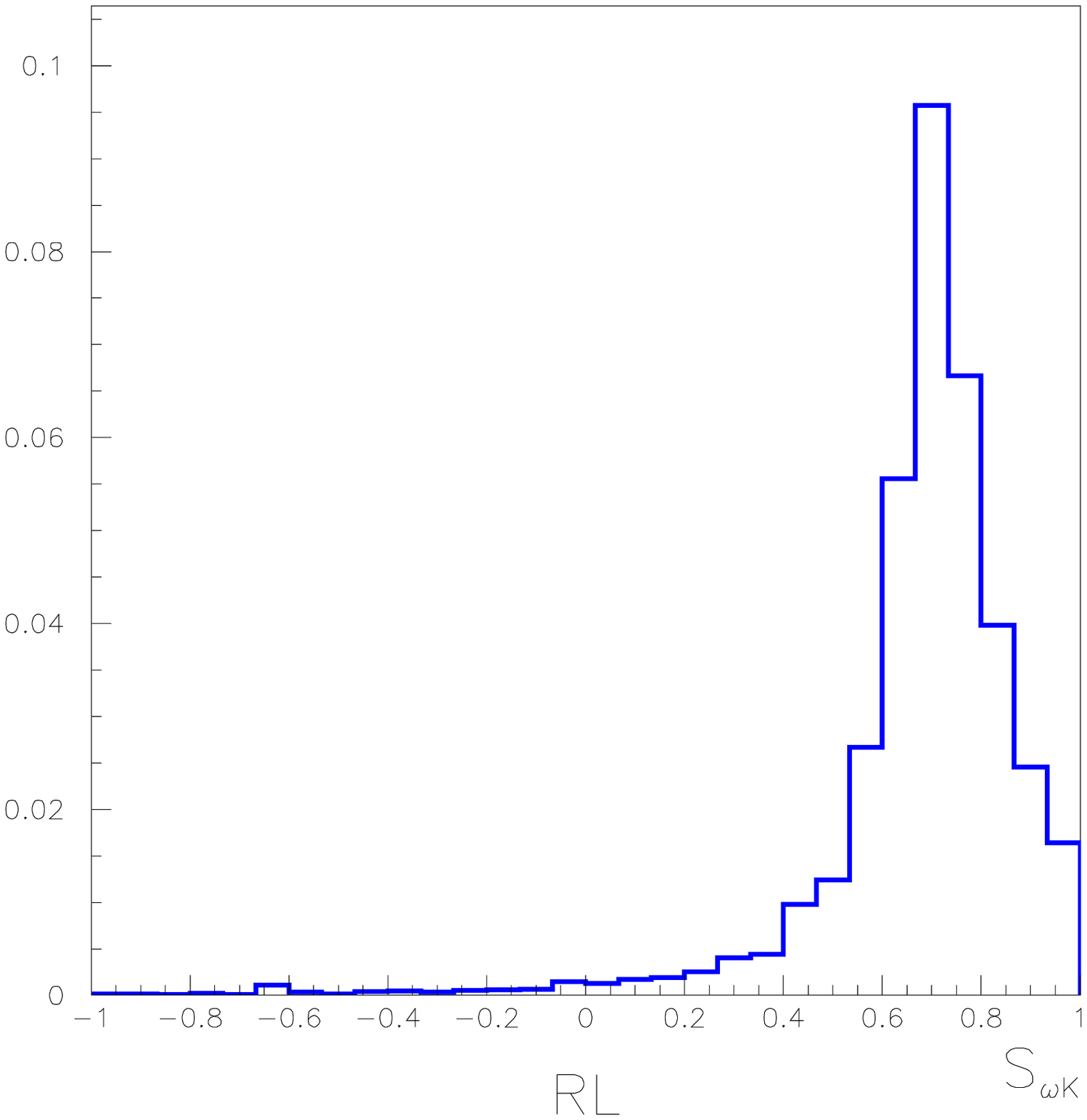} 
\caption{Probability density functions for $S_{\phi K_s}$, $S_{\pi^0
    K_s}$, $S_{\eta^\prime K_s}$ and $S_{\omega K_s}$ induced by
  $(\delta^d_{23})_{\mathrm{RL}}$ for $m_{\tilde g} = m_{\tilde q}
  =\mu=1$ TeV.}
\label{fig:1TeVRL}
\end{center}
\end{figure}

An interesting issue is the scaling of SUSY effects in $\Delta S$ with
squark and gluino masses. We have noticed above that the constraints
from other processes scale linearly with the SUSY masses. Now, it
turns out that also the dominant SUSY contribution to $\Delta S$, the
chromomagnetic one, scales linearly with SUSY masses as long as
$m_{\tilde g} \sim m_{\tilde q} \sim \mu$. This means that there is no
decoupling of SUSY contributions to $\Delta S$ as long as the
constraint from other processes can be saturated for $\delta <1$. From
Figs.~\ref{fig:LL}-\ref{fig:RR} we see that the bounds on $LL$ and
$RR$ mass insertions quickly reach the physical boundary at
$\delta=1$, while $LR$ and $RL$ are safely below that bound. Chirality
flipping $LR$ and $RL$ insertions cannot become too large in order to
ensure the absence of charge and color breaking minima and unbounded
from below directions in the scalar potential~\cite{hep-ph/9606237}.
However, it is easy to check that the flavor bounds given above are
stronger for SUSY masses up to (and above) the TeV scale. We conclude
that $LR$ and $RL$ mass insertions can give observable effects for
SUSY masses within the reach of LHC and even above. This is shown
explicitly in Figs.~\ref{fig:1TeVLR} and \ref{fig:1TeVRL}, where we
present the p.d.f. for $S_{\phi K_s}$, $S_{\pi^0 K_s}$,
$S_{\eta^\prime K_s}$ and $S_{\omega K_s}$ for SUSY masses of $1$ TeV.

\section{NON-SUSY MODELS}
\label{sec:non-SUSY}

In general, one expects sizable values of $\Delta S$ in all models in
which new sources of CP violation are present in $b \to s$ penguins.
In particular, models with a fourth generation, both vectorlike and
sequential, models with warped extra dimensions in which the flavor
structure of the SM is obtained using localization of fermion wave
functions, and models with $Z^\prime$ gauge bosons can all give
potentially large contributions to $b \to s$ penguins
\cite{hep-ph/0307251,hep-ph/0310144,hep-ph/0503151,hep-ph/0509117,hep-ph/0611107}.

In any given NP model, it is possible to perform a detailed analysis
along the lines of Sec.~\ref{sec:SUSY}, considering the constraints
from $B_s - \bar B_s$ mixing and from rare $B$ decays, plus the
constraints from all other sectors if they are correlated with $b \to
s$ transitions in the model. On general grounds, the dominant
contributions to $b \to s$ hadronic decays are expected to come from
electroweak or chromomagnetic penguins. The correlation between the
induced  $\Delta S_{PP}$ and $\Delta S_{PV}$ can give a handle on the
chirality of the NP-generated operators. NP effects in electroweak
penguin contributions are in general correlated with effects in $b \to
s \ell^+ \ell^-$, in $b \to s \gamma$ and possibly in $Z \to b \bar
b$. Depending on the flavor structure of NP, other effects might be
seen in $K \to \pi \nu \bar \nu$ or in
$\varepsilon^\prime/\varepsilon$. NP effects in the chromomagnetic
penguin might also show up in $b \to s \gamma$, in $B_s - \bar B_s$
mixing and, if there is a correlation between the $B$ and $K$ sectors,
in $\varepsilon^\prime/\varepsilon$.

\section{CONCLUSIONS AND OUTLOOK}
\label{sec:concl}

We have reviewed the theoretical status of hadronic $b \to s$ penguin
decays. We have shown that, in spite of the theoretical difficulties
in the evaluation of hadronic matrix elements, in the SM it is
possible to obtain sound theoretical predictions for the coefficient
$S_F$ of time-dependent CP asymmetries, using either models of
hadronic dynamics or data-driven approaches. Experimental data show an
interesting trend of deviations from the SM predictions that
definitely deserves further theoretical and experimental
investigation.

From the point of view of NP, the recent improvements in the
experimental study of other $b \to s$ processes such as $B_s - \bar
B_s$ transitions or $b \to s \gamma$ and $b \to s \ell^+ \ell^-$ have
considerably restricted the NP parameter space. However, there are
still several NP models, in particular SUSY with new sources of $b \to
s$ mixing in squark mass matrices, that can produce deviations from
the SM in the ballpark of experimental values. In any given model, the
study of hadronic $b \to s$ penguins and of the correlation with other
FCNC processes in $B$ and $K$ physics is a very powerful tool to
unravel the flavor structure of NP. 

Any NP model with new sources of CP violation and new particles within
the mass reach of the LHC can potentially produce sizable deviations
from the SM in $b \to s$ penguins. It will be exciting to
combine the direct information from the LHC and the indirect one from
flavor physics to identify the physics beyond the SM that has been
hiding behind the corner for the last decades. In this respect, future
facilities for $B$ physics will provide us with an invaluable tool to
study the origin of fermion masses and of flavor symmetry breaking,
two aspects of elementary particle physics that remain obscure in
spite of the theoretical and experimental efforts in flavor physics.

\section*{Acknowledgments}
I am grateful to M. Ciuchini, E. Franco and M. Pierini for carefully
reading this manuscript and for useful discussions. I acknowledge
partial support from RTN European contracts MRTN-CT-2004-503369 ``The
Quest for Unification'', MRTN-CT-2006-035482 ``FLAVIAnet'' and
MRTN-CT-2006-035505 ``Heptools''.

\end{document}